\documentclass[a4paper,fleqn,usenatbib,natbib]{mnras}
\usepackage{physics}
\usepackage{xspace}
\usepackage{psfrag}
\usepackage{epsfig}
\usepackage{amssymb}
\usepackage{amsmath}
\usepackage{newtxtext,newtxmath}

\newcommand{\be}{\begin{equation}}
\newcommand{\e}{\end{equation}}
\newcommand{\bear}{\begin{eqnarray}}
\newcommand{\ear}{\end{eqnarray}}

\def\xi{x^{i}_{\ion{H}{i}}\,}
\def\xh1{x_{\ion{H}{i}\,}}
\def\xb{\bar{x}_{\ion{H}{i}}}
\def\Ph1{P_{\ion{H}{i}}}
\def\Bh1{B_{\ion{H}{i}}}
\def\eh1{\eta_{\ion{H}{i}}}

\def\HI{\ion{H}{i}\xspace}
\def\HII{\ion{H}{ii}\xspace}
\def\P{\hat{P}}

\begin{document}
\author[Majumdar et al.]{Suman
  Majumdar$^{1}$\thanks{s.majumdar@imperial.ac.uk}, Jonathan
  R. Pritchard$^{1}$, Rajesh Mondal$^{2, 3, 4}$, Catherine
  A. Watkinson$^{1}$, \newauthor Somnath Bharadwaj$^{4}$, Garrelt
  Mellema$^{5}$ \\ $^{1}$Department of Physics, Blackett Laboratory,
  Imperial College, London SW7 2AZ, U.\ K. \\ $^{2}$Astronomy Centre,
  Department of Physics and Astronomy, University of Sussex, Brighton,
  BN19QH, UK \\ $^{3}$National Centre for Radio Astrophysics, Tata
  Institute of Fundamental Research, Post Bag 3, Ganeshkhind, Pune
  411007, India\\ $^{4}$Department of Physics and Centre for
  Theoretical Studies, Indian Institute of Technology Kharagpur,
  Kharagpur - 721 302, India\\ $^{5}$Department of Astronomy and Oskar
  Klein Centre, Stockholm University, AlbaNova, SE-10691 Stockholm,
  Sweden \\}

\title[EoR 21-cm bispectrum]{Quantifying the non-Gaussianity in the EoR 21-cm signal through bispectrum}
\date{Accepted 2018 February 23. Received 2018 February 05; in original form 2017 August 27}

\maketitle

\begin{abstract}
The epoch of reionization (EoR) 21-cm signal is expected to be highly
non-Gaussian in nature and this non-Gaussianity is also expected to
evolve with the progressing state of reionization. Therefore the
signal will be correlated between different Fourier modes ($k$). The
power spectrum will not be able capture this correlation in the
signal. We use a higher-order estimator -- the bispectrum -- to
quantify this evolving non-Gaussianity. We study the bispectrum using
an ensemble of simulated 21-cm signal and with a large variety of $k$
triangles. We observe two competing sources driving the
non-Gaussianity in the signal: fluctuations in the neutral fraction
($\xh1$) field and fluctuations in the matter density field. We find
that the non-Gaussian contribution from these two sources vary,
depending on the stage of reionization and on which $k$ modes are
being studied. We show that the sign of the bispectrum works as a
unique marker to identify which among these two components is driving
the non-Gaussianity. We propose that the sign change in the
bispectrum, when plotted as a function of triangle configuration
$\cos{\theta}$ and at a certain stage of the EoR can be used as a
confirmative test for the detection of the 21-cm signal. We also
propose a new consolidated way to visualize the signal evolution (with
evolving $\xb$ or redshift), through the trajectories of the signal in
a power spectrum and equilateral bispectrum i.e. $P(k)-B(k, k, k)$
space.

\end{abstract}

\begin{keywords}
	cosmology:dark ages, reionization, first stars---methods: numerical
\end{keywords}

\section{Introduction}
\label{sec:intro}
The epoch of reionization (EoR) is one of the least known periods in
the history of our universe. This is the era when the first sources of
light were formed and the high energy UV and X-ray radiation from
these and subsequent population of sources gradually changed the state
of most of the hydrogen in the intergalactic medium (IGM) from neutral
(\HI) to ionized (\HII) (see e.g. \citealt{fan06b, furlanetto06,
  pritchard08,pritchard12,choudhury09} for reviews). Our current
understanding of this epoch is mainly guided by the observations of
the cosmic microwave background radiation (CMBR)
\citep{komatsu11,planck16}, the absorption spectra of high redshift
quasars \citep{becker01,fan03,white03,goto11,becker15,barnett17} and
the luminosity function and clustering properties of Lyman-$\alpha$
emitters \citep{trenti10, ouchi10, jensen13b, choudhury14, bouwens16,
  ota17,zheng17}. These observations suggest that reionization was an
extended process, spanning over the redshift range $6 \lesssim z
\lesssim 15$ (see
e.g. \citealt{alvarez06,mitra13,mitra15,robertson15,bouwens15}
etc.). However, these indirect observations do not resolve many
fundamental questions regarding the EoR, such as the precise duration
and timing of reionization, the properties of major ionizing sources,
and the typical size distribution of the ionized bubbles at different
stages of the EoR etc.

Observations of the redshifted 21-cm line, originating from spin-flip
transitions in neutral hydrogen atoms, could be the key for resolving
many of these long standing issues. The brightness temperature or the
specific intensity of the redshifted 21-cm line directly probes the
\HI distribution at the epoch where the radiation
originated. Observing this line enables us, in principle, to track the
distribution and the state of the hydrogen during the entire
reionization history as it progresses with time or decreasing
redshift.

Motivated by this, a significant number of low frequency radio
interferometers, such as the GMRT \citep{paciga13}, LOFAR
\citep{yatawatta13,haarlem13,jelic14}, MWA \citep{tingay13,bowman13},
PAPER \citep{parsons14} and 21CMA \citep{wang2013}, are attempting to
detect the 21-cm signal from the EoR. These observations are
complicated to a large degree by the presence of foreground emissions,
which can be $\sim\!4-5$ orders of magnitude stronger than the
expected signal (e.g.\ \citealt{dimatteo02,ali08,jelic08,ghosh12}
etc.), and system noise \citep{morales05,mcquinn06}. So far these
first generation interferometers are able to put only weak upper
limits on the expected 21-cm signal at large length scales
\citep{paciga13,dillon14,parsons14,ali15,patil17}.

First generation instruments will probably not be able to directly
image the \HI distribution during this epoch due to their low
sensitivity. Imaging will possibly have to wait for the arrival of the
extremely sensitive (due to its large collecting area) next generation
of telescopes such as the SKA1-LOW
\citep{mellema13,mellema15,koopmans15}. The first generation
interferometers are instead expected to detect and characterize the
signal through statistical estimators such as the variance
(e.g. \citealt{patil14, watkinson14, watkinson15}) and the power
spectrum (e.g. \citealt{pober14, patil17}). The spherically-averaged
power spectrum achieves a higher signal to noise ratio by averaging
the signal over spherical shells in Fourier space, while still
preserving many important features of the signal
(e.g. \citealt{bharadwaj04, barkana05, datta07a, mcquinn07,
  mesinger07, lidz08, choudhury09b, mao12, majumdar13, majumdar16b,
  jensen13}). Thus, in principle it can be used for the EoR parameter
estimation with the upcoming SKA1-LOW
\citep{greig15,greig17,greig15b,koopmans15}.

However, even in case of a claimed detection of the signal through
power spectrum using these first generation interferometers, it will
still be difficult to confirm with absolute certainty that the
measured power spectrum arises from the signal alone and there is no
residual foreground or noise in it. Furthermore, the
spherically-averaged power spectrum can describe a field completely
only when it is a Gaussian random field (i.e. the signal in different
Fourier modes is uncorrelated). This assumption can be true for the
EoR 21-cm signal at sufficiently large length scales during the very
early stages of reionization{\footnote{Assuming that the contribution
    due to the spin temperature fluctuations is very low to the
    signal. If not, it can potentially make the signal significantly
    correlated between different Fourier modes at even these stages
    (see e.g. \citealt{furlanetto06a, pritchard07, fialkov14,
      fialkov15, ghara14, ghara15, watkinson15} etc).}}, when the
ionized regions are significantly small in size. As reionization
progresses, the fluctuations in the redshifted 21-cm signal get
dominated by the fluctuations due to the distribution of ionized
regions, which are gradually growing in size. This makes the EoR 21-cm
signal highly non-Gaussian during the intermediate and the later
stages of reionization. This non-Gaussianity cannot be captured by the
power spectrum of the signal but the error (cosmic) covariance of the
signal power spectrum is significantly effected
\citep{mondal15,mondal16,mondal17}.

To quantify this highly non-Gaussian 21-cm signal one would require
statistics which are of higher order compared to the variance and the
power spectrum. For one-point statistics, the next order estimator of
the signal after variance is skewness. It quantifies the degree of
non-Gaussianity present in the signal at a certain length scale (most
of the cases at the resolution limit of the specific instrument or the
simulation) (see e.g. \citealt{harker09, watkinson14, watkinson15,
  shimabukuro15a, kubota16} etc). While skewness would be able to
capture some broad non-Gaussian features of the signal, being an one
point statistic, it will not be able to quantify the correlation of
the signal between different Fourier modes.

The bispectrum, the Fourier equivalent of the three point correlation
function (estimated for a set of wave numbers which form a closed
triangle in the Fourier space) will be able to quantify the
correlation of the signal between different Fourier modes. One can, in
principle, estimate the bispectrum by calculating the three-visibility
(the basic observable quantity for a radio interferometer)
correlations from a radio interferometric observation of the EoR
\citep{bharadwaj05a, ali06}, similar to the way one can estimate the
power spectrum using the two-visibility correlations (see
e.g. \citealt{bharadwaj01a, bharadwaj05}). It is apparent that a
successful dectection of the three-visibility correlations of the
signal will require more sensitivity compared to it's two-visibility
correlations. Thus understanding the characteristics and evolution of
the signal bispectrum is more relevant at a time when there is a
significant amount of activity going on for building the highly
sensitive next generation radio interferometers (e.g. SKA1-LOW and
HERA \citep{pober14,ewallwice14}). Additionally, measurement of the
signal bispectrum using these future experiments can be treated as a
confirmative detection of the EoR 21-cm signal, which could be
otherwise rather ambiguous in case of a detection through power
spectrum.

Recently, there has been some effort to understand the characteristics
of the EoR 21-cm bispectrum using analytical models
\citep{bharadwaj05a} and semi-numerical simulations
\citep{yoshiura14,shimabukuro16} of the signal. It has also been
proposed that one can possibly constrain the EoR parameters by
studying the evolution of the bispectrum for a specific triangle
configuration \citep{shimabukuro16b}. Though most of these studies,
based on simulations, highlight some broad features of the EoR 21-cm
bispectrum, they lack in providing enough physical interpretation for
the behaviour of the signal bispectrum. It is also worthwhile noting
that, due to the specific definition of the bispectrum estimator used
by \citet{yoshiura14,shimabukuro16} and \citet{shimabukuro16b}, their
estimator is unable to capture the sign of the bispectrum, which is
expected to be an important feature of this statistic
\citep{bharadwaj05a}. In this paper we study the bispectrum of the EoR
21-cm signal using a semi-numerical simulation. We mainly focus on
finding unique signatures in the signal bispectrum at different stages
of reionization and also try to provide some physical interpretation
of it's behaviour using a simple toy model. We also explore various
configurations of wave number triangles that may capture different
unique characteristics of the signal.

The structure of this paper is as follows. In Section
\ref{sec:bispec_th}, we describe the algorithm that we have adopted
here to estimate bispectrum from the simulated signal. Section
\ref{sec:sim} briefly describes our simulation method to generate
variaous realizations of the reionization scenario that we have used
as our mock 21-cm data set. In Section \ref{sec:model}, we present a
simple toy model for interpreting the different observed features in
the bispectrum. Section \ref{sec:results} describes our estimated
21-cm bispectrum for various triangle configurations and the various
components that contribute to the signal and their associated
interpretations. Finally, in Section \ref{sec:summary} we summarise
our findings.

Throughout this paper, we have used the Planck+WP best fit values of
cosmological parameters $h = 0.6704$, $\Omega_{\mathrm{m}} = 0.3183$,
$\Omega_{\Lambda} = 0.6817$, $\Omega_{\mathrm{b}} h^2 = 0.022032$,
$\sigma_8 = 0.8347$ and $n_s = 0.9619$ \citep{planck14}.

\section{Bispectrum Estimator}
\label{sec:bispec_th}
\begin{figure}
  \psfrag{th}[c][c][1][0]{\large{$\theta$}}
  \psfrag{k1}[c][c][1][0]{\large{${\bf k}_1$}}
  \psfrag{k2}[c][c][1][0]{\large{${\bf k}_2$}}
  \psfrag{k3}[c][c][1][0]{\large{${\bf k}_3$}}
\includegraphics[width=.47\textwidth,angle=0]{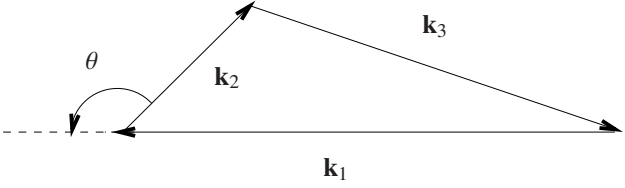}
\caption{Generalized closed triangle configuration in ${\bf k}$ space
  that has been used for bispectrum estimation. This also shows the
  definition of the angle $\theta$ that we have used throughout this
  paper.}
\label{fig:triangle1}
\end{figure}
The 21-cm signal from the epoch of reionization is quantified through
it's brightness temperature fluctuations $ \delta T_{{\rm b}}({\bf x})
= T_{{\rm b}}({\bf x}) - \overline{T}_{{\rm b}}$, where
$\overline{T}_{{\rm b}}$ is the mean brightness temperature at a
specific redshift $z$. It is convinient to use Fourier representation
for the analysis of this paper, as the Fourier transform of the
brightness temperature $\Delta_{{\rm b}}({\bf k})$, contributes to the
observed visibilities in a radio interferometric observation. In this
paper we want to study the non-Gaussian properties of $\Delta_{{\rm
    b}}({\bf k})$ through bispectrum.

We define the bispectrum $B_{{\rm b}}({\bf k}_1, {\bf k}_2, {\bf k}_3)$ of the
21-cm brightness temperature field as,
\begin{equation}
 \langle \Delta_{{\rm b}}({\bf k}_1) \Delta_{{\rm b}}({\bf k}_2)
 \Delta_{{\rm b}}({\bf k}_3)\rangle = V \delta^{{\rm K}}_{{\bf k}_1 +{\bf k}_2 +{\bf k}_3, 0}\, B_{{\rm b}}({\bf k}_1, {\bf k}_2, {\bf k}_3) \, ,
\end{equation}
where $\delta^{{\rm K}}_{{\bf k}_1 +{\bf k}_2 +{\bf k}_3, 0}$ is the
Kroneker delta function in the discrete Fourier space and is $1$ if
${\bf k}_1 +{\bf k}_2 +{\bf k}_3 = 0$ and $0$ otherwise. This ensures
that only those ${\bf k}$ triplets contribute to the bispectrum which
forms a closed triangle (see Figure \ref{fig:triangle1}) in the
Fourier space. The angular brackets represent an ensemble
average. Note that all of the above equations are written for a fixed
redshift $z$. Here we do not consider the redshift evolution of the
signal along the line-of-sight direction \citep{barkana06, datta12,
  datta14, zawada14, laplante14, ghara15, mondal17b} of a comoving
observed or simulated volume $V$. For brevity we drop the subscript
``b'' when describing the brightness temperature from this point
onwards.

The binned bispectrum estimator that one can use to compute this
quantity from the observed data or simulations, can be defined for the
$m^{\rm th}$ triangle configuration bin as
\begin{equation}
  \hat{B}_{m}({\bf k}_1, {\bf k}_2, {\bf k}_3) = \frac{1}{N_{{\rm tri}}V} \sum_{[{\bf k}_1 +{\bf k}_2 +{\bf k}_3 =0] \in m }\Delta ({\bf k}_1) \Delta ({\bf k}_2) \Delta ({\bf k}_3) \, ,
\label{eq:bispec_est}
\end{equation}
where $N_{{\rm tri}}$ is the number of closed triangles contributing
to the $m^{\rm th}$ triangle bin for which we estimate the
bispectrum. An important property of any polyspectrum of a real field
is that the polyspectrum will also be a real quantity. In this paper
we are interested in the bispectrum of 21-cm signal, which is a real
field, so the bispectrum of 21-cm signal will also be real. This
property of the bispectrum can be understood in the following manner:
For every closed triangle, consisting of vectors ${\bf k}_1,\, {\bf
  k}_2\,{\rm and}\, {\bf k}_3$ (triangle A), one can construct another
closed triangle having vectors $-{\bf k}_1,\, -{\bf k}_2\,{\rm and}\,
-{\bf k}_3$ (triangle B). It is very straight forward to show that
these two triangle configurations are essentially same and will
therefore measure the same bispectrum and will contribute to the same
bispectrum triangle bin. As the Fourier transform of a real field is
Hermitian in nature [i.e. $\Delta^{\dagger}({\bf k}) = \Delta (-{\bf
    k})$], so only half of the Fourier space will contain unique
information about the field and the other half can be created by using
the Hermitian property. Using this Hermitian property of the field one
can show that the bispectrum estimate for triangle A (or the
corresponding product of three $\Delta$s) will have a real component
which is identical in both sign and amplitude to that of the triangle
B. Whereas the bispectrum estimate for triangle A will have an
imaginary component which is identical in amplitude but opposite in
sign to that of the triangle B, thus they will exactly cancel each
other out. Therefore any binned estimate of the bispectrum of 21-cm
signal will always be real.

One can directly implement Equation \eqref{eq:bispec_est} on the
Fourier transform of the simulated brightness temperature data cube to
estimate the bispectrum. We consider a data cube of Fourier
transformed brightness temperature having $N_{{\rm G}}^3/2$ grid
points{\footnote{The Hermitian properties of the field implies that
    only half of the Fourier space have unique information about the
    field. Thus the actual number of grid points where the data is
    unique in the Fourier space would be $N_{{\rm G}}^3/2$ instead of
    $N_{{\rm G}}^3$. Thus two axes of ${\bf k}$ vector space will have
    $N_{{\rm G}}$ number of intervals and one would have $N_{{\rm
        G}}/2$ intervals.}}. If ${\bf k}$ vectors are three
dimensional (which is the case here), to estimate all possible
bispectra, one would expect to go through nine nested {\it for} loops
in the discretized Fourier space. However, the Kroneker delta function
in the estimator introduces a vector equation of constraint ${\bf k}_1
+{\bf k}_2 +{\bf k}_3 = 0$ for the closure of the triangle, which
removes three nested {\it for} loops from the algorithm. Still six
nested loops are computationally very expensive to execute. To reduce
the computation time further we introduce two constraints on ${\bf
  k}_1$ and ${\bf k}_2$. For a specific kind of triangle
configuration, the ratio between the two arms of the triangle must
remain constant, {\it i.e.}
\begin{equation}
k_2/k_1 = n\,,
\label{eq:ratio}
\end{equation}
and cosine of the angle ($\theta$, see Figure \ref{fig:triangle1} for
it's definition) between the two arms of the triangle also remains
fixed to
\begin{equation}
\frac{{\bf k}_1 . {\bf k}_2}{k_1 k_2} = \cos\theta\, .
\label{eq:alpha}
\end{equation}
This reduces the total number of steps in the algorithm to $N_{{\rm
    G}}^4/2$ from $N_{{\rm G}}^6/4$.

The four nested {\it for} loops in this algorithm determines all
possible combinations of three components of ${\bf k}_1$
vector{\footnote{This implies, for a specific triangle type we
    consider all possible orientations of the ${\bf k}_1$ vector (and
    subsequent orientations of the other two ${\bf k}$s) in the
    $k$-space and thus the estimated bispectrum is essentially
    spherically averaged in $k$-space.}} and one component of ${\bf
  k}_2$ vector. The other two components of ${\bf k}_2$ vector are
determined by Equations \eqref{eq:ratio} and \eqref{eq:alpha}. All
three components of ${\bf k}_3$ vector are determined using the closer
of the triangle condition {\it i.e.} ${\bf k}_1 +{\bf k}_2 +{\bf k}_3
= 0$. Once all components of ${\bf k}_1$, ${\bf k}_2$ and ${\bf k}_3$
are determined, we take the product of three $\Delta({\bf k})$
corresponding to these three vectors, which will give us a complex
number, as all $\Delta({\bf k})$s are complex.

This particular approach to estimate bispectrum is very restrictive in
nature. It does not allow any arbitrary bin width around ${\bf k}_2$
and ${\bf k}_3$ vectors. One can put a precise bin width around the
${\bf k}_1$ vector, but for a specific set of components of ${\bf
  k}_1$ and one component of ${\bf k}_2$, other two components of
${\bf k}_2$ and all three components of ${\bf k}_3$ are determined
precisely by Equations \eqref{eq:ratio}, \eqref{eq:alpha} and the
Kroneker delta function in \eqref{eq:bispec_est}. These equations of
constraints do not allow any variation in the two components of ${\bf
  k}_2$ and all three components of ${\bf k}_3$. This reduces the
number of triangles contributing to every triangle bin, which affects
more severely the triangle bins containing small ${\bf k}$ modes. One
should thus be aware of the effect of the sample variance, which will
be more significant at triangle bins containing small ${\bf k}$ modes,
while interpreting the bispectrum estimated using this method.

\section{Simulating the redshifted 21-cm signal from the EoR}
\label{sec:sim}
\begin{figure}
\includegraphics[width=.47\textwidth,angle=0]{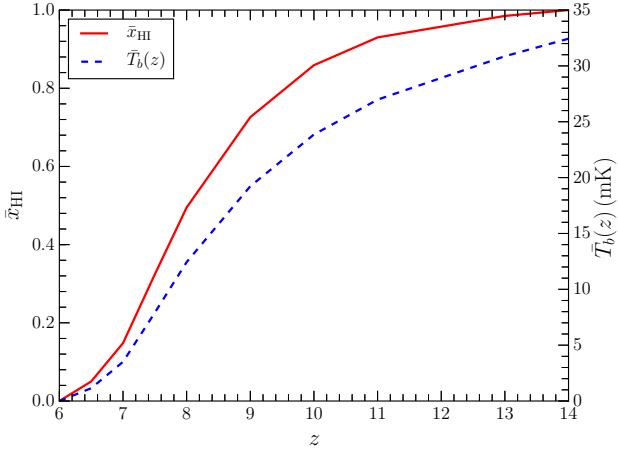}
\caption{This shows the reionization history obtained from our
  simulations through the redshift evolution of average neutral
  fraction and average brightness temperature of the 21-cm signal.}
\label{fig:reion_his}
\end{figure}
\begin{figure*}
\includegraphics[width=1.00\textwidth,angle=0]{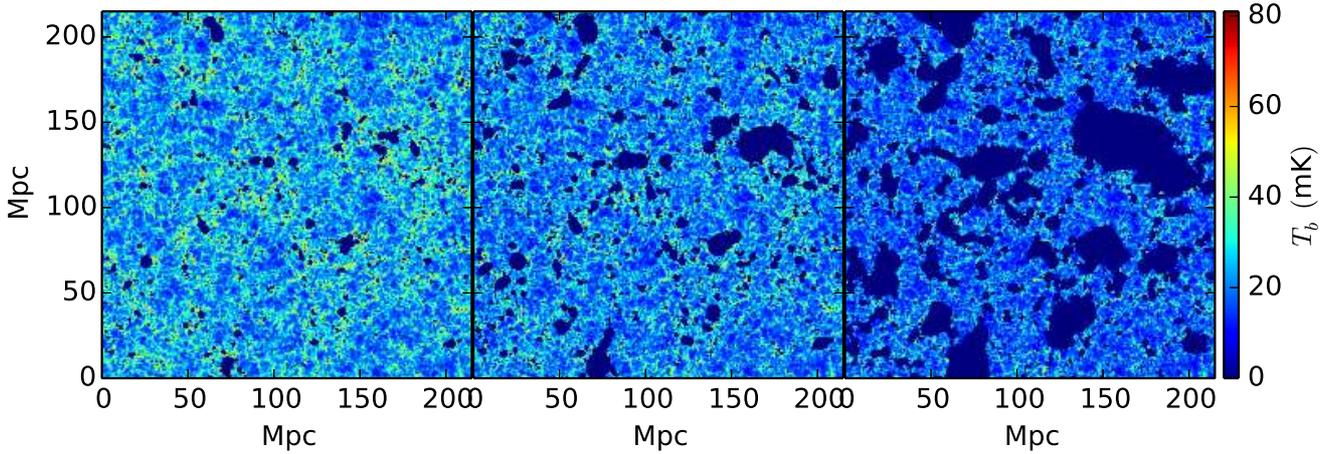}
\caption{One realization of the real space 21-cm map at three
  representative stages of the reionization, $\xb = 0.86,\, 0.73\,
  {\rm and}\, 0.49$ (from left to right), respectively.}
\label{fig:21_cm_maps}
\end{figure*}
We briefly describe here our method for simulating the redshifted
21-cm signal, which is identical to the approach of
\citet{mondal17}. The three main steps in this method are -- a)
generating the matter distribution at different redshifts using a
particle-mesh $N$-body code \citep{bharadwaj04b, mondal15}; b)
identifying collapsed halos in that matter density field using a
Friends-of-Friends (FoF) algorithm \citep{davis85}; and finally c)
generating an ionization field from the matter and halo distribution
following an algorithm similar to the excursion set formalism of
\citet{furlanetto04b}. Finally these ionization fields are converted
into the redshifted 21-cm brightness temperature fields at their
respective redshifts.

Our $N$-body simulation have a comoving volume of $V = [{\rm 215\,
  Mpc}]^3$ with $3072^3$ grids of spacing $0.07$ Mpc and a mass
resolution of $1.09 \times 10^8 {\rm M_{\odot}}$. To identify halos
from this matter distribution, we have set the criteria that a halo
should have at least 10 dark matter particles (which implies a minimum
halo mass of $M_{{\rm min}} = 1.09 \times 10^9 {\rm M_{\odot}}$) and
used a fixed linking length of $0.2$ times the mean interparticle
distance in our FoF halo finder.

We assume that the hydrogen traces the dark matter distribution. We
also assume that the collapsed halos host the sources of ionizing
photons and the number of ionizing photons emitted by these sources
are proportional to their host halo mass with a constant of
proportionality $N_{{\rm ion}}$, which is a dimensionless
parameter. The $N_{{\rm ion}}$ is a combination of several unknown
degenerate reionization parameters, such as, the star formation
efficiency of the first galaxies, their UV photon production
efficiency and escape fraction of UV photons from them. To implement
the excursion set formalism, we first map the matter and the ionizing
photon density fields to a $384^3$ grid with spacing $0.56$ Mpc. In
this coarser grid (compared to the $N$-body grid), whether a grid
point is neutral or ionized at a certain stage of reionization is
determined by smoothing the hydrogen density and the ionizing photon
density fields using spheres of different radii starting from a
minimum radius of $R_{{\rm min}}$ (the coarse grid spacing) to
$R_{{\rm mfp}}$. The $R_{{\rm mfp}}$ is another free parameter of our
simulation, which represents the mean free path of the ionizing
photons. A specific grid point in the simulation box is considered to
be ionized if for any smoothing radius $R$ ($R_{{\rm min}} \leq R \leq
R_{{\rm mfp}}$) the photon density exceeds the neutral hydrogen
density at that grid point. This simulated ionization map or the
corresponding neutral hydrogen density map is then converted into the
21-cm brightness temperature map. Our method of simulating the
redshifted 21-cm signal is similar to that of \citet{choudhury09b} and
\citet{majumdar14}. Note that we do not include redshift space
distortions caused by peculiar velocities in our simulations, which in
principle can have a significant effect on any estimator of the
redshifted 21-cm signal from the EoR \citep{bharadwaj04, barkana05,
  mao12,majumdar13,jensen13,fialkov15,ghara15,majumdar16b,majumdar16}. We plan to
study this effect in a follow up work.

One can generate different reionization histories (i.e. mass averaged
neutral fraction $\xb$ as a function of $z$) by varying parameters
$N_{{\rm ion}}$ and $R_{{\rm mfp}}$. We use $R_{{\rm mfp}} = 20$ Mpc
for all redshifts which is consistent with the findings of
\citet{songaila10} from the study of Lyman limit systems at low
redshifts. We keep $N_{{\rm ion}} = 23.21$ fixed for all redshifts, so
that $\xb \approx 0.5$ at $z = 8$. It also ensures that reionization
ends at $z \sim 6$ and we obtain Thomson scattering optical depth
$\tau = 0.057$, which is consistent with $\tau = 0.058 \pm 0.012$
reported by \citet{planck16}. The resulting reionization history is
shown in Figure \ref{fig:reion_his}. We generate five statistically
independent realizations with the same reionization history to
quantify the effect of cosmic variance on the signal
bispectrum. Figure \ref{fig:21_cm_maps} shows a visual representation
of one slice of the signal cubes at three representative stages of
reionization.

\section{Modelling the redshifted 21-cm bispectrum from the EoR}
\label{sec:model}
It is convenient to consider a model for the \HI fluctuations to
interpret the behaviour of the bispectrum estimated from the simulated
redshifted 21-cm signal. The 21-cm brightness temperature fluctuations
from the EoR can be expressed as $\delta T_{{\rm b}}({\bf x},z) =
\overline{T}_{{\rm b}}(z) \eta_{\HI}({\bf x},z)$ (where
$\overline{T}_{{\rm b}}(z) = 4.0\, {\rm mK}\, [1+z]^2 [\Omega_{{\rm
      b}}h^2/0.02] [0.7/h][ H_0/H(z)]$) and considering the stage of
the EoR when the spin temperature is much larger than the CMBR
temperature ($T_S \gg T_{{\rm CMBR}}$) one has (similar to
\citealt{bharadwaj04,bharadwaj05,zaldarriaga04,barkana05} etc.):
\begin{equation}
  \eta_{\HI}({\bf x},z) = \frac{\rho_{\HI}({\bf
      x},z)}{\bar{\rho}_{H}(z)} = \xh1({\bf x},z) [1 + \delta({\bf
      x},z)] \, ,
  \label{eq:Tb_comp}
\end{equation}
where $\bar{\rho}_{H}(z)$ and $\rho_{\HI}({\bf x},z)$ are the mean
hydrogen density at redshift $z$ and neutral hydrogen density at
location ${\bf x}$ and redshift $z$, respectively. Further, assuming
that the hydrogen density fluctuations at large length scales to be
same as matter density fluctuations one can arrive at the final
expression of Equation \eqref{eq:Tb_comp}. One can further
express the neutral fraction as $\xh1 = \xb (1+ \delta_x)$, where
$\xb$ is the average neutral fraction and $\delta_x$ is the
fluctuation in the neutral fraction distribution. Note that we have
not considered an apparent enhancement in the fluctuation of the
signal due to the peculiar velocities of the matter particles. This
will introduce an additional term, $-[(1+z)/H(z)](\partial
v_{\parallel}/\partial r)$ (where $r$ is the comoving distance from
the observer and $v_{\parallel}$ is the line-of-sight component of the
peculiar velocity), at the right hand side of Equation
\eqref{eq:Tb_comp}. We plan to study the effect of peculiar velocity
on the bispectrum in a follow up work. Assuming that $\delta_x\,,
\delta \ll 1$, one can drop all quadratic and higher order terms
involving $\delta_x$ and $\delta$ to express the $\eta_{\HI}$ in
Fourier space as:
\begin{equation}
  \tilde{\eta}_{\HI}({\bf k},z) = \xb(z) [\Delta_x({\bf k},z) +
    \Delta({\bf k},z)]\,,
  \label{eq:eta}
\end{equation}
where $\tilde{\eta}_{\HI}$, $\Delta$ and $\Delta_x$ are the Fourier
transform of $\eta_{\HI}$, $\delta$ and $\delta_x$ respectively.

Following Equation \eqref{eq:eta} one can write the \HI bispectrum (of
$\eta_{\HI}$) as:
\begin{align}
  \Bh1({\bf k}_1, {\bf k}_2, {\bf k}_3) &= B_{\Delta \Delta \Delta} + B_{x x x} +
  B_{x \Delta \Delta} + B_{\Delta x \Delta} + B_{\Delta \Delta x}
  \nonumber \\ &+ B_{x x \Delta} + B_{x \Delta x} + B_{\Delta x x}
  \label{eq:B_h1}
\end{align}
where the first, second and third subscript in each component
bispectrum on the right hand side correspond to the ${\bf k}_1,\, {\bf
  k}_2$ and ${\bf k}_3$ vector arms of a closed triangle in the
Fourier space. We will use this model of the \HI bispectrum (built
using the linear approximation) to interpret and analyse the 21-cm
bispectrum estimated from our simulations.

\subsection{A toy model for \HI fluctuation during the EoR}
\label{sec:toy_model}
\begin{figure}
  \includegraphics[width=.47\textwidth,angle=0]{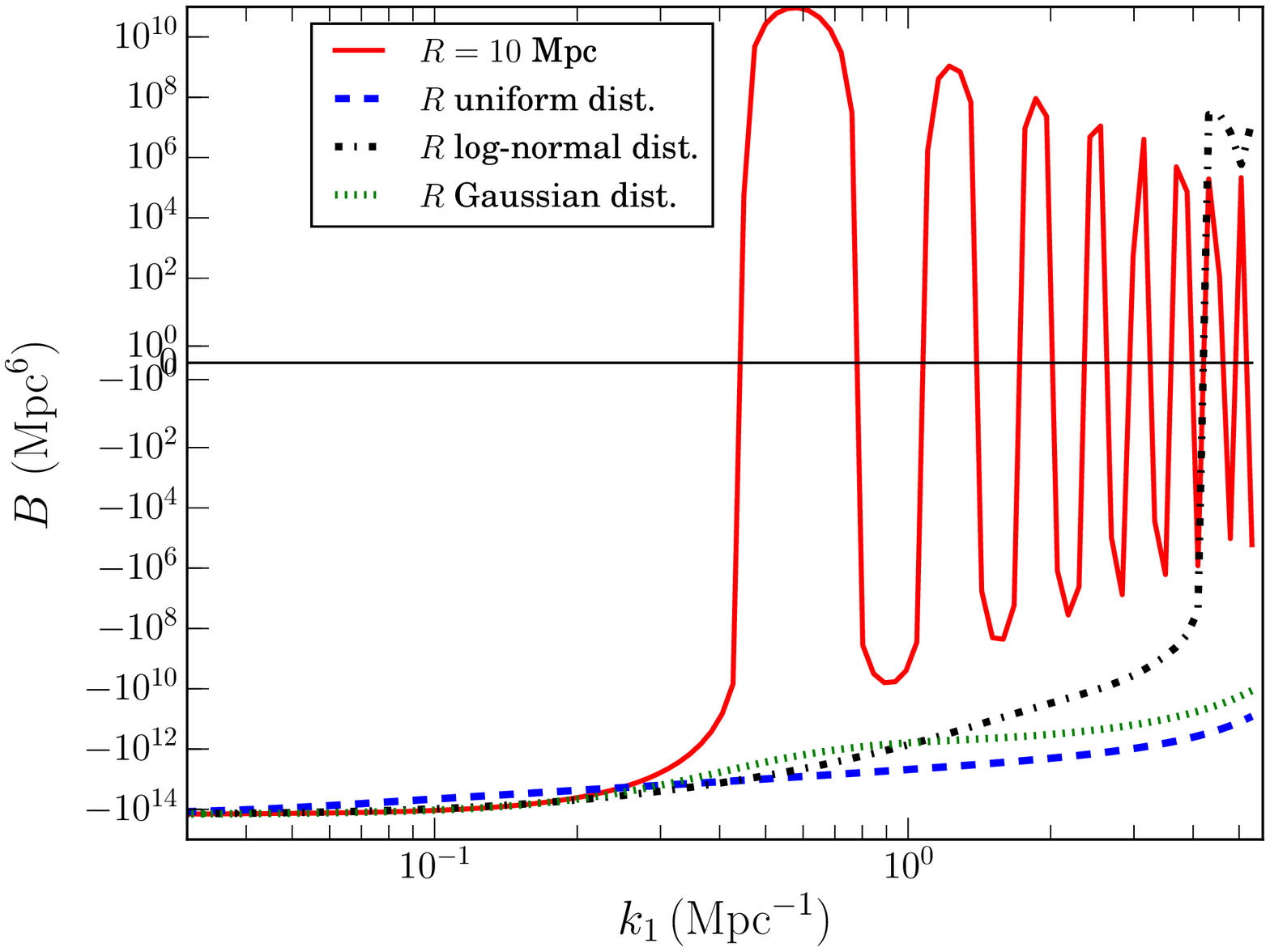}
  \includegraphics[width=.47\textwidth,angle=0]{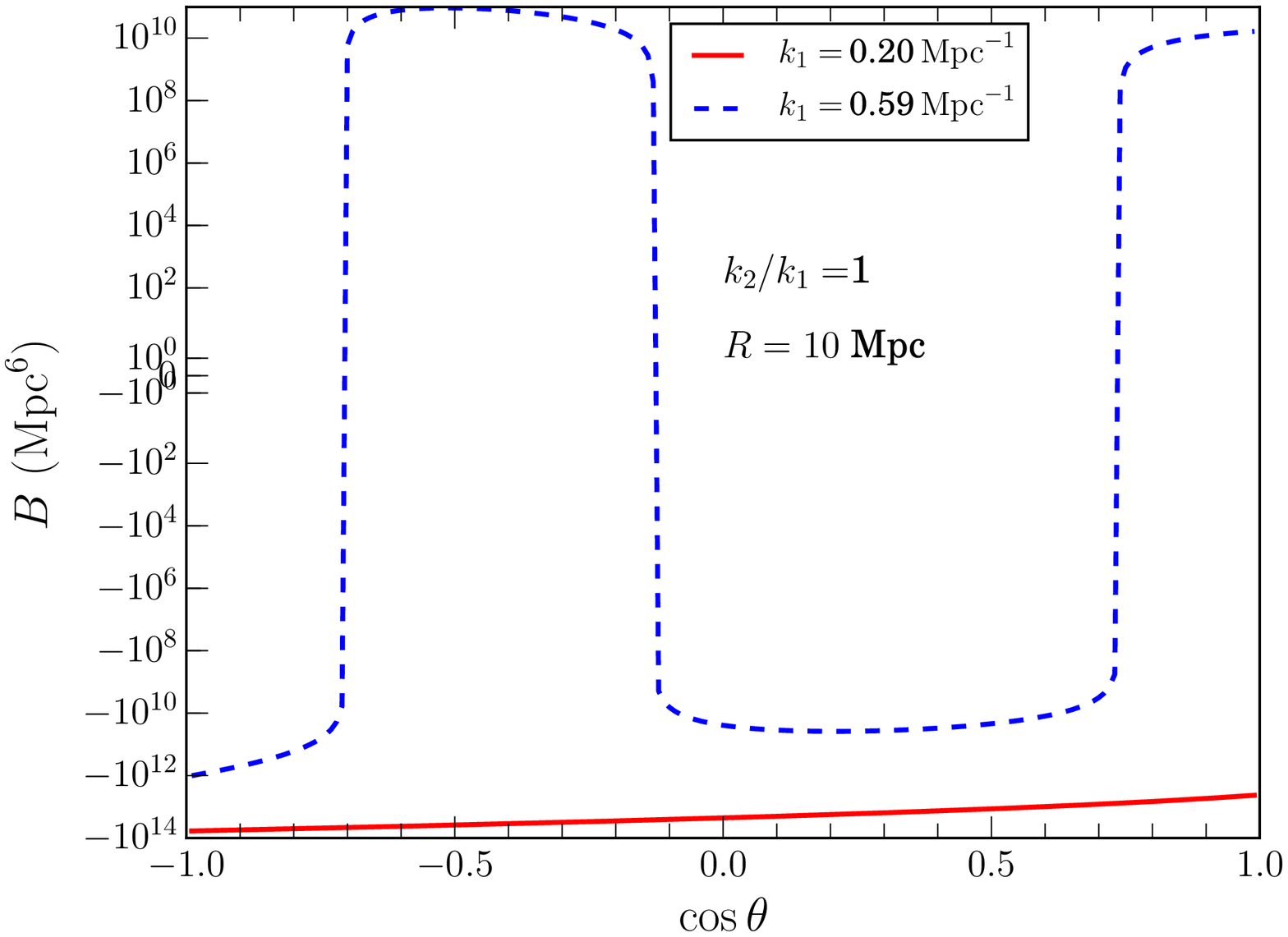}
  \includegraphics[width=.47\textwidth,angle=0]{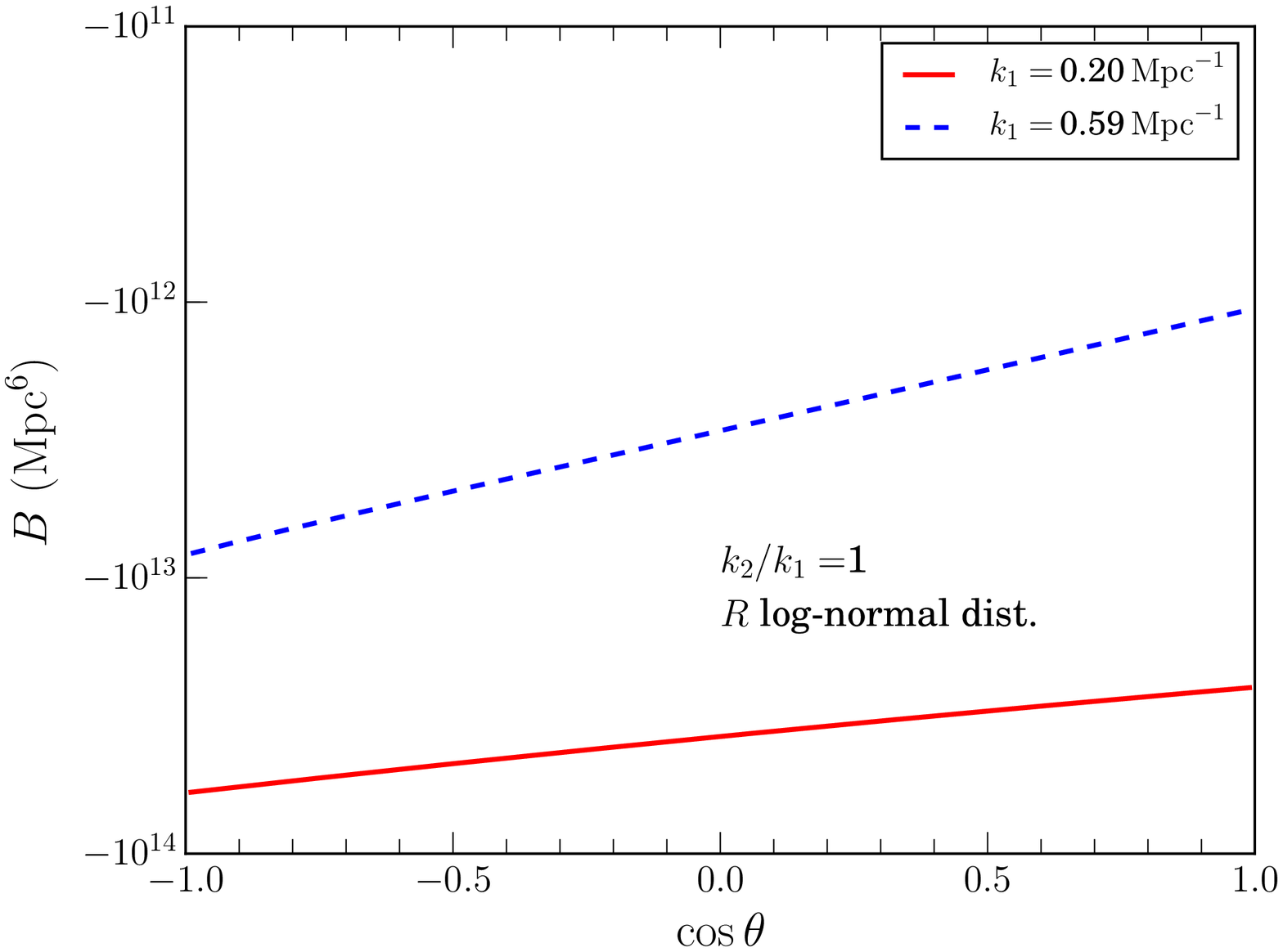}
\caption{{\bf Top panel:} Model bispectrum (in arbitrary units) for
  equilateral triangles as function of $k$. Solid, dashed, dash-dotted
  and dotted lines represent the model bispectra for a fixed bubbles
  radius $R = 10$ Mpc, uniformly distributed bubble radii in the range
  $0.56 \leq R \leq 25$ Mpc, log-normal bubble radii distribution with
  $\mu =2.3$ and $\sigma = 0.3$, Gaussian bubble radii distribution
  with $R_{{\rm mean}} = 10$ Mpc and $\sigma = 3.0$,
  respectively. {\bf Central panel:} Model bispectrum for isosceles
  triangles ($k_2/k_1 = 1$) as a function of $\cos{\theta}$. The model
  bispectra shown here is for a fixed bubbles radius $R = 10$. Solid
  and dashed lines represent bispectra for two different values of
  $k_1$ ($0.20$ and $0.59\,{\rm Mpc}^{-1}$, respectively). {\bf Bottom
    panel:} Same as the central panel but for the model with
  log-normal bubble radii distribution. Note that in all three panels,
  the y-axis is shown in log scale, using the {\it symlog} function of
  {\it matplotlib} in {\it python}, which is linear in between $-1$ to
  $1$ and log in the rest of the range.}
\label{fig:model_sig}
\end{figure}
The \HI fluctuation described by $\eta_{\HI}$ in Equation
\eqref{eq:Tb_comp} has contributions from two distinct components. One
is the gravitational clustering of the hydrogen which follows the
underlying dark matter distribution and the other is the spatial
fluctuations in the neutral fraction of the hydrogen, which is
determined by the size and spatial distribution of ionized regions at
a certain stage of reionization. As a rudimentary model of
reionization one can assume that the ionized regions at any stage of
the EoR are non-overlapping spheres of radius $R$ and their centres
are distributed randomly in the space. If the mean comoving number
density of such ionized spheres at a certain stage of reionization is
$\bar{n}_{\HI}$, then the mean ionization fraction at that stage will
be $x_{\rm i} = 1 - \xb = 4\pi R^3 \bar{n}_{\HI}/3$ and under this
model the Equation \eqref{eq:Tb_comp} can be rewritten as:
\begin{equation}
  \eta_{\HI}({\bf x},z) = [1 + \delta({\bf x},z)] \left[ 1 - \sum_{a} \theta \left( \frac{\mathopen|{\bf  x - x_a} \mathclose|}{R} \right) \right]\, ,
  \label{eq:Tb_comp_model1}
\end{equation}
where $a$ represents different ionized spheres with centres at ${\bf
  x_a}$ and $\theta (y)$ is the Heaviside step function defined such
that $\theta (y) = 1$ for $0 \leq y \leq 1$ and zero
otherwise. Several authors (e.g. \citealt{bharadwaj04,bharadwaj05} and
\citealt{zaldarriaga04} etc.) have shown that, under such a model, at
length scales larger than the typical ionized region size, the \HI
21-cm signal fluctuation is dominated by the fluctuations coming from
these individual ionized regions. If one assumes that the dark matter
density fluctuations ($\delta$) at high redshifts and large length
scales have negligible non-Gaussianity and thus their contribution to
the \HI bispectrum can be ignored, then one can rewrite the \HI
fluctuations for this purpose as:
\begin{equation}
  \eta_{\HI}({\bf x},z) = \left[ 1 - \sum_{a} \theta \left(
    \frac{\mathopen|{\bf x - x_a} \mathclose|}{R} \right) \right]\, .
  \label{eq:Tb_comp_model2}
\end{equation}

The Fourier transform of Equation \eqref{eq:Tb_comp_model2} for $k>0$
is
\begin{equation}
  \tilde{\eta}_{\HI}({\bf k},z) = -\frac{x_{\rm i} W(kR)}{\bar{n}_{\HI}}
  \sum_{a} e^{i{\bf k.x_a}}\,,
  \label{eq:eta_model}
\end{equation}
where $W(y) = (3/y^3)[\sin (y) - y \cos (y)]$ is the spherical top hat
window function. The \HI bispectrum for such a model can be then
expressed as \citep{bharadwaj05a}:
\begin{equation}
  \Bh1({\bf k}_1, {\bf k}_2, {\bf k}_3) = - \frac{x_{\rm i}^{3} W(k_1 R) W(k_2 R)  W(k_3 R)}{\bar{n}_{\HI}^2}\, .
  \label{eq:B_h1_model}
\end{equation}

As we plan to use this toy model of \HI bispectrum very extensively to
interpret our simulated 21-cm bispectra, thus we discuss some of the
important and relevant features of it in next few
paragraphs. \citet{bharadwaj05a} have discussed some of the
shortcomings of this model when considered in the context of a
realistic reionization scenario. First of all it assumes that the
ionized spheres do not overlap with each other. This can be true at
the very early stages of reionization when there are only a few
ionized regions centered around the first ionizing sources. As
reionization progresses, these ionized regions gradually grow in size
and start overlapping with each other. Thus this model for the
topology of ionization field is expected to break down during the
later stages of the EoR or in other words when the average ionization
fraction is relatively high ({\it i.e.}  $x_{\rm i} > 0.5$). Another
limitation of this model is that, it assumes that these ionized
regions are randomly distributed in space, which is probably also not
true as one would expect the ionized regions to be centered around the
ionizing sources and these sources are expected to be located
specifically in the matter density peaks. Thus one should expect a
gravitational clustering in the distribution of the centre of these
\HII regions.

The other limitation of the above model is that it assumes all ionized
regions at a specific stage of reionization to have the same radius
$R$. This implies that the \HI bispectrum [Equation
  \eqref{eq:B_h1_model}] for an equilateral triangle (i.e. $k_1 = k_2
= k_3$) will be an oscillatory function of $k_1$ with the first zero
crossing appearing at $k_1R \approx 4.49$. As an example, for spheres
of radius $R = 10$ Mpc this translates into first zero crossing to
appear at $k_1 \approx 0.45\, {\rm Mpc}^{-1}$. However, various
realistic simulations of reionization predict that the ionized regions
at a certain stage of reionization will be of different shapes and
volume (see e.g. \citealt{zahn07, friedrich11, majumdar14, iliev15}
etc.). Some recent studies (see e.g. \citealt{furlanetto16} and
\citealt{bag18}) even suggest that the ionized regions are not
spherical but rather filamentary in shape. Additionally, there is no
such characteristic `size' that can be assigned to the ionized regions
as they percolate with each other even at very early stages of the
EoR. As a first order improvement of the \citet{bharadwaj05a} model,
we propose a phenomenological model for the \HI bispectrum where we
assume that $B_{\HI}({\bf k}_1, {\bf k}_2, {\bf k}_3)\propto -
\sum_{a} W(k_1 R_a) W(k_2 R_a) W(k_3 R_a)$. Under this model $R_a$s
are drawn from a distribution rather than having a fixed value. Note
that, though this modified toy model is inspired by the model of
\citet{bharadwaj05a} [i.e. Equation \eqref{eq:B_h1_model}] but is not
a direct extension or generalization of the above.

Figure \ref{fig:model_sig} shows bispectra for a set of triangle
configurations for the two models of \HI fluctuations introduced
earlier. The top panel of this figure shows the \HI bispectra for
equilateral triangles as a function of $k_1$. The solid line
correspond to the model with a fixed bubble radius ($R = 10$ Mpc). As
expected the bispectrum for this model is an oscillatory function of
$k_1$ and its first zero corssing appears at $k_1 \approx 0.45\, {\rm
  Mpc}^{-1}$. The dashed, dash-dotted and dotted lines represent
models with an uniform (with $R$ in the range $0.56 \leq R \leq 25$
Mpc), log-normal (with parameters $\mu = 2.3$ and $\sigma =0.6$) and
Gaussian (with parameters $\mu = 10$ Mpc and $\sigma = 3.0$) bubble
size distributions. We have chosen the paramter values for the
log-normal and the Gaussian bubble size distributions such that both
of the distributions (corresponding histograms not shown here) peak
around $R \approx 10$ Mpc. In contrast with the bispectrum for fixed
$R$, the bispectra for models with a distribution in $R_a$ turn out to
be a power law like smooth functions of $k_1$. For the uniform and
Gaussian bubble distribution of $R$ we do not observe any zero
crossing in the bispectra. The bispectrum for the log-normal
distribution in $R$ shows a zero crossing at significantly large $k_1$
mode ($k_1 \sim 4.00\,{\rm Mpc}^{-1}$), possibly reflects the presence
of a large number of smaller $R$ values in this distribution. The
other main features of the bispectrum from this modified toy model,
some of which are same as the model with fixed bubble radius, are the
following: a) this bispectrum is negative; b) it remains almost
constant as a function of $k$, for a $k$ range corresponding to
significantly large length scales; c) the amplitude of the bispectrum
rapidly falls off and shows almost a power law like feature for $0.1 <
k \lesssim 4.00 \,{\rm Mpc}^{-1}$.

We expect some of these features of the model \HI bispectra that are
observed for the equilateral triangles to hold true for other triangle
configurations as well. To demonstrate it further, we show the
bispectrum for isosceles triangles ({\it i.e.}  $k_1 = k_2$) in the
central and bottom panels of Figure \ref{fig:model_sig} for models
with a fixed bubble radius and with a log-normal distribution in
bubble sizes, respectively. We choose two fixed $k_1$ values ($k_1 =
0.20$ and $0.59\,{\rm Mpc}^{-1}$) and plot the bispectrum as a
function of $\cos{\theta}$ [defined by Equation \eqref{eq:alpha}]. It
is apparent from Figure \ref{fig:triangle1} that $\cos{\theta} \sim
-1$ represents `squeezed' limit of triangles and $\cos{\theta} \sim 1$
represents `stretched' limit of triangles. Further, $\cos{\theta} =
-0.5$ in this plot correspond to equilateral triangles. For the \HI
fluctuation model with a fixed bubble radius (central panel of Figure
\ref{fig:model_sig}), with triangles consisting of sufficiently large
length scales ($k_1 = 0.20 \,{\rm Mpc}^{-1}$) the bispectrum (solid
red line) is negative and a smooth function of $\cos{\theta}$. It
attains maximum amplitude at the `squeezed' limit of triangles
($\cos{\theta} \sim -1$) and minima at the `stretched' limit
($\cos{\theta} \sim 1$). Note that the value of $k_3$ for this type of
triangles, in the entire $\cos{\theta}$ range stays below $k_3 \leq
0.40 \,{\rm Mpc}^{-1}$, which is smaller than the first zero crossing
$k$ mode for this \HI fluctuation model. This ensures that the model
bispectrum is a smooth function of $\cos{\theta}$ at large length
scales. However, the bispectrum for $k_1 = 0.59 \,{\rm Mpc}^{-1}$
(dashed blue line) becomes an oscillatory function of $\cos{\theta}$
as all three $k$ modes involved in this type of triangles are beyond
the first zero crossing limit. The bispectrum for our modified model
with a log-normal bubble size distrubution, turns out to be a smooth
function of $\cos{\theta}$ for both $k_1 = 0.20 \,{\rm Mpc}^{-1}$ and
$0.59 \,{\rm Mpc}^{-1}$ (bottom panel of Figure \ref{fig:model_sig}),
provided that the range of the bubble radii has been sampled
sufficiently enough that the oscillations due to different bubble
radii washes out each other. This model also shows a larger bispectrum
amplitude at smaller $k_1$ modes and for any fixed $k_1$ mode
bispectrum amplitude is maximum for $\cos{\theta} \sim -1$ and minimum
for $\cos{\theta} \sim 1$.

In the next few sections of this paper we will observe that the
behaviour of the bispectrum estimated from the simulated 21-cm signal
shows behaviour that are similar to this model \HI bispectrum for a
wide variety of triangles. We thus discuss and interpret the 21-cm
bispectrum estimated from our simulations in the light of this
particular modified model of \HI bispectrum.

\section{Results}
\label{sec:results}
The bispectrum estimator described in Section \ref{sec:bispec_th} has
been tested extensively for different ${\bf k}$-triangle
configurations using a matter density field (from a particle-mesh
$N$-body simulation) and a toy non-Gaussian field made of randomly
distributed non-overlapping ionized bubbles in a uniformly-dense
neutral medium in \citet{watkinson17}. \citet{watkinson17} has also
compared its performance against another estimator of bispectrum. We
thus do not repeat those tests in this paper and direct the interested
reader to \citet{watkinson17} for a detailed description of these
tests.

While studying the bispectrum for a highly non-Gaussian field such as
the EoR 21-cm signal, one can in principle consider a large variety of
triangle configurations using different values of the wave number
triplets ${\bf k}_1, {\bf k}_2,\,{\rm and}\, {\bf k}_3$. Bispectra for
different unique ${\bf k}$-triangle configurations can potentially
probe different unique non-Gaussian features present in the signal. We
can vary the ${\bf k}$-triangle configuration for our analysis using
the two equation of constraints [Equations \eqref{eq:ratio} and
  \eqref{eq:alpha}] of our algorithm. To keep the length of our
analysis within a reasonable limit, we choose a set of triangle
configurations which probes the different extremes of the possible
triangle configurations. Following Equation \eqref{eq:ratio}, we
choose four different values for the ratio $k_2/k_1$, namely $n = 1,\,
2,\, 5,\,{\rm and}\, 10$. For each value of $n$ we vary $\cos{\theta}$
within the range $-1 \leq \cos{\theta} \leq 1$ with steps of $\Delta
\cos{\theta} = 0.01$. Compared to power spectrum, the binning for
bispectrum can be tricky, as instead of only one $k$ mode here we have
three $k$ modes associated with each bispectrum value. We bin our
bispectrum estimates in the following way: for a specific value of $n$
and $\cos{\theta}$ we divide the entire $k_1$ range ($k_{{\rm max}} =
\pi/[{\rm grid\, spacing}] = 5.61\,{\rm Mpc}^{-1}$ to $k_{{\rm min}} =
2\pi/[{\rm box\, size}] = 0.03\,{\rm Mpc}^{-1}$) into $15$ logarithmic
bins and label each bin with the value of $k_1$ at the centre of that
bin. Note that the effective mean value of $k_1$ vector in each such
bin can be different from central value of the bin. Throughout this
paper, unless otherwise stated, we show the bispectrum as a function
of $(k_1, n,\cos{\theta})$. Also note that for a specific triangle
type we consider all possible orientations of the ${\bf k}_1$ vector
(and subsequent orientations of the other two ${\bf k}$s) in the
$k$-space and thus all of the bispectra that we describe here are
essentially spherically averaged in $k$-space.

\begin{figure}
  \includegraphics[width=.47\textwidth,angle=0]{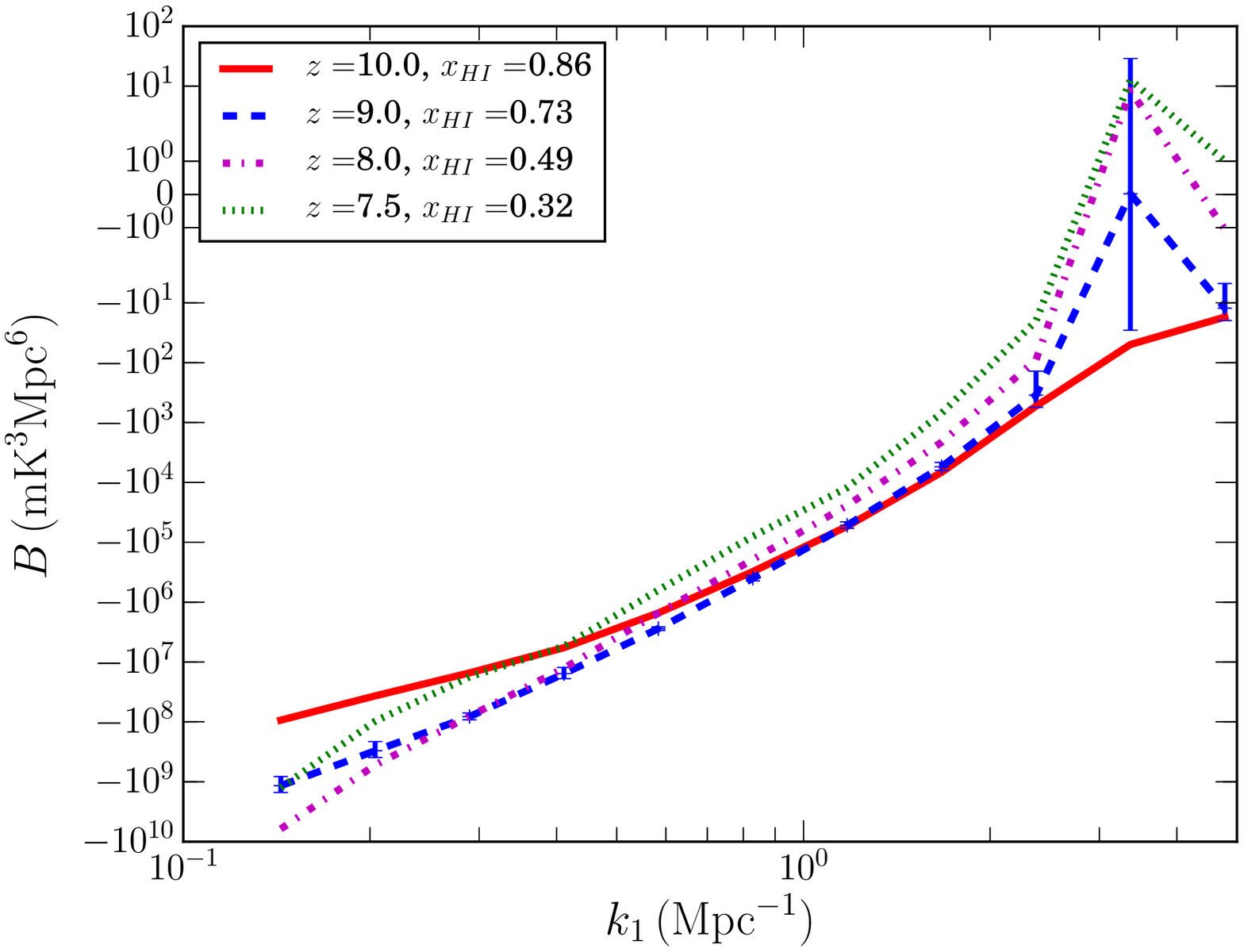}
  \includegraphics[width=.47\textwidth,angle=0]{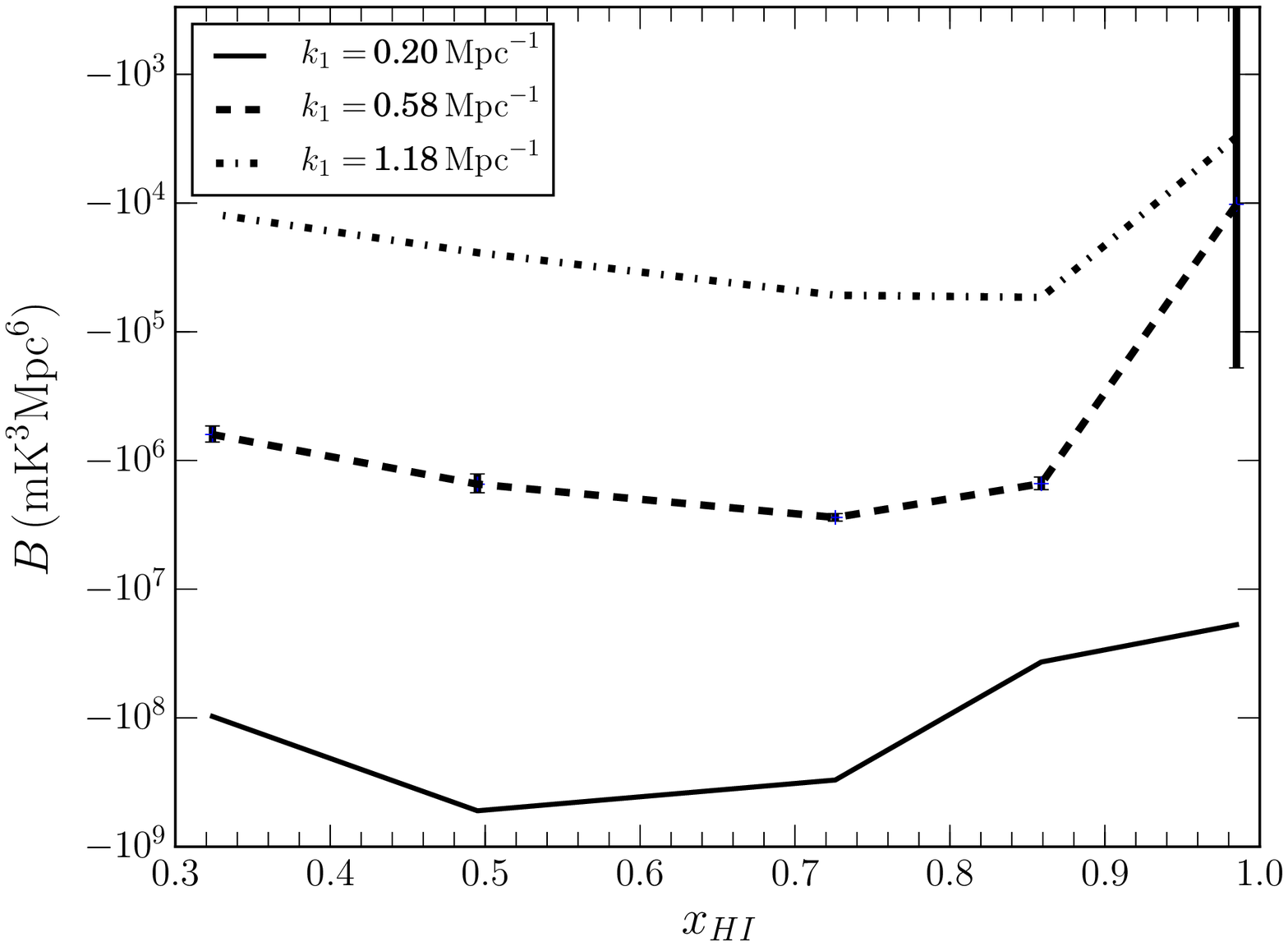}
  \includegraphics[width=.47\textwidth,angle=0]{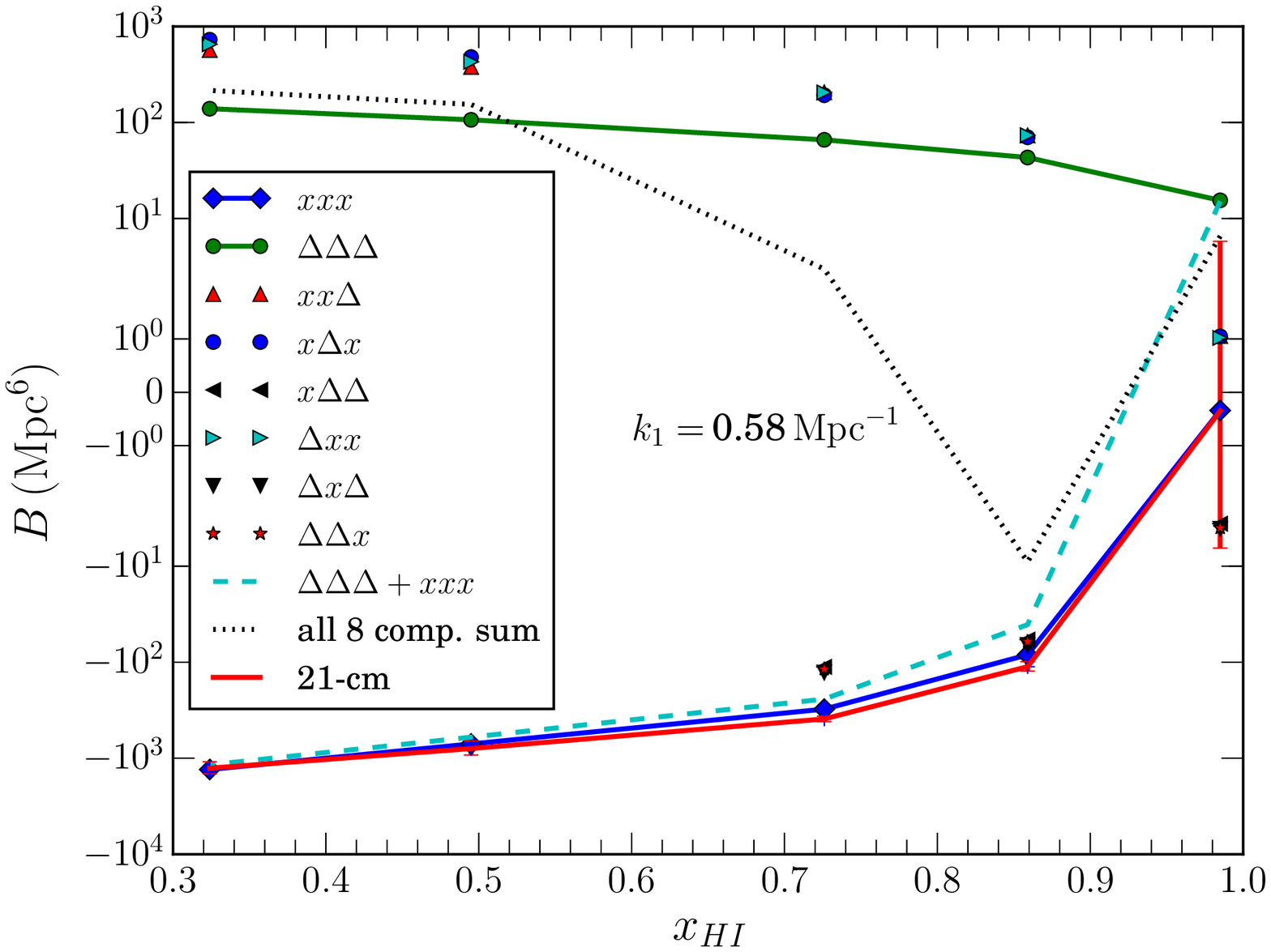}
  \caption{{\bf Top panel:} Bispectra for equilateral triangles as a
    function of $k$ at four representative stages ($\xb = 0.86,\,
    0.73,\, 0.49\, {\rm and}\, 0.32$) of the EoR. The error bars
    represent $1\sigma$ uncertainty estimated from the five
    statistically independent realizations of the simulated
    signal. For clarity, we show error bars only for one $\xb$
    value. {\bf Central panel:} 21-cm bispectrum for equilateral
    triangles as a function of $\xb$ at three representative length
    scales ($k = 0.20,\, 0.58,\, {\rm and}\, 1.18\,{\rm
      Mpc}^{-1}$). {\bf Bottom panel:} Eight component [following
      Equation \eqref{eq:B_h1}] bispectra estimated from the
    simulations as a function of $\xb$ for $k = 0.58\,{\rm Mpc}^{-1}$
    ($x$ represents $\xh1$ field and $\Delta$ represents matter
    density field). This also shows the sum of all eight components
    and two main component bispectra ($B_{\Delta \Delta \Delta}$ and
    $B_{xxx}$), separately, along with the 21-cm bispectrum
    [$B/\overline{T}^3_b(z)$] for comparison. Note that in all three
    panels, the y-axis is shown using the {\it symlog} function of
    {\it matplotlib}, which is linear in between $-1$ to $1$ and log
    in the rest of the range.}
  \label{fig:bispec_equilat}
\end{figure}
\subsection{Equilateral triangles}
The first obvious triangle configuration to study is the equilateral
triangle (i.e. $k_1 = k_2 = k_3$). Just as the power spectrum, which
can be expressed as a function of the amplitude of ${\bf k}$ alone,
the equilateral bispectrum can also be expressed as a amplitude of
just ${\bf k}_1$, as all arms of the triangle have the same
amplitude. It will capture the correlation present in the signal
between three different point in $k$ space equidistant from each
other. Figure \ref{fig:triangle1} suggests that for equilateral
triangles $\cos{\theta} = -0.5$. However, as discussed in Section
\ref{sec:bispec_th}, due to the very restrictive nature of the
algorithm (which does not allow a finite width in the $\cos{\theta}$
bin but rather estimates bispectra for very precise sharp values of
$\cos{\theta}$) that we have employed to estimate bipsecturm, it will
be prone to sample variance, specifically for triangles involving
small $k$ modes. To reduce the effect of this sample variance for
equilateral triangle configurations, we average over all bispectra
estimated within the $\cos{\theta}$ range $-0.52 \leq \cos{\theta}
\leq -0.48$.

The top panel of Figure \ref{fig:bispec_equilat} shows the equilateral
bispectra estimated at four representative stages of the EoR (in terms
of $\xb$ values) as a function of $k_1$. The error bars on the curves
in this figure represent the $1\sigma$ uncertainty estimated from five
statistically independent realizations of the simulated signal. For
the sake of clarity we show error bars only for a single value of
$\xb$ (namely $0.73$). The estimated bispectrum becomes sample
variance dominated for $k_1$ bins corresponding to $k_1 \lesssim 0.1\,
{\rm Mpc}^{-1}$ as the number of closed triangles becomes
significantly small at these values due to the very restrictive nature
of our estimator. Thus we do not show the bispectrum beyond $k_1
\lesssim 0.1\, {\rm Mpc}^{-1}$. For a better understanding of the
evolution of $B$ with $\xb$ at different length scales we also plot
$B$ as a function of $\xb$ for three representative $k_1$ values ($k_1
= 0.20,\, 0.58\, {\rm and}\, 1.18\,{\rm Mpc}^{-1}$) in the central
panel of Figure \ref{fig:bispec_equilat}.

The first obvious observation that one can make from these plots is
that the bispectrum is non-zero and for smaller $k$ modes its
amplitude increases with increasing global ionization fraction. This
establishes the fact that the 21-cm signal is highly non-Gaussian in
nature and its degree of non-Gaussianity increases with the progress
of reionization. The other important and obvious feature of the
bispectrum for the equilateral triangle configuration is that it has a
negative sign for a large range of $k_1$ values ($0.1 \lesssim k_1
\lesssim 3.0 \, {\rm Mpc}^{-1}$), during almost the entire period of
reionization. This is in agreement with the predictions from the toy
model of \HI fluctuations discussed in Section \ref{sec:toy_model} and
shown in the top panel of Figure \ref{fig:model_sig}. We note that
this particular feature of the bispectrum was not observed in the
analysis done by \citet{yoshiura14,shimabukuro16,shimabukuro16b}. This
is due to the fact that they define their bispectrum estimator as the
absolute value (or modulus) of the product of three complex $\Delta
({\bf k})$s, i.e. their estimator is essentially $\langle
\{{\Re}[\Delta ({\bf k}_1) \Delta ({\bf k}_2) \Delta ({\bf k}_3)]^2 +
  {\Im}[\Delta ({\bf k}_1) \Delta ({\bf k}_2) \Delta ({\bf
      k}_3)]^2\}^{1/2} \rangle$. Note that, they have an imaginary
  component in their bispectrum estimation which appears simply
  because for a bispectrum estimation they consider closed triangles
  from only one half of the $k$-space. As we discuss in Section
  \ref{sec:bispec_th} that when triangles from the entire $k$-space is
  considered this imaginary component will become exactly
  zero. Therefore the imaginary component do not have any physical
  siginicance. Further, it is clear that, by construction their
  estimator will always be positive and will not be able to capture
  the sign of the bispectrum or any change in its sign either. The
  sign of the bispectrum is an imporatnt feature of the signal and we
  will discuss it further in the context of other ${\bf k}$-triangle
  configurations in the later part of this paper.

A further visual inspection of the different $B(k_1)$ curves in the
top panel of Figure \ref{fig:bispec_equilat} reveals that $B(k_1)$
shows a power-law like decline in amplitude with increasing $k_1$ for
$0.1 \lesssim k_1 \lesssim 2.0 \, {\rm Mpc}^{-1}$ during almost the
entire period of reionization, which is again in agreement with the
toy model of Section \ref{sec:toy_model}. For the later stages of
reionization, $B(k_1)$ shows a further sharp decline beyond $k_1
\gtrsim 2.0 \, {\rm Mpc}^{-1}$ and even reaches positive values at
$\xb \leq 0.5$ for $k_1 \gtrsim 3.0 \, {\rm Mpc}^{-1}$. The central
panel of the Figure \ref{fig:bispec_equilat} further shows that for
triangles involving large length scales i.e. small $k_1$ values ($k_1
= 0.20\,{\rm Mpc}^{-1}$) the amplitude of $B(k_1)$ increases
significantly with the decreasing $\xb$ (or increasing $x_i$) until
reionization is half way through (i.e.  $\xb \sim 0.5$). The toy model
also predicts an increase in amplitude with the increasing $x_{\rm
  i}$. For $\xb < 0.5$ the amplitude of $B(k_1)$ gradually
decreases. Note that this is also the regime during the EoR when the
overlap of different ionized regions becomes significant in a
realistic ionization topology and the actual shapes of the ionized
regions depart further from that of an ideal sphere. Thus it is
unlikely that the toy model of individual non-overlapping spherical
ionized regions will capture the true nature of the bispectrum in this
regime. For triangles involving intermediate ($k_1 = 0.58\,{\rm
  Mpc}^{-1}$) and large ($k_1 = 1.18\,{\rm Mpc}^{-1}$) $k$ modes the
increase in amplitude is observed until $\xb \sim 0.7$ and for the
later stages of reionization the amplitude gradually decreases. The
next simulation snapshot that we have available for $\xb$ values below
$\xb = 0.32$ is at $\xb \approx 0.15$. We find that the amplitude of
the bispectrum goes down significantly for $\xb \approx 0.15$ and the
sample variance for our bispectrum estimate is also very high at this
stage. We therefore do not show the results for any neutral fraction
values below $\xb < 0.32$.
\begin{figure}
  \includegraphics[width=.47\textwidth,angle=0]{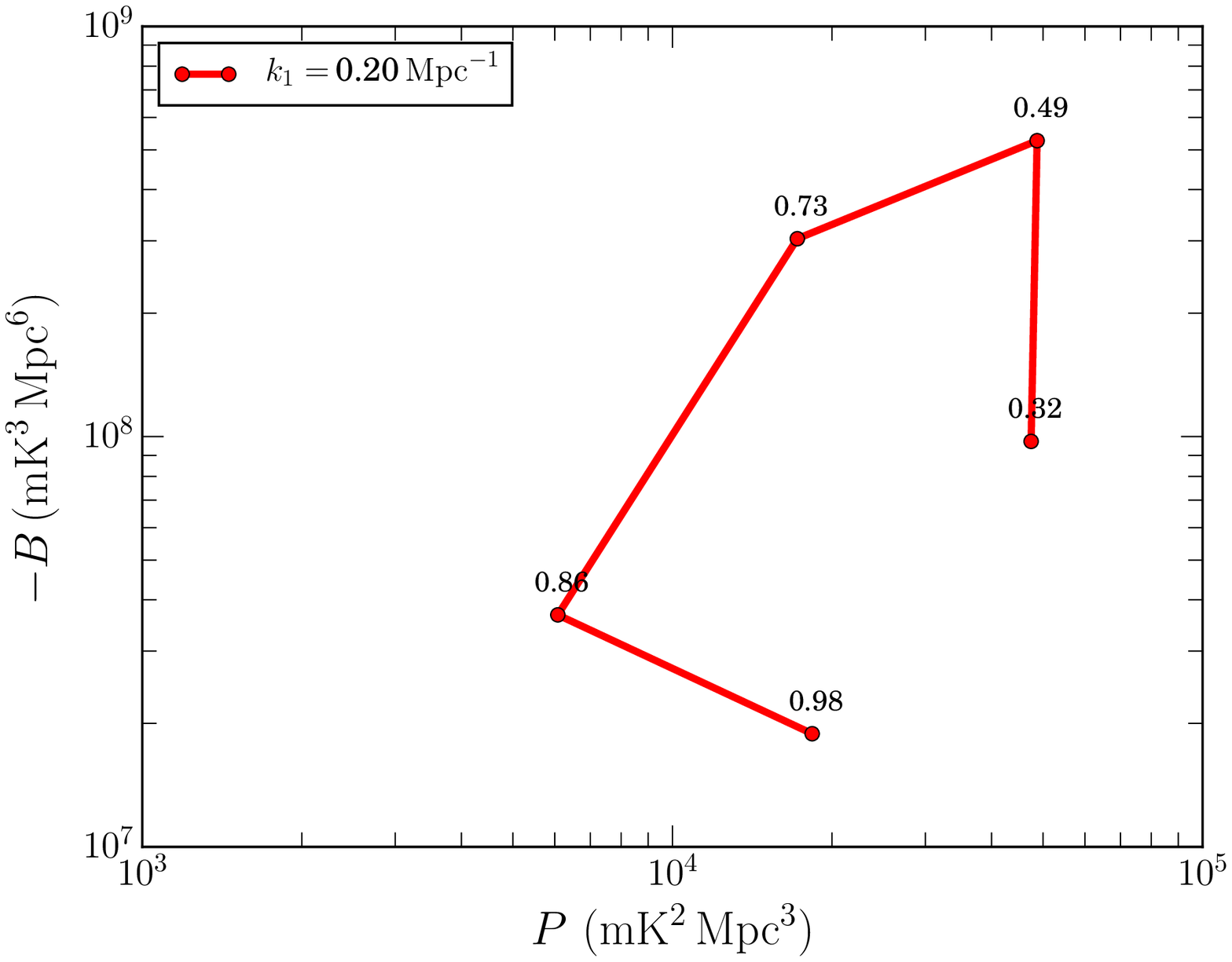}
  \includegraphics[width=.47\textwidth,angle=0]{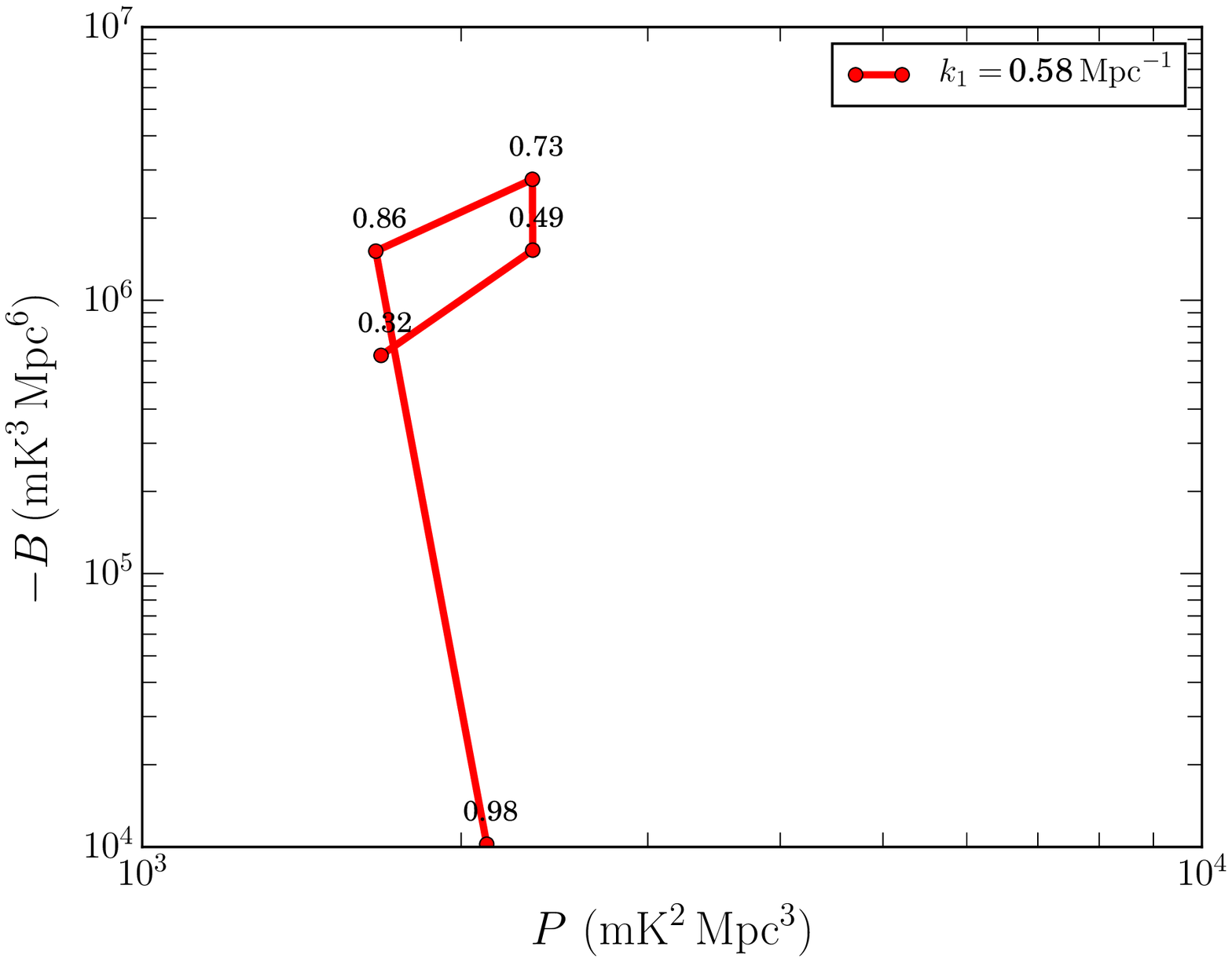}
\caption{Signal trajectories in the $P(k) - B(k, k, k)$ phase space
  for $k = 0.20$ and $0.58 \, {\rm Mpc}^{-1}$. Note that here we show
  $-B$ instead of $B$. The corresponding values of $\xb$ have been
  printed on the respective points of the trajectories.}
\label{fig:Bk_Pk}
\end{figure}

The bottom panel of Figure \ref{fig:bispec_equilat} shows the
evolution of the various component bispectra (against $\xb$, for $k_1
= 0.58\,{\rm Mpc}^{-1}$), estimated from the matter density and $\xh1$
fields that are used to simulate the 21-cm signal [see Equation
  \eqref{eq:B_h1}]. In this figure we have scaled the 21-cm bispectrum
with $1/\overline{T}^3_{{\rm b}}(z)$ to keep its dimension similar to
all the component bispectra. It is evident from this plot that the
major component contributing to the 21-cm bispectrum for equilateral
triangles (solid red line with error bars) for almost the entire
period of reionization is the neutral fraction bispectrum (solid blue
line with diamonds). We also observe that the sum of all of the eight
components of bispectrum (black dotted curve) does not follow the
evolution of the 21-cm bispectrum, even qualitatively. However, the
sum of the neutral fraction and density bispectra (cyan dashed line)
does follow the evolution of the 21-cm bispectrum in terms of its
shape, if not exactly in terms of its amplitude, especially during the
intermediate and late stages of reionization. We observe a similar
behaviour, i.e. the 21-cm bispectrum closely following the $\xh1$
field bispectrum, for a wide range of length scales i.e. $0.20\, \leq
k_1 \leq 1.18\, {\rm Mpc}^{-1}$ (not shown in the figure). The $\xh1$
bispectrum, which is the main contributor to the 21-cm bispectrum for
this type of triangle, also follows (at least qualitatively) the shape
and evolution predicted by the toy model of $\xh1$ fluctuations
discussed in Section \ref{sec:toy_model}. This further establishes
that this toy model is a reasonably good tool for interpreting the
qualitative behaviour of the 21-cm bispectrum, at least for this type
of triangles, as this type of bispectra is sensitive to the
fluctuations generated by the distribution of \HII regions. However,
the sum of all eight components of Equation \eqref{eq:B_h1} does not
follow the 21-cm bispectrum (even qualitatively) which can be
attributed to the fact that the Equation \eqref{eq:B_h1} has been
obtained under the linear approximation limit of the Fourier transform
of Equation \eqref{eq:Tb_comp}. If one considers that the higher order
terms in density fluctuations and neutral fraction also have
significant contributions to the 21-cm signal, then there will be many
higher order component bispectra (e.g. $B_{\Delta x (\Delta x)},\,
B_{\Delta (\Delta x) (\Delta x)}, ...$ etc.)  contributing to the
21-cm bispectrum. Thus a correction to the Equation \eqref{eq:B_h1} by
including higher order contributions of two constituent fields may
provide us a more accurate model for the EoR 21-cm bispectrum. A
similar model in the context of the EoR 21-cm power spectrum has been
discussed in \citet{lidz07} through their equation (2) and (3). We
plan to study the effectiveness of such an higher order model in the
context of interpreting the EoR 21-cm bispectrum in a future follow up
work.

\subsection{Evolution of the 21-cm signal in $P(k) - B(k, k , k)$ space}
Since the spherically averaged power spectrum and the bispectrum for
equilateral triangles are two independent but complementary measures
of the 21-cm signal and both can be represented as a function of the
amplitude of just one wave number $k$, one can thus visualize the
evolution of the 21-cm signal as a trajectory in the phase
space{\footnote{Note that \citet{majumdar16} has demonstrated the
    evolution of the EoR 21-cm signal using a similar phase space of
    monopole and quadrupole moments of the redshift space 21-cm power
    spectrum.}} of $P(k) - B(k, k , k)$. In Figure \ref{fig:Bk_Pk} we
show the evolution of the signal in such a phase space diagram for two
representative values of $k$ ($k = 0.20$ and $0.58 \, {\rm
  Mpc}^{-1}$). As, for these two wavenumbers, the $B$ stays negative
for the entire range of $\xb$ values that we consider, we plot
$-{\Re}[B]$ instead of $B$ in this figure for a better understanding
of the signal evolution. A close inspection of Figure \ref{fig:Bk_Pk}
reveals many similarities in the qualitative behaviour of the signal
trajectories at different length scales along with some
differences. For both length scales, the signal initially shows a rise
in amplitude in the bispectrum but a decline in amplitude in the power
spectrum. The power spectrum reaches a minimum around $\xb \approx
0.85$ beyond which it starts to grow in amplitude with decreasing
$\xb$. Both power spectra and bispectra continues to grow in amplitude
until the trajectory reaches a turn-around point. For relatively
larger length scales (i.e.  $k = 0.20 \, {\rm Mpc}^{-1}$, top panel of
Figure \ref{fig:Bk_Pk}) this turn-around point appears at $\xb \approx
0.50$ and for intermediate length scales (i.e.  $k = 0.58 \,{\rm
  Mpc}^{-1}$, bottom panel of Figure \ref{fig:Bk_Pk}) it appears at
$\xb \approx 0.70$. Beyond this turn-around point both power spectra
and bispectra decrease in amplitude with decreasing neutral
fraction. The bispectrum decreases relatively sharply compared to the
power spectrum. If different reionization source models lead to
different 21-cm topologies, one would then expect their 21-cm signal
trajectories in $P(k) - B(k, k, k)$ phase space to be different from
each other. These trajectories provide us a cosolidated view of the
signal by combining both power spectrum and bispectrum. Each of which
is expected to probe a different characteristic of the
signal. Similarly, one would expect a joint MCMC analysis (or a
similar sort of likelihood analysis) of power spectrum and bispectrum
to provide a more robust constraint on the reionization model
parameters compared to the similar MCMC analysis done with a single
signal estimator (either power spectrum or bispectrum) alone
(e.g. \citealt{greig15, greig17, shimabukuro16b, schmit17} etc).
  
\subsection{Isosceles triangles}
\begin{figure*}
\includegraphics[width=.47\textwidth,angle=0]{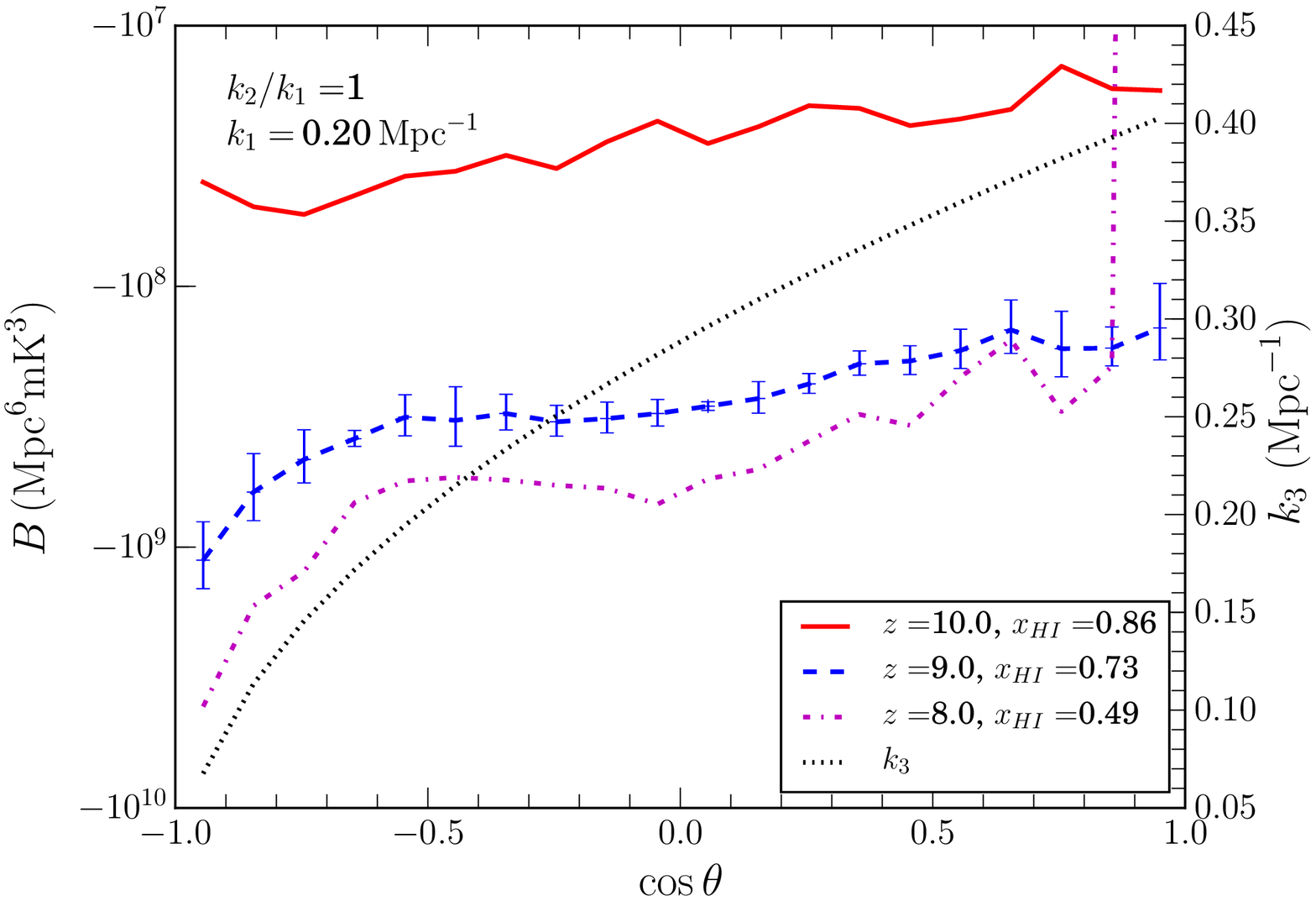}
\includegraphics[width=.47\textwidth,angle=0]{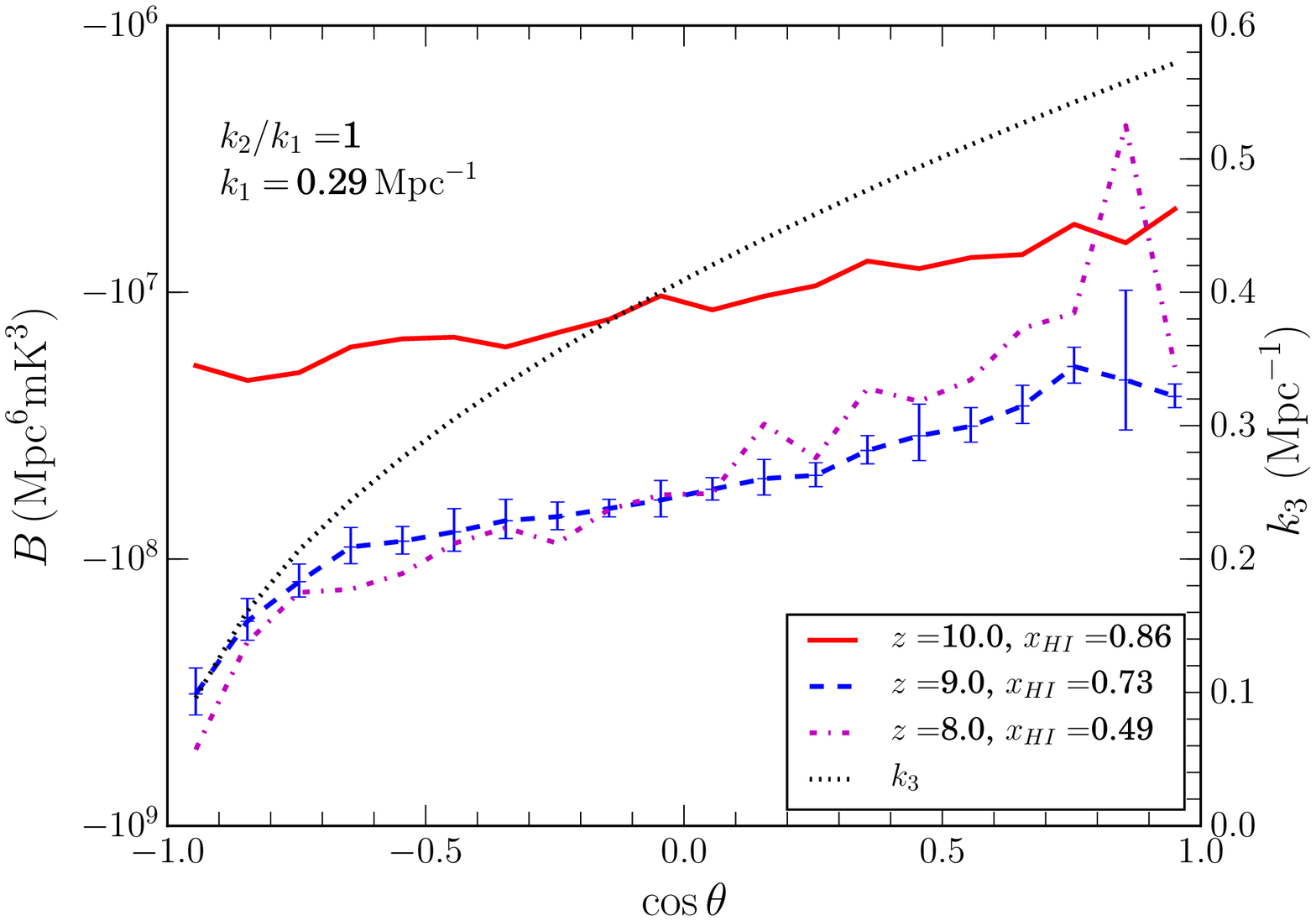}
\includegraphics[width=.47\textwidth,angle=0]{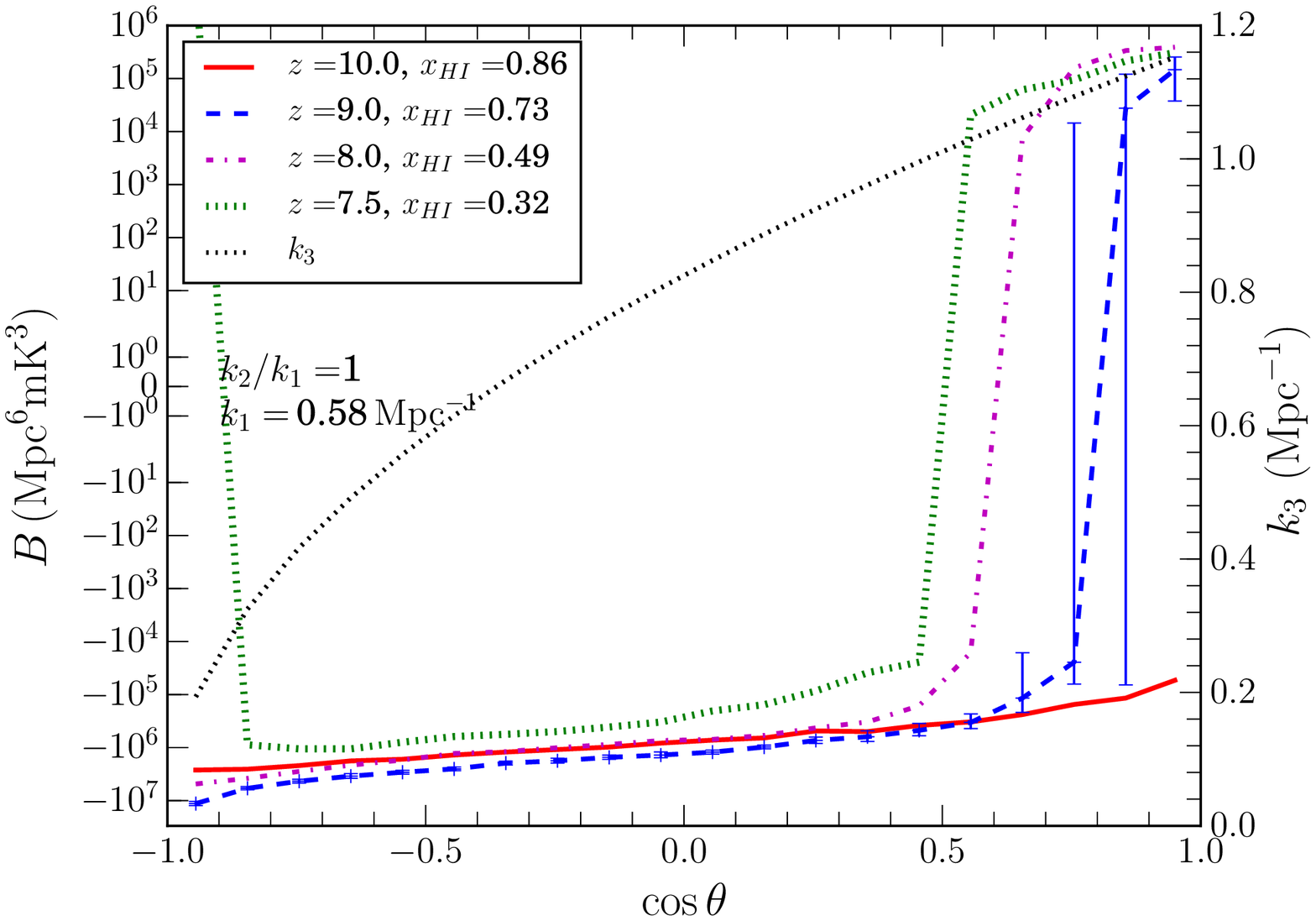}
\includegraphics[width=.47\textwidth,angle=0]{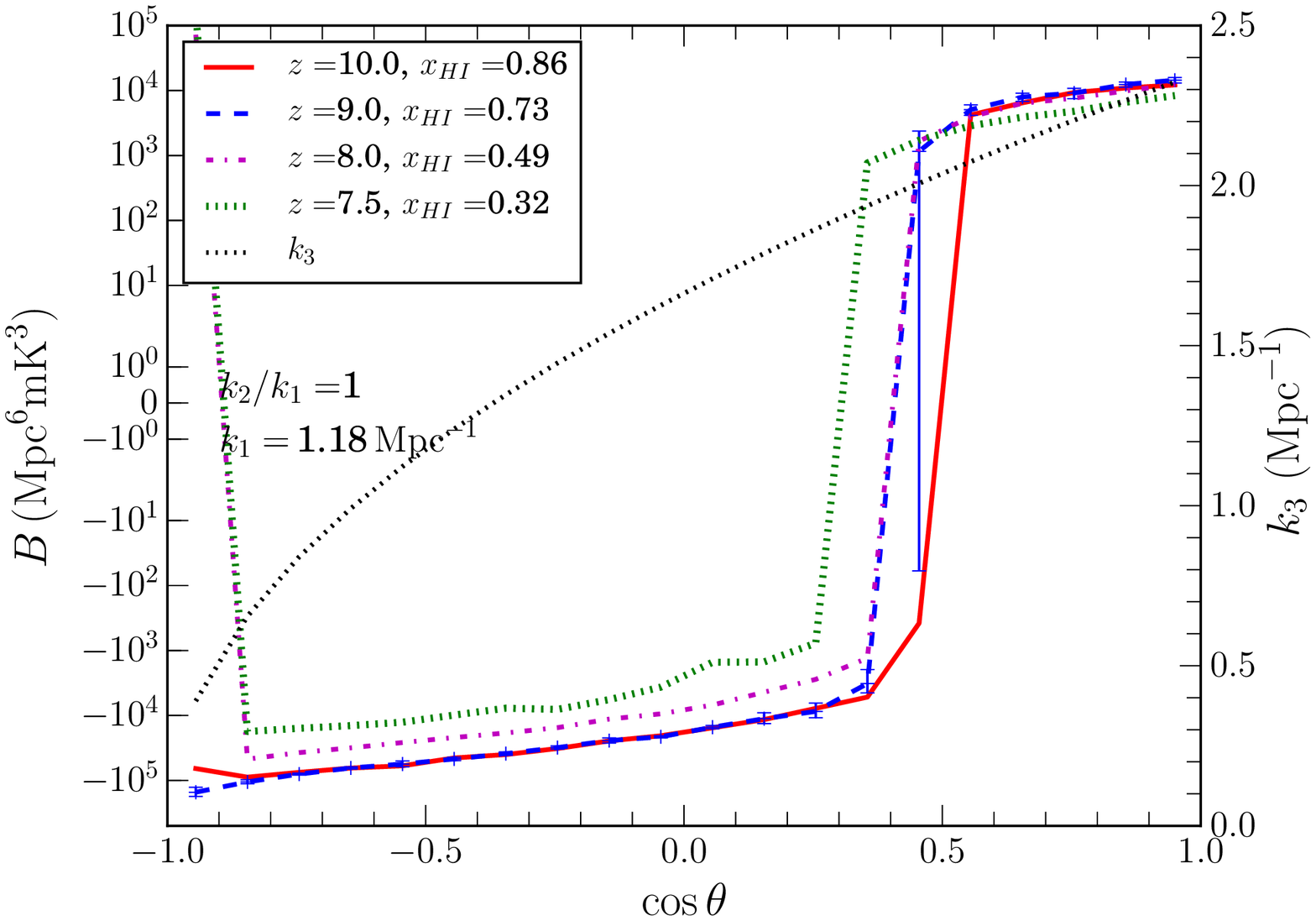}
\includegraphics[width=.47\textwidth,angle=0]{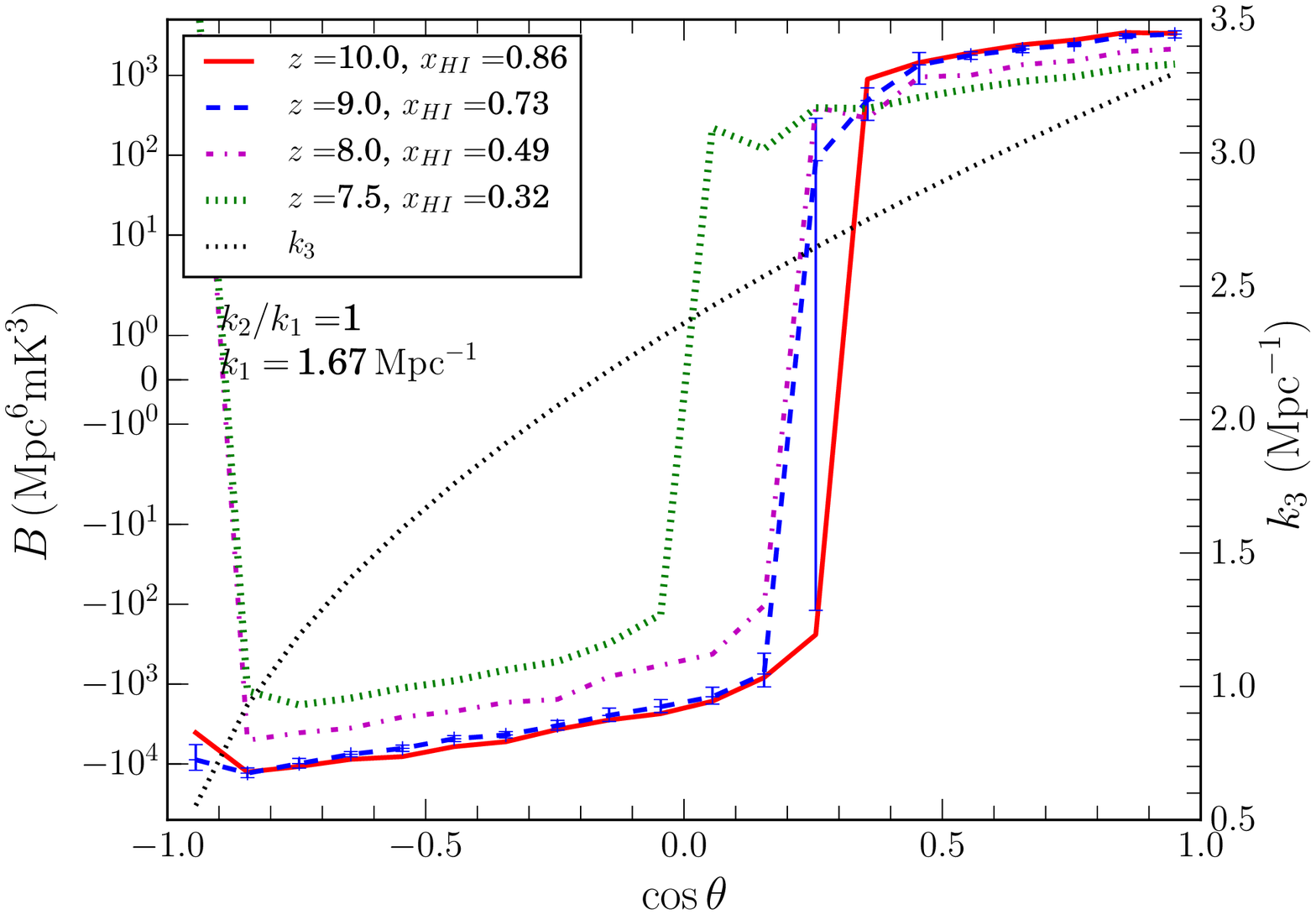}
\includegraphics[width=.47\textwidth,angle=0]{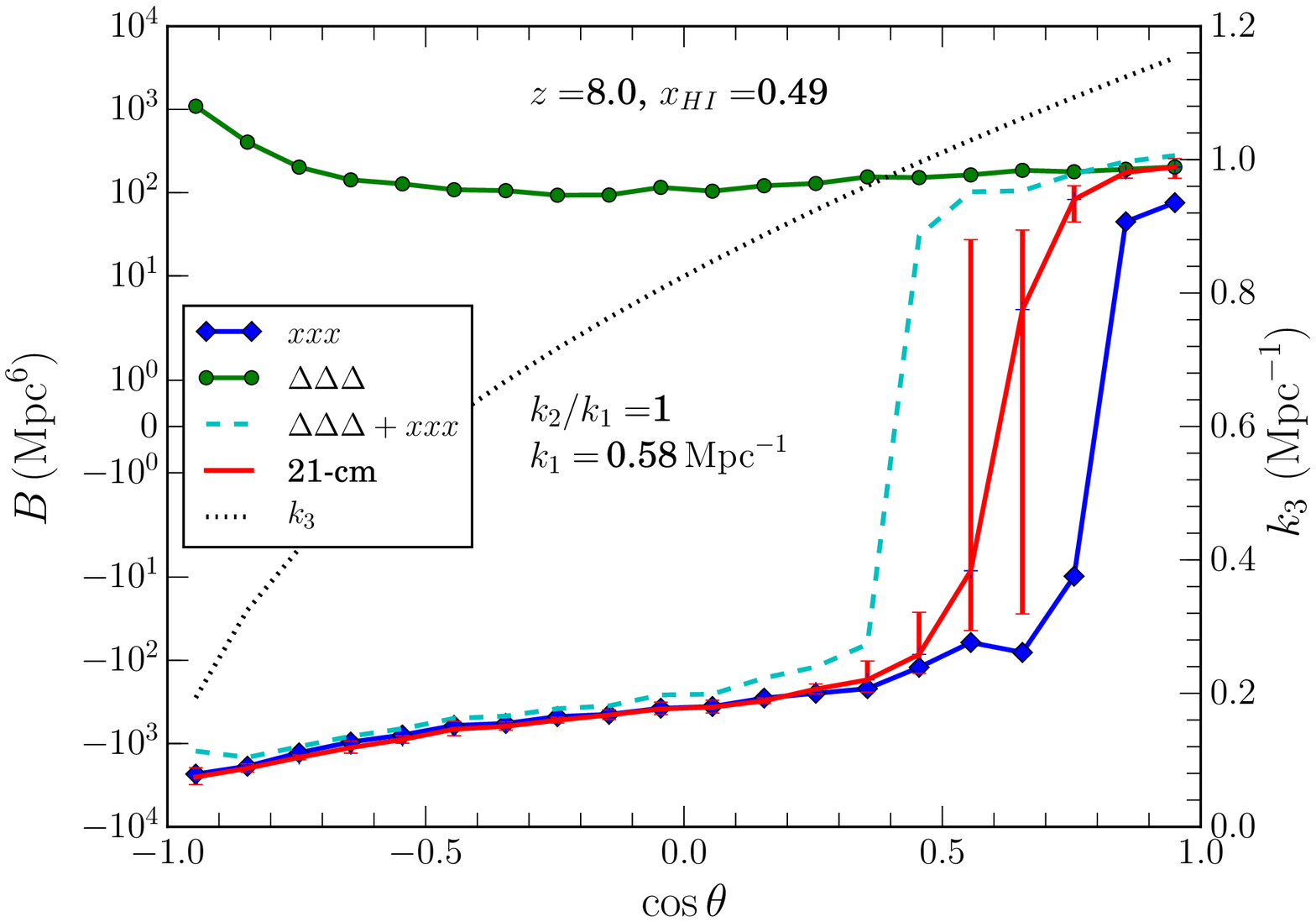}
\caption{21-cm bispectra for isosceles triangles as a function of
  $\cos{\theta}$ at different stages of reionization marked by their
  corresponding $\xb$ values. Different panels show bispectra for
  different values of $k_1$ ($k_1 = 0.20,\, 0.29,\, 0.58,\, 1.18\,
  {\rm and}\, 1.67\,{\rm Mpc}^{-1}$). The entire $\cos{\theta}$ range
  has been divided into $20$ linearly spaced bins. Points in each
  bispectrum curve represent the mid point of the corresponding
  $\cos{\theta}$ bin. The error bars shown on the bispectrum for one
  $\xb$ value represent $1\sigma$ uncertainty estimated from the five
  statistically independent realizations of the signal. To have a
  better understanding of all the length scales involved in the
  bispectrum estimation we also show the amplitude of ${\bf k}_3$ as a
  function of $\cos{\theta}$ through the dotted black curve in all
  panels (refer to the right y-axis in each panel to evaluate the
  $k_3$ curve at a specific value of $\cos{\theta}$). The bottom right
  panel shows the two major component bispectra (matter density --
  $B_{\Delta\Delta\Delta}$ and neutral fraction -- $B_{xxx}$) and
  their sum as functions of $\cos{\theta}$ for $k_1 = 0.58\,{\rm
    Mpc}^{-1}$ and $\xb = 0.49$. We also show the corresponding 21-cm
  bispectrum [$B/\overline{T}^3_b(z)$] in the same panel for
  comparison. Note that in all panels the entire left y-axis is shown
  in logarithmic scale, except the range $-1$ to $1$, where the scale
  is linear. The right y-axis is shown in linear scale in all panels.}
\label{fig:bispec_k2k1_1}
\end{figure*}

The first order generalization of equilateral triangles ($n = 1$ and
$\cos{\theta} = -0.5$) will lead to isosceles triangles ($n = 1$ and
$-1 \leq \cos{\theta} \leq 1$). Compared to equilateral triangles,
which show correlations in the signal for three equidistant points in
the Fourier space, isosceles triangles will have two equal length $k$
arms and the third arm will be different in length (determined by the
respective $\cos{\theta}$ value). Thus the bispectrum for isosceles
triangles will effectively show the correlation in the signal between
two different $k$ modes. Figure \ref{fig:bispec_k2k1_1} shows the
isosceles bispectra for the simulated 21-cm signal at different stages
of reionization as a function of $\cos{\theta}$.

For relatively small ${\bf k}$-triangles (i.e. $k_1 = k_2 = 0.20\,{\rm
  Mpc}^{-1}$, top left panel of Figure \ref{fig:bispec_k2k1_1}) and
during the early and the intermediate stages of reionization ($\xb =
0.86$ and $0.73$) the 21-cm bispectrum is negative. At these stages of
reionization, the amplitude of the bispectrum is maximum in the
`squeezed' limit ($\cos{\theta} \sim -1$) and it decreases with the
increasing $\cos{\theta}$ (as one moves towards the `stretched'
limit). This decrease is rapid for $-1 \leq \cos{\theta} \leq -0.5$
(while $k_3$ stays within the range $0.05 \leq k_3 \leq 0.20 \,{\rm
  Mpc}^{-1}$) and relatively slow for the rest of the $\cos{\theta}$
range (i.e. $0.20 \leq k_3 \leq 0.40 \,{\rm Mpc}^{-1}$). The overall
amplitude of $B$ also increases with decreasing neutral
fraction and this behaviour holds true for average neutral fraction
values as low as $\xb \approx 0.50$ and for $\cos{\theta}$ values up
to $\approx 0.8$ (i.e.  $k_3 \leq 0.40\,{\rm Mpc}^{-1}$). Most of
these characteristics qualitatively follow the predictions of our toy
model in Section \ref{sec:toy_model} (see bottom panel of Figure
\ref{fig:model_sig}).

As we move towards triangles with slightly larger $k$ values
(e.g. $k_1 = k_2 = 0.29\,{\rm Mpc}^{-1}$, top right panel of Figure
\ref{fig:bispec_k2k1_1}) we observe that the increse in the overall
amplitude of $B$ with decreasing $\xb$ predicted by our toy model
breaks down. We observe that for $k_1 = 0.29\,{\rm Mpc}^{-1}$ the
bispectrum for $\xb = 0.49$ becomes equal or smaller in amplitude than
the bispectrum for $\xb = 0.73$ for $\cos{\theta} \geq 0$ (i.e. $k_3
\geq 0.40\,{\rm Mpc}^{-1}$). This deviation of the signal bispectrum
from the toy model prediction continues even in the later stages of
reionization (not shown in this panel of Figure
\ref{fig:bispec_k2k1_1}). Note that we do not show the bispectrum for
$\xb = 0.32$ in the top panels of Figure \ref{fig:bispec_k2k1_1}
(i.e. for $k_1 = 0.20\, {\rm and}\, 0.29\, {\rm Mpc}^{-1}$) as it has
a large dynamic range compared to the bispectrum for other neutral
fractions. We discuss the characteristics of bispectrum for $\xb =
0.32$ corresponding to these $k_1$ modes later in this section.

As we take our ananlysis to triangles with intermediate and large
values of $k_1$ ($k_1 = k_2 = 0.58,\, 1.18$ and $1.67\,{\rm
  Mpc}^{-1}$) we observe several other interesting features in the
bispectrum. For $k_1 = 0.58\,{\rm Mpc}^{-1}$ (central left panel of
Figure \ref{fig:bispec_k2k1_1}), the `squeezed' limit
(i.e. $\cos{\theta} \approx -1$, $k_3 \approx 0.20\,{\rm Mpc}^{-1}$)
triangles follow the trend of rise in bispectrum amplitude with
decreasing $\xb$ up to $\xb \geq 0.7$. Beyond this, we observe a
gradual reversal in this treand for lower neutral fraction values. At
these stages of reionization the signal amplitude tends to decrease
with decreasing $\xb$ values.

The most interesting feature that we observe for triangles with $k_1 =
0.58\,{\rm Mpc}^{-1}$ is a change of sign in the bispectrum (from
negative to positive) as we approach the `stretched' limit of these
triangles. This change in sign is observed first for $\xb = 0.73$
prominently around $\cos{\theta} \approx 1$ (though the actual
transition happens somewhere in between $ 0.7 \lesssim \cos{\theta}
\lesssim 0.9$, i.e. $ 1.08 \lesssim k_3 \lesssim 1.16\,{\rm
  Mpc}^{-1}$, but difficult to ascertain the exact point of transition
due to the large cosmic variance in this range). Interestingly enough,
during the later stages of reionization this transition (change in
sign) happens at gradually lower values of $\cos{\theta}$. For $\xb =
0.32$ the $B$ becomes positive at around $\cos{\theta} \approx 0.5$
(i.e. $k_3 \approx 1.0 \, {\rm Mpc}^{-1}$), beyond which it remains
positive. To confirm these features are real and not due to any
numerical artifact in our algorithm, we have used the bispectrum
estimator of \citet{watkinson17} and obtained the same results. We
have observed similar features in the 21-cm bispectrum for $k_1 =
0.20\, {\rm and}\, 0.29\,{\rm Mpc}^{-1}$ (not shown in these figures)
at very late stages of reionization ($\xh1 = 0.32$).

The fact that this change of sign in $B$ is a systematic behaviour,
and is strongly dependent on the stage of reionization (i.e. $\xb$)
and the length scales involved for the bispectrum estimation, becomes
very apparent as one observes the results for triangles with $k_1 =
1.18 \, {\rm and}\, 1.67 \, {\rm Mpc}^{-1}$ (central right and bottom
left panels of Figure \ref{fig:bispec_k2k1_1}). As one approaches
triangles with larger $k_1$ values (which implies larger $k_3$
amplitudes as well), one observes this sign reversal to occur at very
early stages of reionization ($\xb = 0.86$) and at even lower values
of $\cos{\theta}$. For $k_1 = 1.18 \, {\rm Mpc}^{-1}$ it occurs at
$\cos{\theta} \approx 0.5$ (i.e. $k_3 \approx 2.10 \, {\rm Mpc}^{-1}$)
for $\xb = 0.86$ and at $\cos{\theta} \approx 0.3$ (i.e. $k_3 \approx
1.95 \, {\rm Mpc}^{-1}$) for $\xb = 0.32$. Similarly, for $k_1 = 1.67
\, {\rm Mpc}^{-1}$ it occurs at $\cos{\theta} \approx 0.3$ (i.e. $k_3
\approx 2.8 \, {\rm Mpc}^{-1}$) for $\xb = 0.86$ and at $\cos{\theta}
\approx 0$ ({\it i.e.} $k_3 \approx 2.4 \, {\rm Mpc}^{-1}$) for $\xb
=0.32$. Further, for the triangle configurations with the largest
$k_1$ amplitude we observe that the overall amplitude of the
bispectrum (irrespective of its sign) gradually decreases with
decreasing $\xb$. We also observe that for triangles with these two
large $k_1$ values, during the later stages of reionization ($\xb \leq
0.5$), the bispectrum at the `squeezed' limit ($\cos{\theta} \approx
-1$) also becomes positive.

To get some physical insight in the trends for isosceles triangles, we
plot in the bottom right panel of Figure \ref{fig:bispec_k2k1_1} the
two main component bispectra: $B_{\Delta \Delta \Delta}$ and $B_{x x
  x}$, and the sum of these two components, along with the 21-cm
bispectra at $\xb = 0.49$ for $k_1 = 0.58 \, {\rm Mpc}^{-1}$. We
choose this set of $k_1$ and $\xb$ values because this combination of
$k_1$ and $\xb$ clearly shows most of the main features of the 21-cm
isosceles bispectrum that we have described. The above panel shows
that for most of the $\cos{\theta}$ range ($-1 \lesssim \cos{\theta}
\lesssim 0.5$, i.e. $0.2 \lesssim k_3 \lesssim 1 \, {\rm Mpc}^{-1}$)
where it is negative, the 21-cm bispectrum very closely follows $B_{x
  x x}$. This is also the $\cos{\theta}$ range where the shape of the
$B$ curve as a function of $\cos{\theta}$ follows the prediction from
our toy model for the $\xh1$ fluctuation bispectrum (see bottom panel
of Figure \ref{fig:model_sig}). Note that the $k$ modes that form
closed triangles within this $\cos{\theta}$ range are either small or
intermediate in amplitude ($0.2 \lesssim k \lesssim 1 \, {\rm
  Mpc}^{-1}$). The contribution from $B_{\Delta \Delta \Delta}$ (which
is, as expected, always positive in the entire $\cos{\theta}$ range)
is negligible compared to $B_{x x x}$ at this range, as their sum
($B_{\Delta \Delta \Delta} + B_{x x x}$) also largely follows the
$\xh1$ bispectrum. During the early and intermediate stages of the EoR
(i.e. $\xb \gtrsim 0.5$), the 21-cm signal fluctuations involving
these $k$ modes will be dominated by the size distribution of the
individual isolated ionized regions. A simplified version of which
formulates our toy model for $\xh1$ fluctuations. Thus one would
expect the toy model to be able to qualitatively describe the 21-cm as
well as the $\xh1$ bispectrum in this regime.

As reionization progresses the ionization topology deviates from this
simplified model. A recent study by \citet{bag18} (done with an almost
identical simulation as the one presented here; see also
\citealt{furlanetto16}) suggest that most of the isolated ionized
regions percolate and lead to a single ionized region as early as $\xb
\approx 0.73$. They also find that this interconnected `infinitely'
long ionized region is not spherical but filamentary in shape (see
figure 2, 3 and 7 of \citealt{bag18}). As reionization progresses this
large ionized region which is spread accross the entire box, grows in
length, whereas its breadth and thickness remains almost constant. It
maintains this filamentary nature until the very late stages of the
EoR (i.e. $\xb \geq 0.1$). After percolation, a continued ionization
of the universe lead to the production of more ionized tunnels of
different lengths within this filamentary ionized region. It implies
that effectively there is no such characteristic ionized bubble size
after percolation. During this period the neutral IGM also stays
interconnectd and filamentary in nature, until the very end of the EoR
(i.e. $\xb \leq 0.1$), when only small isolated islands of lower
densities remain neutral. Thus during most of reionization the neutral
and ionized gas resides in just two distinct but delicately
intertwined regions. One would expect the complicated evolution of
these two topologies (i.e. the number and length distribution of
ionized tunnels and neutral bridges) to affect the behaviour and
evolution of the 21-cm bispectrum for relevant $k$-triangles.

In the the bottom right panel of Figure \ref{fig:bispec_k2k1_1} for
$\cos{\theta} \geq 0.5$ ($1.0 \lesssim k_3 \lesssim 1.2 \,{\rm
  Mpc}^{-1}$) we observe a sharp decline in the amplitude of both
21-cm and $\xh1$ bispectra. Finally around $\cos{\theta} \approx 0.85$
both of these bispectra change sign and become positive. However,
$B_{\Delta \Delta \Delta}$ is significantly larger in amplitude
compared to $B_{x x x}$ for these $k$ modes and thus the 21-cm
bispectrum follows $B_{\Delta \Delta \Delta}$ here. The reason of this
behaviour of the 21-cm bispectrum can be understood in the following
manner. During the entire period of reionization, $B_{\Delta \Delta
  \Delta}$ evolves rather slowly. However, the correlation between
different length scales (or $k$ modes) in the $\xh1$ field evolves
rather rapidly, as it depends on the complicated filamentary
ionization topology discussed earlier. Our simplified toy model for
the $\xh1$ fluctuations is not based on this kind of complicated
topology and thus fails to predict the observed $B_{x x x}$ evolution
at this stage. The strength and the nature of the correlation between
different $k$ modes in the $\xh1$ field will depend on evolution of
the sizes and distribution length scales of the ionized tunnels and
neutral bridges in the IGM. As signal originates from the neutral part
of the IGM, at late stages of reionization at relevant $k$ modes the
21-cm fluctuations are expected to be driven by these neutral bridges
distributed inside a large filamentary \HII region. The 21-cm
bispectrum thus changes its sign and starts to probe $B_{\Delta \Delta
  \Delta}$ corresponding to those $k$ modes. Thus, depending on the
stage of the EoR and the $k$ modes involved, one would be either
probing the $B_{x x x}$ or the $B_{\Delta \Delta \Delta}$, through the
21-cm bispectrum. We discuss the possible physical significance of
this behaviour in more detail in Section
\ref{sec:sign_change}. Further, this transition between these two
states of the 21-cm bispectrum can actually be used as a confirmative
test of the 21-cm signal detection using the upcoming radio
interferometers such as the SKA and HERA.

\begin{figure*}
\includegraphics[width=.47\textwidth,angle=0]{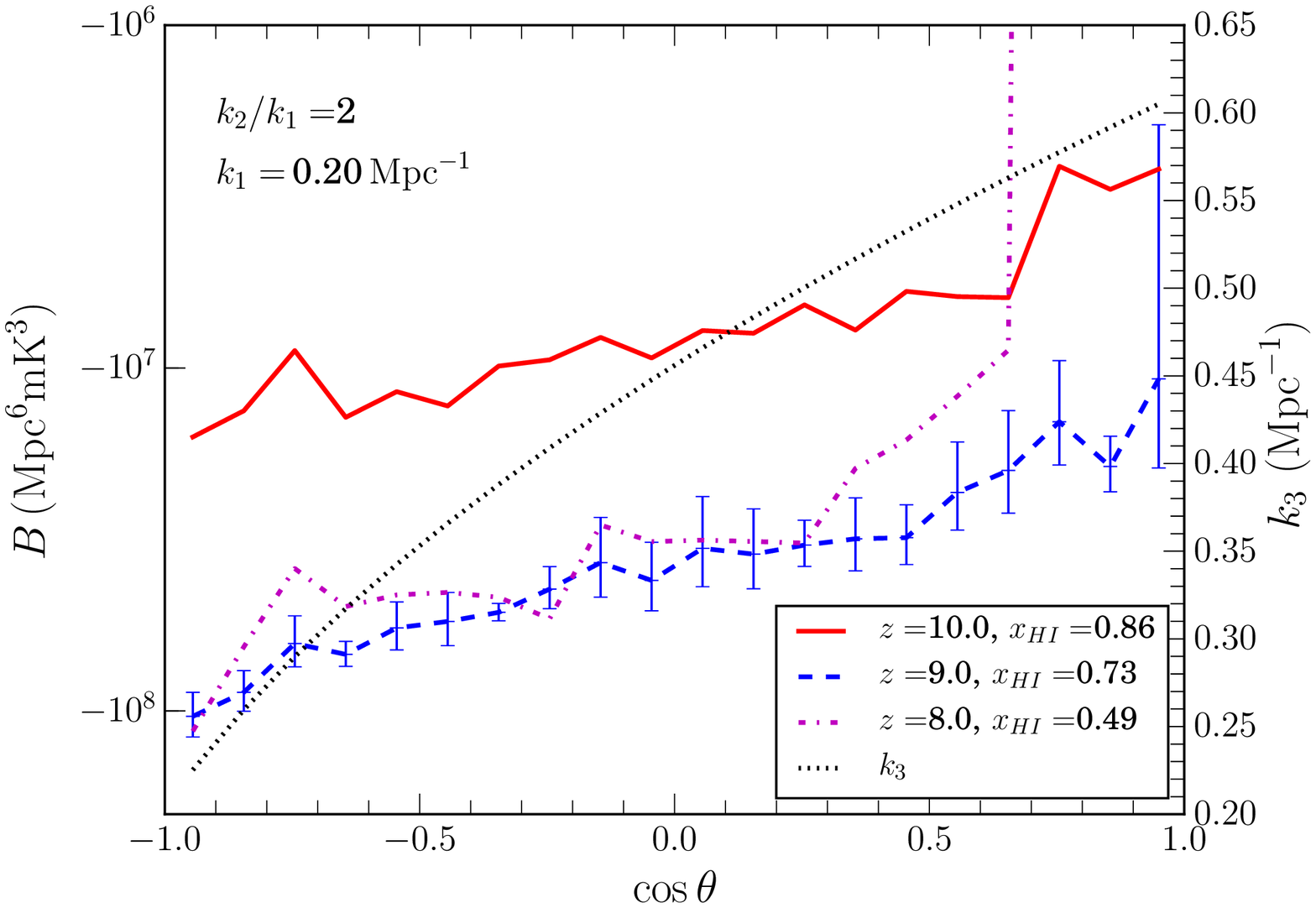}
\includegraphics[width=.47\textwidth,angle=0]{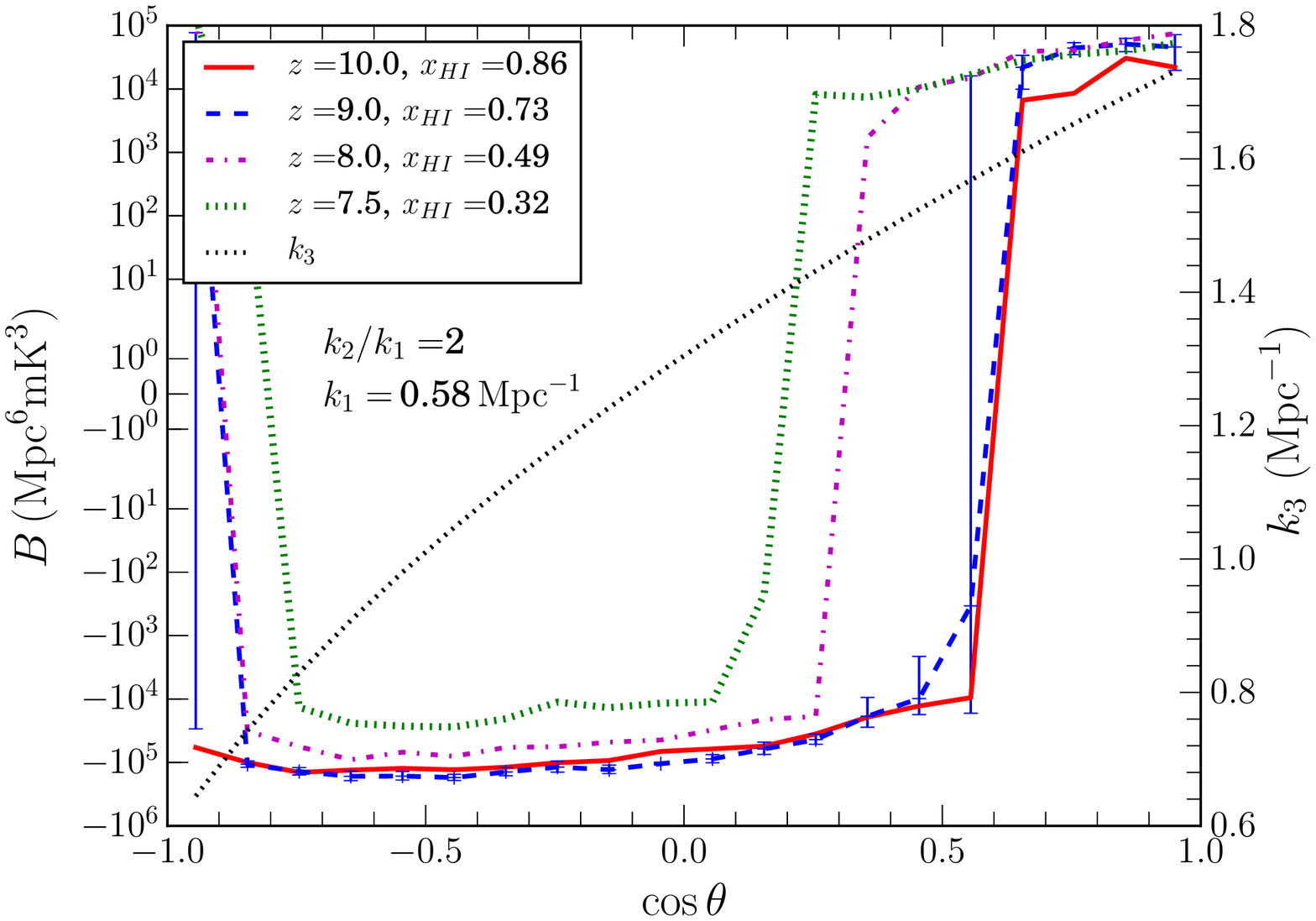}
\includegraphics[width=.47\textwidth,angle=0]{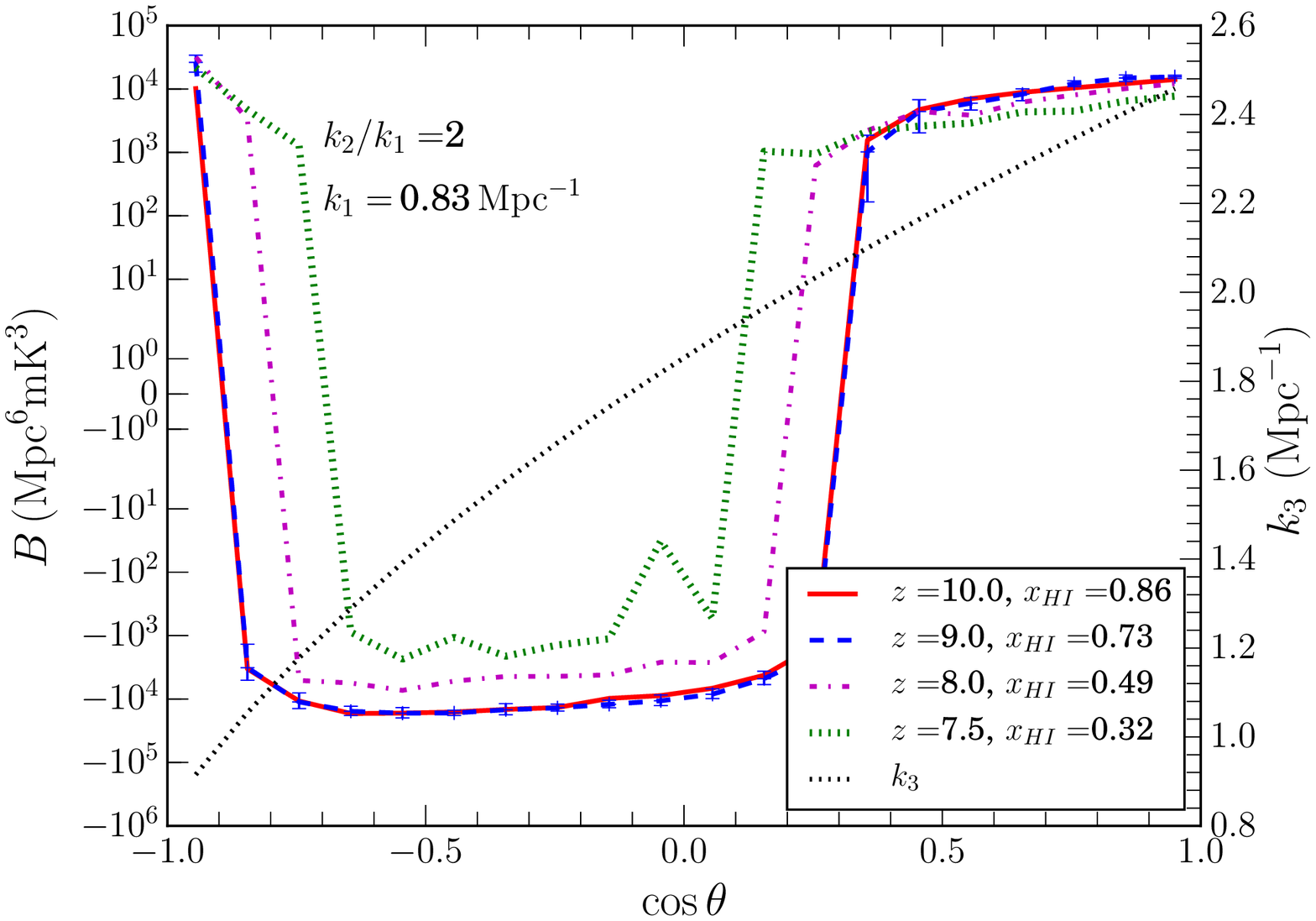}
\includegraphics[width=.47\textwidth,angle=0]{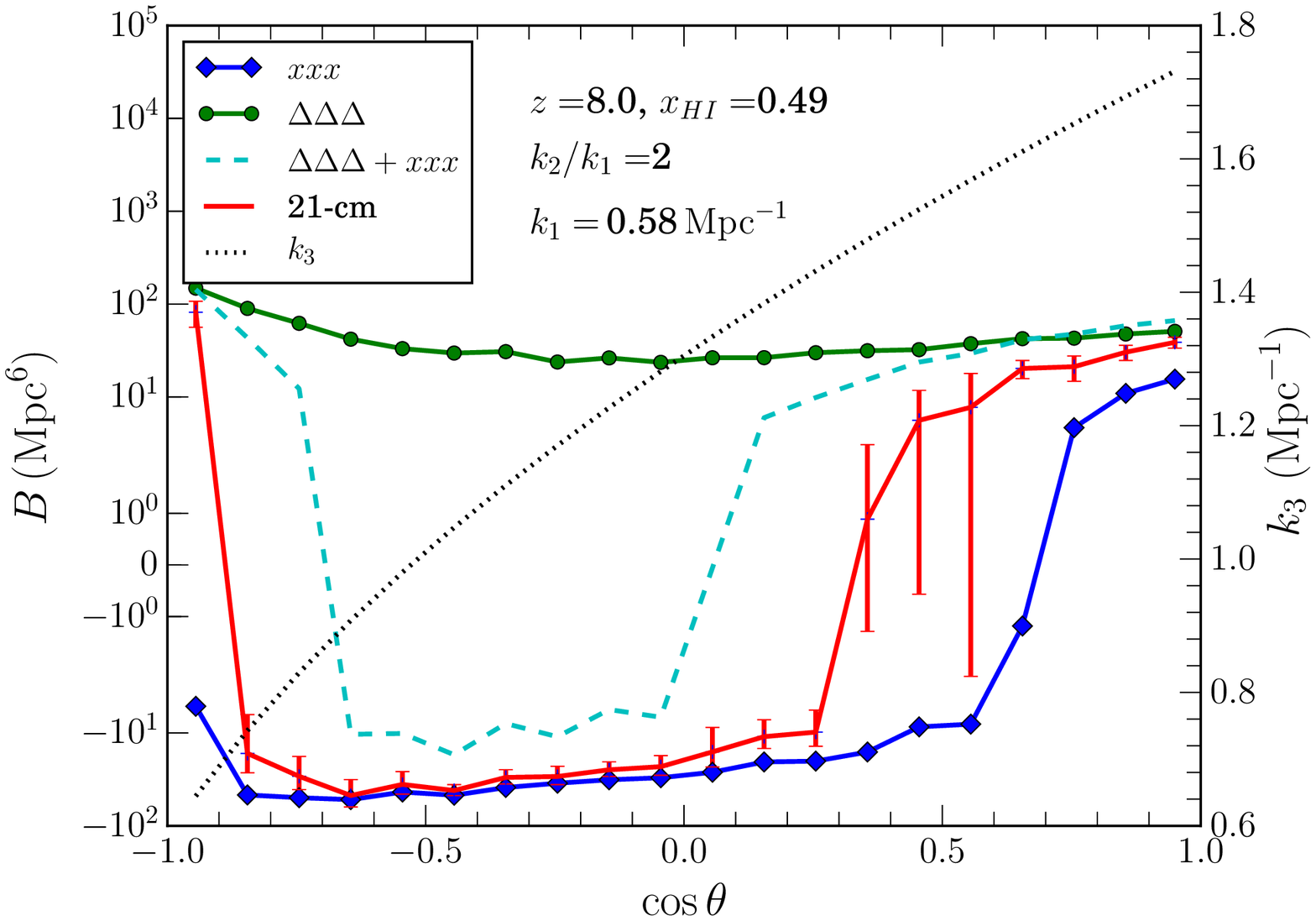}
\caption{Same as Figure \ref{fig:bispec_k2k1_1} but for $k_2/k_1 = 2$
  and $k_1 = 0.20,\, 0.58\, {\rm and}\, 0.83\,{\rm Mpc}^{-1}$. Similar
  to Figure \ref{fig:bispec_k2k1_1}, bottom right panel shows
  $B_{\Delta\Delta\Delta}$ and $B_{xxx}$, their sum, along with the
  21-cm bispectra [$B/\overline{T}^3_b(z)$] for $k_1 = 0.58 \,{\rm
    Mpc}^{-1}$ and $\xb = 0.49$ for comparison.}
\label{fig:bispec_k2k1_2}
\end{figure*}
\subsection{Triangles with $n =2,\, 5$ and $10$}
Both equilateral and isosceles triangles, for which we have discussed
the bispectrum so far, are symmetric in nature. Next we investigate
the bispectrum for a set of asymmetric triangle configurations, which
probe the correlation present in the signal mostly among small and
very large $k$ modes. We start our discussion with triangles having
$k_2/k_1$ or $n = 2$.  Figure \ref{fig:bispec_k2k1_2} shows the
bispectra for this triangle configuration with three representative
$k_1$ values ($k_1 = 0.20,\, 0.58\, {\rm and}\, 0.83 \,{\rm
  Mpc}^{-1}$).  We observe that for triangles with small and
intermediate $k$ modes ($k_1 = 0.2,\, k_2 = 0.4\, {\rm and}\, 0.2
\lesssim k_3 \lesssim 0.6\,{\rm Mpc}^{-1}$, top left panel of Figure
\ref{fig:bispec_k2k1_2}), similar to the behaviour for isosceles
triangles, here also bispectrum remains negative and its amplitude
gradually decreases with increasing $\cos{\theta}$. Furthermore the
overall amplitude of the bispectrum also increases with the decreasing
$\xb$ for $\xb \geq 0.73$, which is in agreement with our toy
model. We observe a deviation from this trend when reionization is
half way through ($\xb \approx 0.5$). For most of the $\cos{\theta}$
range ($-1 \lesssim \cos{\theta} \lesssim 0.5$) it follows the
bispectrum for $\xb = 0.73$ and for $\cos{\theta} \geq 0.5$ we observe
a sharp decline in its amplitude and finally it reaches a positive
value (not shown in the figure). This is the same sign change that we
have observed in the bispectrum for isosceles triangles but here it
occurs at even smaller $k_1$ modes, though the corresponding $k_3$
mode is larger here. This sign change is more prominent at even lower
neutral fractions ($\xb = 0.32$, not shown in the figure).

The bispectrum for $k_1 = 0.58,\, k_2 = 1.16\, {\rm and}\, 0.6
\lesssim k_3 \lesssim 1.72\,{\rm Mpc}^{-1}$ (top right panel of Figure
\ref{fig:bispec_k2k1_2}) also follows the broad features that are
expected from the study of isosceles triangles. In this case as well,
the bispectrum shows a sharp transition from negative to positive
value at higher $\cos{\theta}$. At early stages, $\xb = 0.86$, this
transition happens at $\cos{\theta} \approx 0.65$ ($k_3 = 1.6\,{\rm
  Mpc}^{-1}$) and at late stages, $\xb = 0.32$, it appears around
$\cos{\theta} \approx 0.25$ ($k_3 = 1.4\,{\rm Mpc}^{-1}$).  As
expected, a similar transition is observed for triangles with $k_1 =
0.83\,{\rm Mpc}^{-1}$ (bottom left panel of Figure
\ref{fig:bispec_k2k1_2}) as well. We also observe that the bispectrum
becomes positive at all stages of reionization for the `squeezed'
limit ($\cos{\theta} \approx -1$) of this triangle configuration.

We show the major components contributing to the 21-cm bispectrum for
this triangle type as well (bottom right panel of Figure
\ref{fig:bispec_k2k1_2}) at $\xb = 0.49$ for $k_1 = 0.58\,{\rm
  Mpc}^{-1}$. Just as for isosceles triangles, in this case also the
21-cm bispectrum follows $B_{x x x}$ within the $\cos{\theta}$ range
($0.65 \lesssim k_3 \lesssim 1.40\,{\rm Mpc}^{-1}$) where the 21-cm
bispectrum is negative. After its sharp transition to a positive value
(for $k_3 \geq 1.60\,{\rm Mpc}^{-1} $), it closely follows $B_{\Delta
  \Delta \Delta}$ in shape (during this regime $B_{x x x}$ also become
positive). One can thus draw similar physical inferences as for the
isosceles triangles, as discussed in the previous section. We find
that for extremely `squeezed' triangles ($\cos{\theta} \approx -1$),
the positive 21-cm bispectrum, follows $B_{\Delta \Delta
  \Delta}$. However, we do not have any physical interpretation for
this behaviour of the positive `squeezed' limit bispectrum.

\begin{figure*}
\includegraphics[width=.47\textwidth,angle=0]{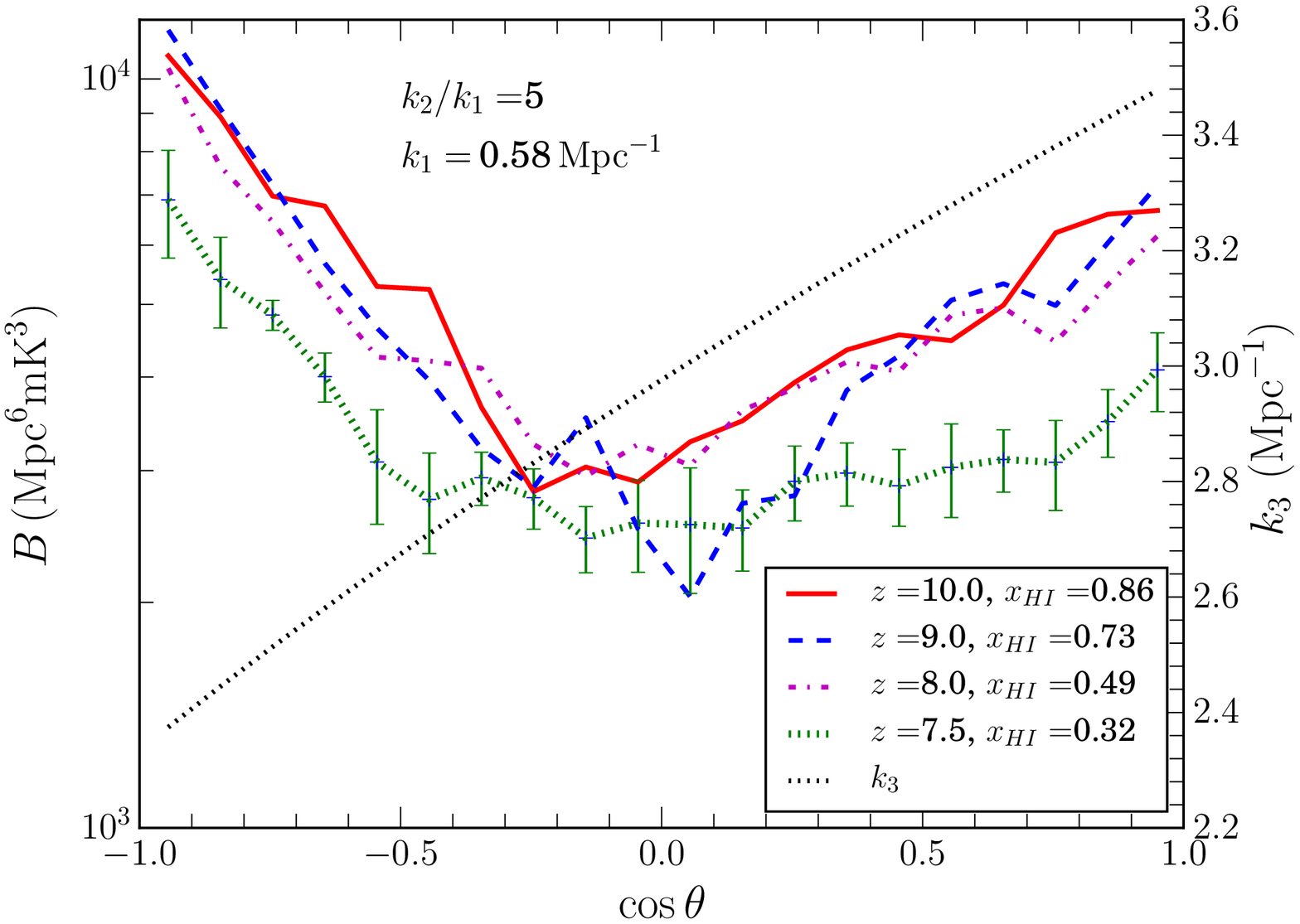}
\includegraphics[width=.47\textwidth,angle=0]{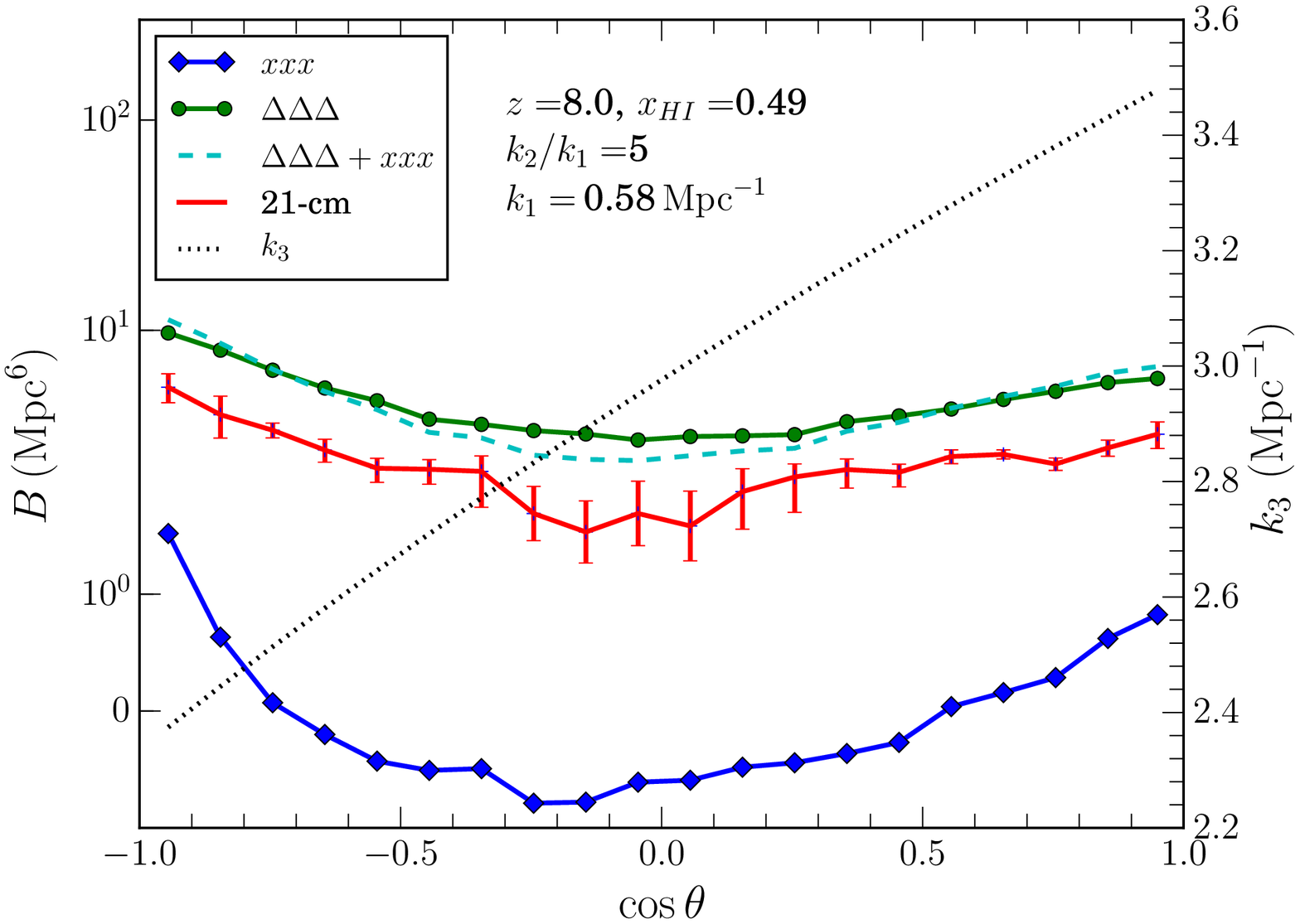}
\caption{Same as Figure \ref{fig:bispec_k2k1_1} but for $k_2/k_1 = 5$
  and $k_1 = 0.58\,{\rm Mpc}^{-1}$. Similar to Figure
  \ref{fig:bispec_k2k1_1}, right panel shows $B_{\Delta\Delta\Delta}$
  and $B_{xxx}$, their sum, along with the 21-cm bispectra
  [$B/\overline{T}^3_b(z)$] for $k_1 = 0.58 \,{\rm Mpc}^{-1}$ and $\xb
  = 0.49$ for comparison.}
\label{fig:bispec_k2k1_5}
\end{figure*}
As many of the broad characteristics of $n=2$ and $n=1$ bispectra are
similar in nature (as they depend on the stage of reionization and the
specific $k$ modes involved), we do not discuss the similar
characteristics for $n =5$ and $10$ triangles. For $n = 5$ triangles
we choose to show bispectra for only one $k_1$ value, to highlight the
unique features that has not been observed for triangle types that we
have studied earlier. The left panel of Figure \ref{fig:bispec_k2k1_5}
shows the bispectrum for $n = 5$ triangles with $k_1 = 0.58,\, k_2 =
2.9\, {\rm and}\, 2.4 \lesssim k_3 \lesssim 3.5\, {\rm Mpc}^{-1}$. The
most obvious feature that one can notice is that the signal bispectra
are positive and have a characteristic `U' shape as functions of
$\cos{\theta}$, irrespective of the stage of reionization. The right
panel of Figure \ref{fig:bispec_k2k1_5}, which shows the major
components of this 21-cm bispectrum at $\xb = 0.49$, highlights the
reason for this `U' shape. It is clearly due to the dominant
contribution from $B_{\Delta \Delta \Delta}$ which also has the same
characteristic shape. The contributions from $B_{x x x}$ and other
higher order cross bispectra (some of which may have negative signs)
mainly reduces the amplitude of the 21-cm bispectrum slightly from
that of $B_{\Delta \Delta \Delta}$, without introducing any
significant change in its shape. It is expected that the overall
amplitude of $B_{\Delta \Delta \Delta}$ will increase with decreasing
redshift, as the matter density fluctuations become more non-linear
(thus non-Gaussian) at lower redshifts. One thus may expect that the
21-cm bispectrum will behave in the similar fashion. However, we do
not see any such pattern in these 21-cm bispectra simply because 21-cm
bispectrum is proportional to $\overline{T}^3_{{\rm b}}(z)$ as well
and $\overline{T}_{{\rm b}}(z)$ decreases with the decreasing redshift
(see Figure \ref{fig:reion_his}).

\begin{figure*}
\includegraphics[width=.47\textwidth,angle=0]{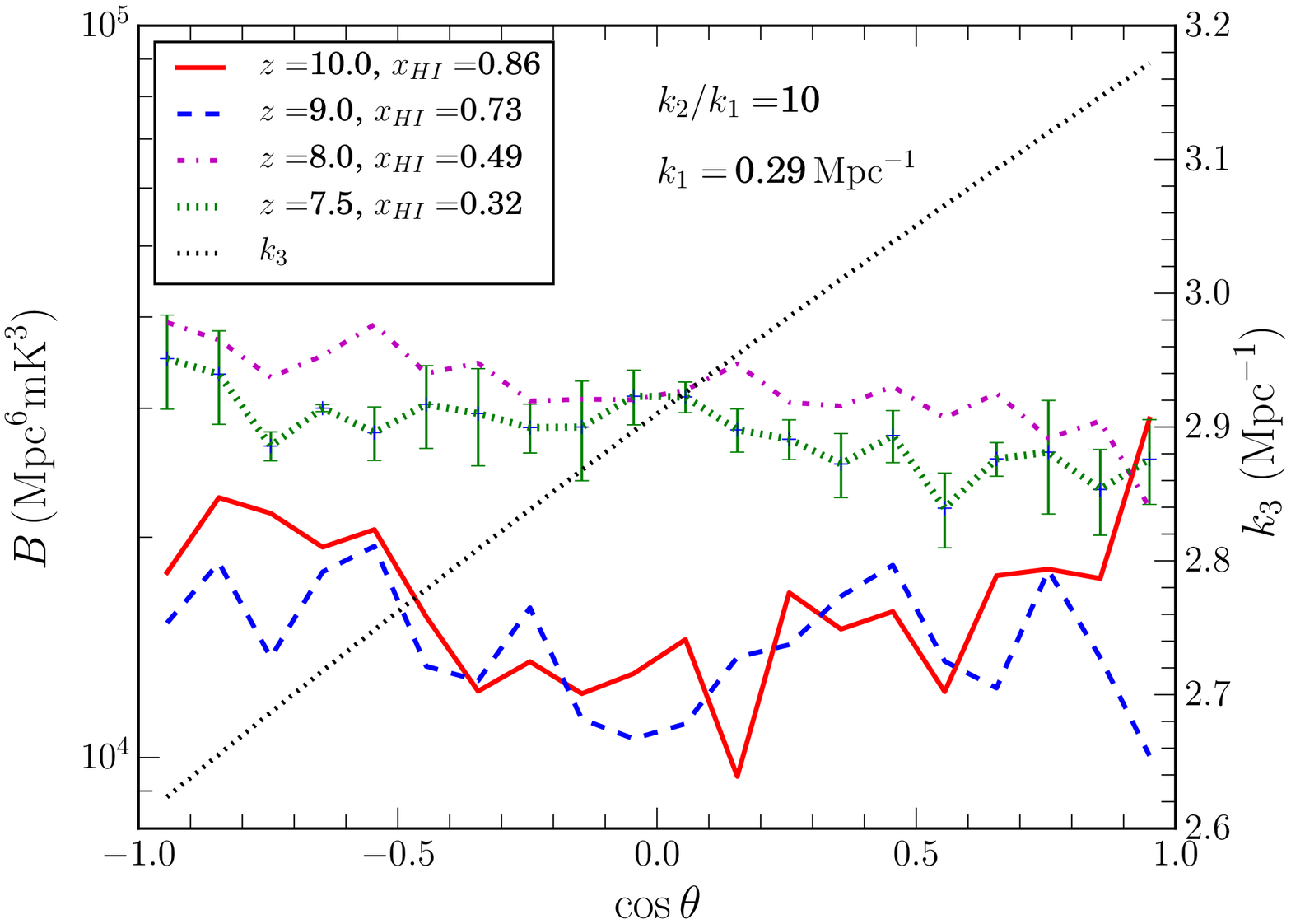}
\includegraphics[width=.47\textwidth,angle=0]{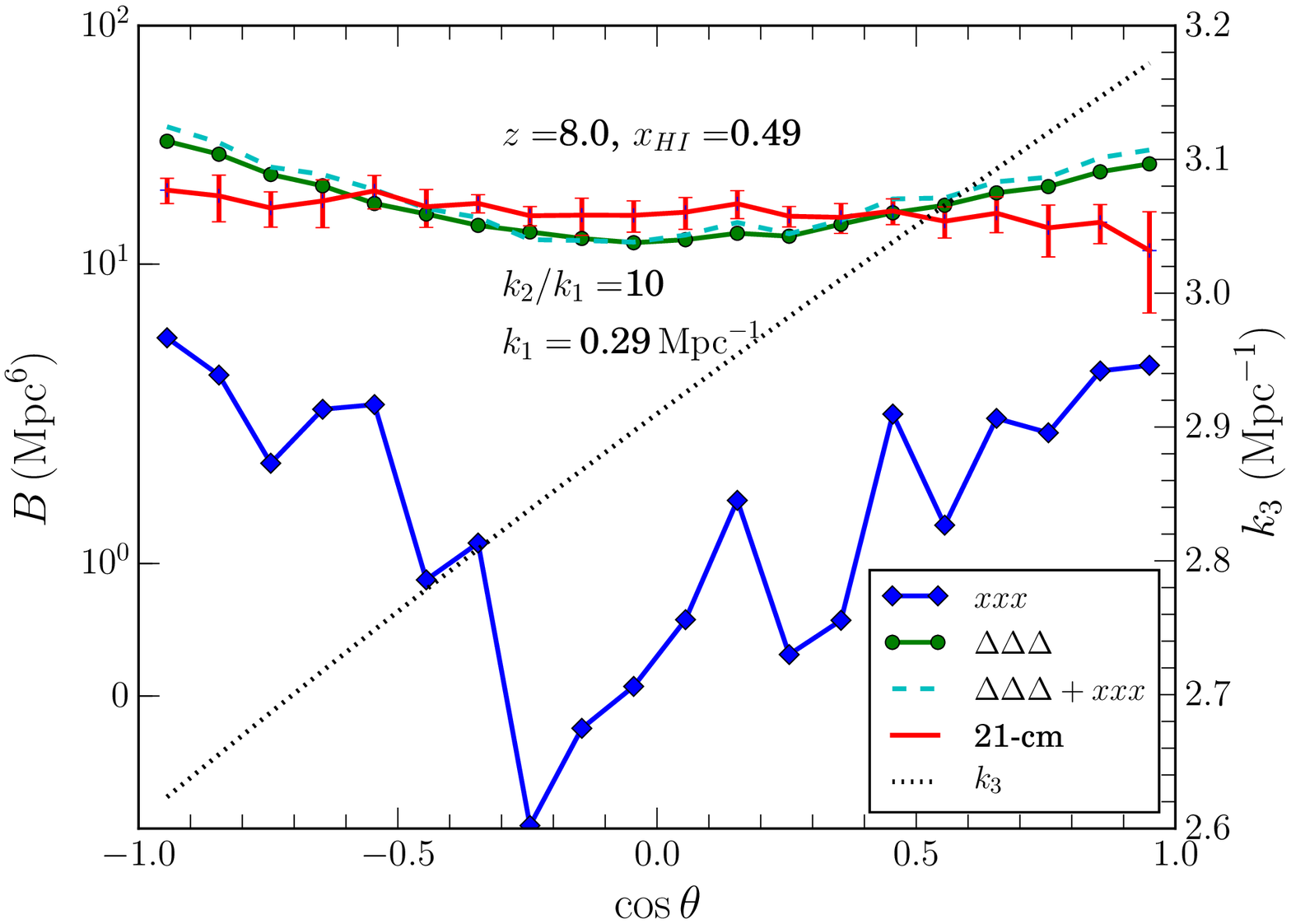}
\caption{Same as Figure \ref{fig:bispec_k2k1_1} but for $k_2/k_1 = 10$
  and $k_1 = 0.29\,{\rm Mpc}^{-1}$. Similar to Figure
  \ref{fig:bispec_k2k1_1}, right panel shows $B_{\Delta\Delta\Delta}$
  and $B_{xxx}$, their sum, along with the 21-cm bispectra
  [$B/\overline{T}^3_b(z)$] for $k_1 = 0.29 \,{\rm Mpc}^{-1}$ and $\xb
  = 0.49$ for comparison.}
\label{fig:bispec_k2k1_10}
\end{figure*}
We next analyse the bispectra for triangle type $n = 10$. Here we show
the bispectra for triangles with $k_1 = 0.29,\, k_2 = 2.9\, {\rm
  and}\, 2.6 \lesssim k_3 \lesssim 3.2\, {\rm Mpc}^{-1}$ (left panel
of Figure \ref{fig:bispec_k2k1_10}). Notice that these triangles probe
the same $k_2$ and approximately same $k_3$ as the $n=5$ triangles. As
expected from the discussion in the previous paragraph, in this case
also we find that the bispectra is positive for the entire range of
$\cos{\theta}$ irrespective of the state of reionization. The
components of this type of bispectra for $\xb \approx 0.49$ (right
panel of Figure \ref{fig:bispec_k2k1_10}) demonstrates that again the
major contributing component is $B_{\Delta \Delta \Delta}$. The 21-cm
bispectrum in this stage follows $B_{\Delta \Delta \Delta}$ in both
shape and amplitude for a wide range of $\cos{\theta}$. Only at the
squeezed and stretched limits of triangles we observe a deviation in
shape and amplitude in the 21-cm bispectra from that of the matter
density field. These deviations are possibly caused by the higher
order cross bispectra contributions. We find that the amplitudes of
the 21-cm bispectra are more or less comparable at different stages of
reionization.

\subsection{Sign of the EoR 21-cm bispectrum}
\label{sec:sign_change}
\begin{figure*}
  \includegraphics[width=.47\textwidth,angle=0]{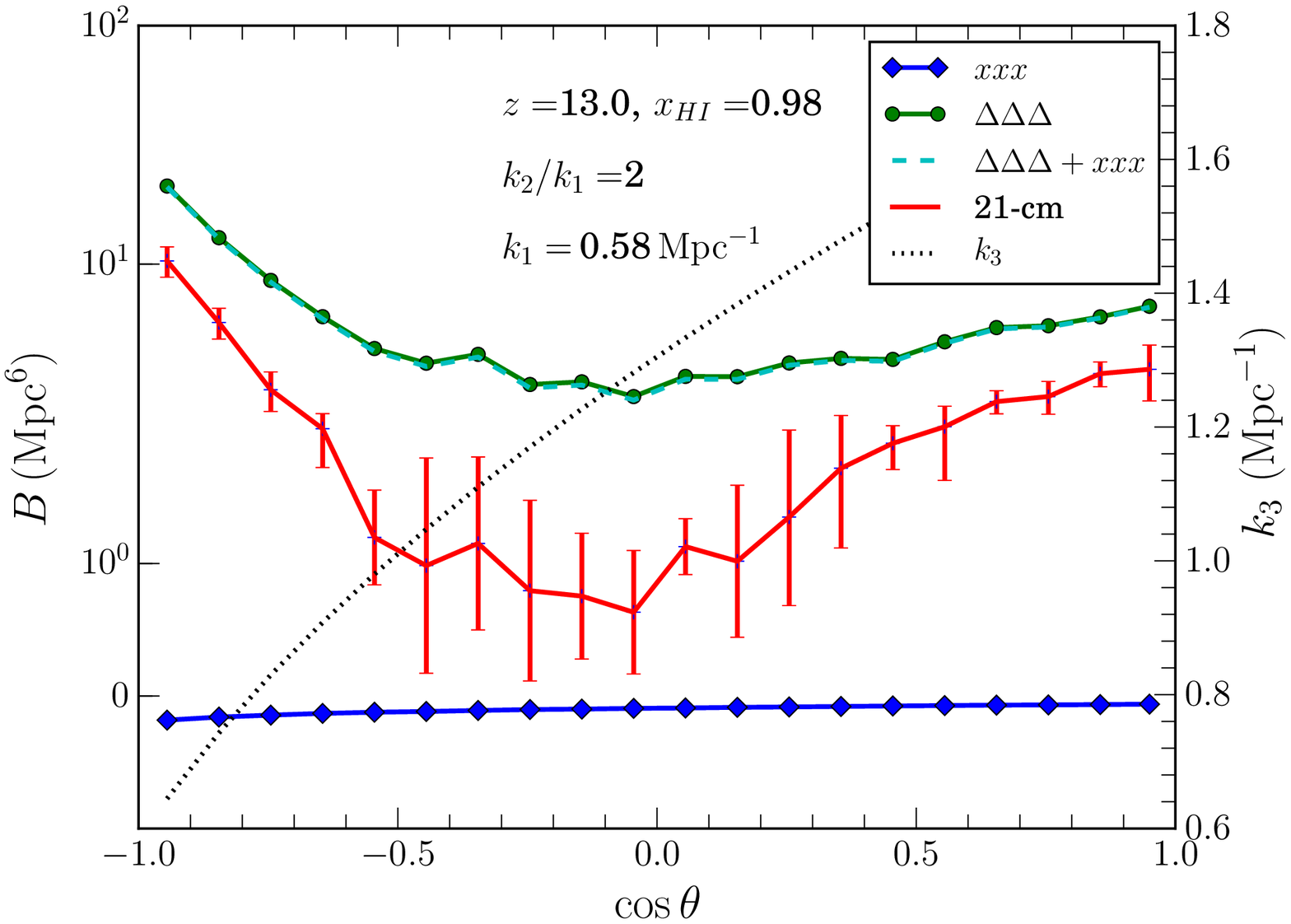}
  \includegraphics[width=.47\textwidth,angle=0]{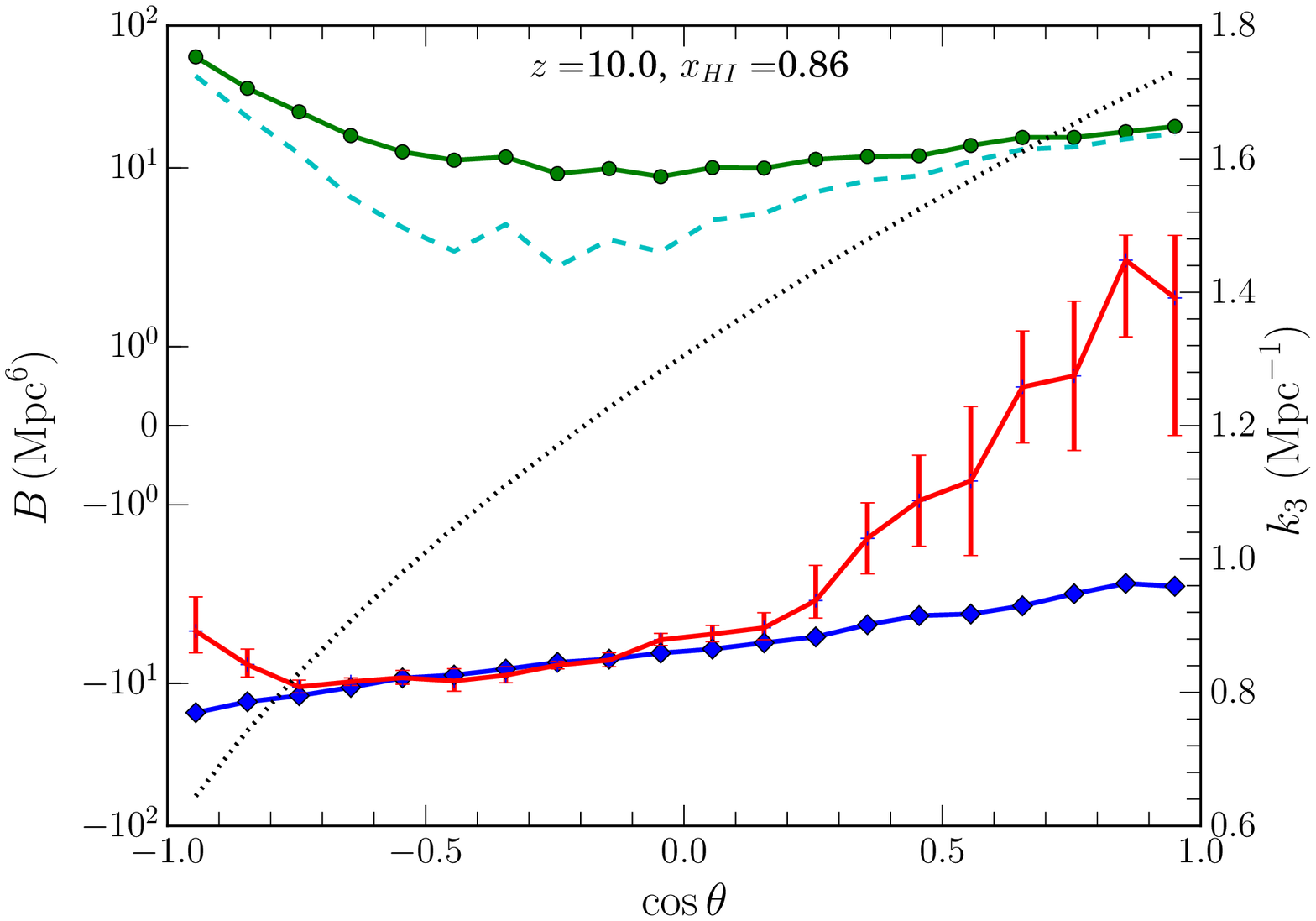}
  \includegraphics[width=.47\textwidth,angle=0]{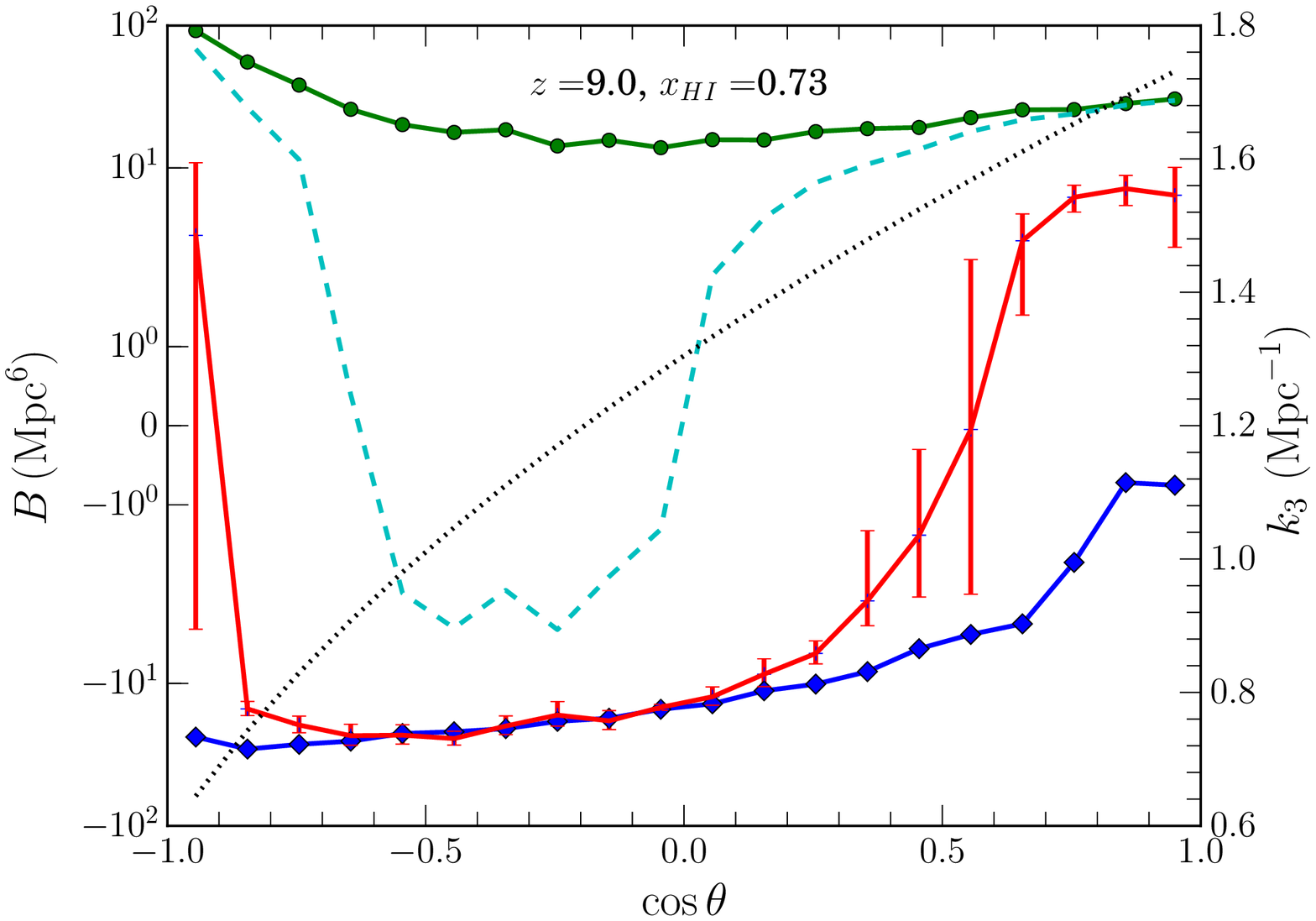}
  \includegraphics[width=.47\textwidth,angle=0]{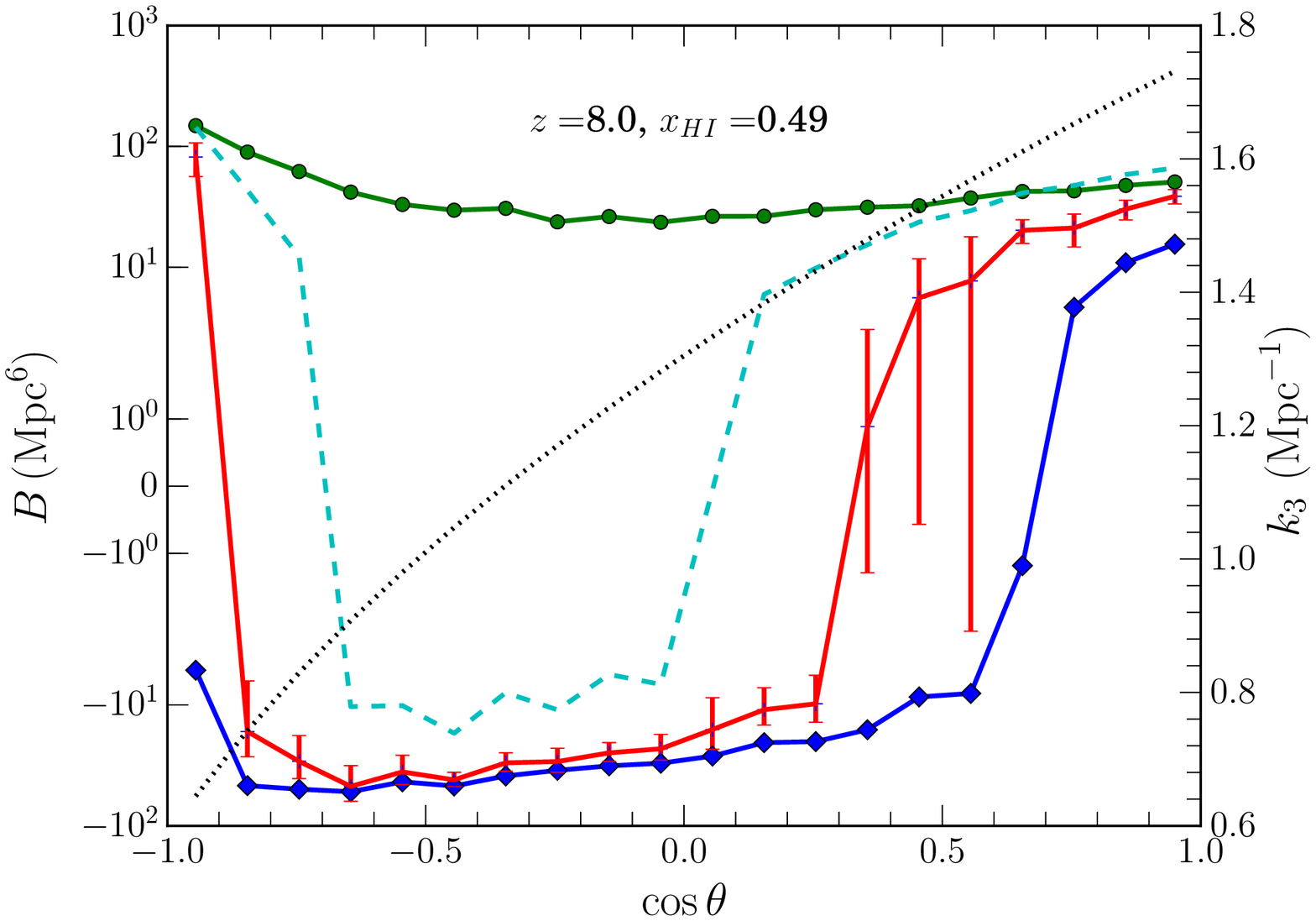}
  \includegraphics[width=.47\textwidth,angle=0]{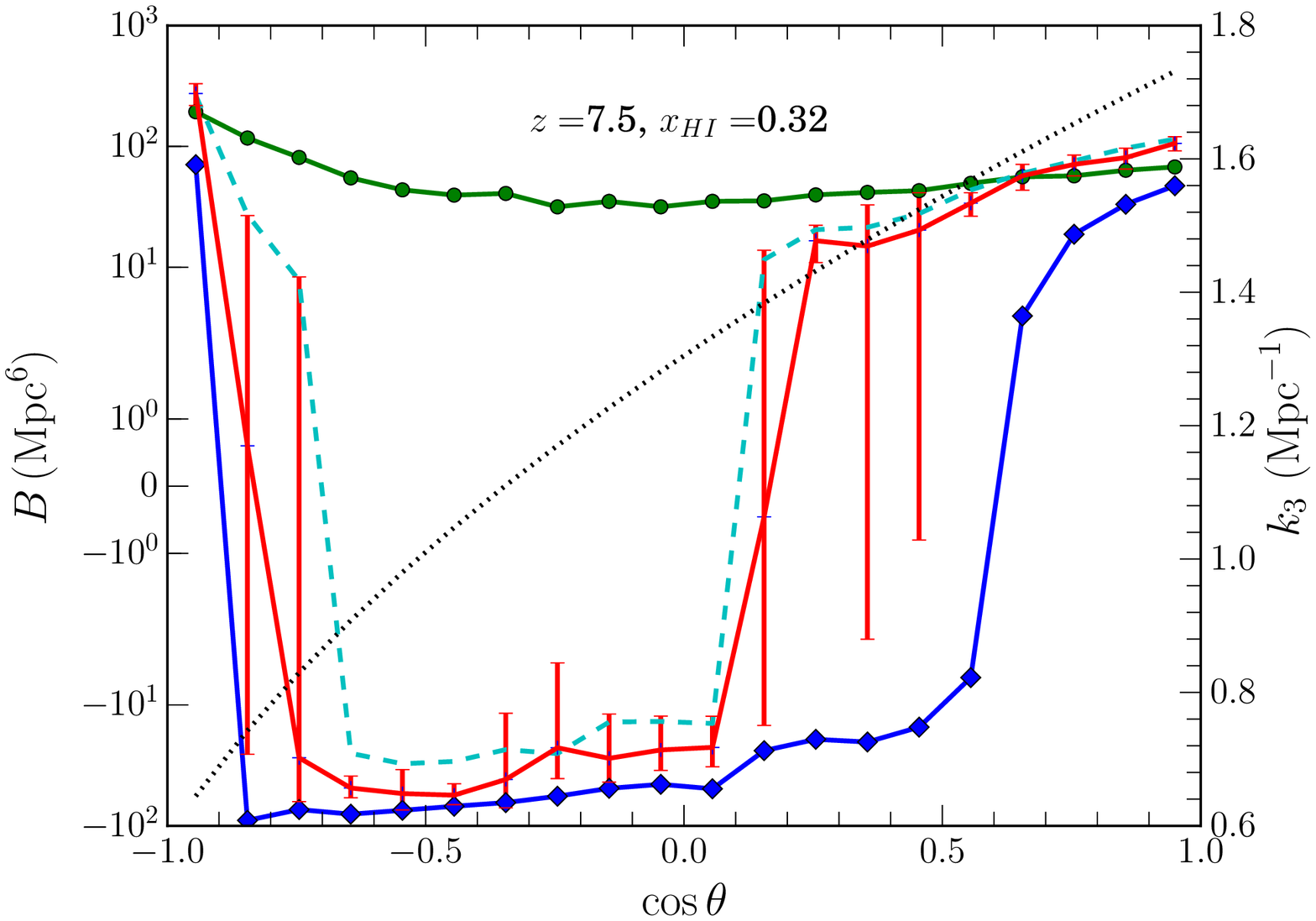}
\caption{Components of 21-cm bispectra [$B/\overline{T}^3_b(z)$] for
  triangles with $k_2/k_1 = 2$ and $k_1 = 0.58\,{\rm Mpc}^{-1}$ at
  five different stages of the EoR.}
\label{fig:bispec_k2k1_2_comp}
\end{figure*}
Our results indicate that the sign of the bispectrum (and its change)
is an important signature of the EoR 21-cm signal. However, this has
not been observed or reported in any of the previous simulation based
studies \citep{yoshiura14,shimabukuro16,shimabukuro16b}, due to the
definition of the bispectrum estimator used in those
studies. \citet{bharadwaj05a} predicted a negative sign for the EoR
21-cm bispectrum through their analytical model. This model does not
predict a sign change in the bispectrum due to the variation in the
${\bf k}$-triangle configurations and with the changing state of
ionization of the universe, which we observe in our analysis here. To
understand the physical significance of the sign of the bispectrum we
revisit this issue in the context of an asymetric triangle
configuration ($k_1 = 0.58\, {\rm Mpc}^{-1}$ and $n = 2$) with the
evolving ionization state of the universe. Our specific choice of
triangle configuration ensures that it prominantly captures the
transition from negative to positive bispectrum as a function of
$\cos{\theta}$ and $\xb$. Being an asymetric triangle configuration,
it will also capture the correlation in the signal present in between
three different wave numbers or $k$ modes. Figure
\ref{fig:bispec_k2k1_2_comp} shows the 21-cm bispectra at five
different stages of the reionization for these triangle
configurations, along with their two major component bispectra
($B_{\Delta \Delta \Delta}$ and $B_{x x x}$).

Our discussion in Section \ref{sec:model} suggests that the
fluctuations in the 21-cm signal and thus its inherent non-Gaussianity
will be determined by the fluctuations in the underlying matter
density and $\xh1$ fields and the interplay between them. At very
early stages of reionization ($\xb = 0.98$), when ionized regions are
minuscule in both number and size, one would expect the 21-cm
fluctuations and its non-Gaussianity to be driven by the fluctuations
present in the matter density field. We observe this expected
behaviour in our simulated 21-cm signal bispectrum (top left panel of
Figure \ref{fig:bispec_k2k1_2_comp}). The 21-cm bispectrum is positive
and follows the shape (the characteristic `U' shape) of $B_{\Delta
  \Delta \Delta}$ at this stage. $B_{x x x}$ is rather negligible in
amplitude (close to zero) throughout the entire $\cos{\theta}$ range
during this period. The correlation between different length scales
(in this case intermediate and small length scales; $k_1 = 0.58\,$,
$k_2 = 1.16\,$, $0.7 \leq k_3 \leq 1.7\, {\rm Mpc}^{-1}$) in the
matter density thus determines the non-Gaussianity in the signal. The
shape of the 21-cm bispectrum is very close to that of the $B_{\Delta
  \Delta \Delta}$, and the difference in their amplitude can possibly
be attributed to the further contribution in the 21-cm bispectrum from
the various cross-bispectra of the two constituent fields.

At a later stage but still early on during reionization ($\xb = 0.86$,
top right panel of Figure \ref{fig:bispec_k2k1_2_comp}), we observe
the first change in sign in the 21-cm bispectrum. The signal
bispectrum becomes negative and starts to follow $B_{x x x}$ very
closely for a wide range in $\cos{\theta}$ (i.e.  $-1 \leq
\cos{\theta} \leq 0.6$ or $0.7 \leq k_3 \leq 1.6\, {\rm
  Mpc}^{-1}$). This is the stage when one would expect the \HII
regions to have grown larger in size and more numerous (compared to
the very early stages of the EoR) but still mostly
isolated. Furthermore, being an inside-out reionization model, these
ionized regions are also expected to cluster around the peaks of the
matter density field. Therefore, the non-Gaussianity of \HI will be
driven by the distribution of these small but numerous, clustered and
isolated \HII regions rather than the non-linearities in the matter
density field. This is approximately the main assumption of our toy
model for \HI fluctuations (see Section \ref{sec:toy_model}), which
also predicts a negative signal bispectrum. In this regime both
$B_{\Delta \Delta \Delta}$ and $B_{x x x}$ are comparable in
amplitude, which implies some of the cross bispectra between $\Delta$
and $\Delta_x$ are able to counter the contribution from the matter
density field, such that the 21-cm bispectrum follows $B_{x x
  x}$. However, when one focuses on the correlations present in the
signal between intermediate and relatively smaller length scales ({\it
  i.e. } $\cos{\theta} \geq 0.6$ or $k_3 \gtrsim 1.6\, {\rm Mpc}^{-1}
$), the 21-cm bispectrum deviates significantly away from $B_{x x x}$
and even changes its sign to positive. This is the regime where two
($k_2$ and $k_3$) out of the three arms of the bispectrum triangle
probes the signal at length scales smaller than the charateristic size
of the \HII regions. Thus one is essentially looking at the
correlations present in the signal at intermediate-small-small scales,
which is expected to be driven by the non-Gaussianities in the matter
density.

As we go to $\xb = 0.73$ (central left panel of Figure
\ref{fig:bispec_k2k1_2_comp}), the characteristics of the signal
bispectrum that we have discussed for $\xb = 0.86$, becomes even more
prominent. By this time most of the \HII bubbles have grown further in
size and at the brink of percolating with each other but are possibly
still somewhat isolated in their spatial distribution. Thus we find
the 21-cm bispectrum to follow $B_{x x x}$ largely within the same
$\cos{\theta}$ range. However, $B_{x x x}$ shows a rapid decline in
amplitude for $\cos{\theta} \geq 0.7$ and the 21-cm bispectrum reaches
a higher positive value in the same range. Note that this increase in
the amplitude of 21-cm bispectrum with decreasing $\xb$, observed in
the panels of Figure \ref{fig:bispec_k2k1_2_comp}, are not a rise in
terms of absolute amplitude of the 21-cm signal (Figure
\ref{fig:bispec_k2k1_2} shows the actual amplitude of the signal
bispectrum), as we have divided the 21-cm bispectrum here by the
factor $\overline{T}^3_b(z)$.

Next at $\xb = 0.49$ (central right panel of Figure
\ref{fig:bispec_k2k1_2_comp}) we are in the percolation phase where
most of the \HII regions are connected with each other through ionized
tunnels. At this stage most of the neutral and ionized gas reside in
two distinct but interconnected and intertwined filamentary structures
(see figure 3 of \citealt{bag18}). The length scales associated with
sizes and distribution of the ionized tunnels and neutral bridges in
these structures determine whether $B_{x x x}$ or $B_{\Delta \Delta
  \Delta}$ will dominate in terms of their contribution to the 21-cm
bispectrum. This will in turn affect the values of $k_3$ at which one
would see the transition from a negative to positive bispectrum. We
observe that at this stage the transition in sign happens at a smaller
value of $k_3$ ($\cos{\theta} \approx 0.4-0.5$ i.e. $k_3 \approx 1.5
\,{\rm Mpc}^{-1}$) compared to the stage when $\xb = 0.73$. We also
observe a positive bispectrum for the `squeezed' limits triangles at
this stage, for which we presently do not have a proper physical
interpretation.

With further ionization of the IGM ($\xb = 0.32$, bottom panel of
Figure \ref{fig:bispec_k2k1_2_comp}), the topology of the low density
interconnected neutral region evolves further, thus affecting the
particular length scale for the sign change of the bispectrum. This
makes the change in sign to appear at $k_3 \gtrsim 1.45\,{\rm
  Mpc}^{-1}$. $B_{x x x}$ also reaches a significantly high positive
value for $k_3 \approx 1.7\,{\rm Mpc}^{-1}$, comparable to that of the
$B_{\Delta \Delta \Delta}$. The sum of $B_{x x x}$ and $B_{\Delta
  \Delta \Delta}$ also provides a rather good description of 21-cm
bispectrum for the entire range of $\cos{\theta}$ at this stage.

We conclude that the sign of the bispectrum is a crucial
characteristic of the EoR 21-cm signal. For a specific triangle
configuration, the shape of the bispectrum (as a function
$\cos{\theta}$) along with its sign can potentially tell us about the
topology of the \HI distribution and the state of the
reionization. The sign can thus also reveal whether the bispectrum at
that stage is driven by the non-Gaussianities in the matter density or
the $\xh1$ fluctuations.

\section{Summary}
\label{sec:summary}
The redshifted 21-cm signal from the epoch of reionization is expected
to be highly non-Gaussian in nature. This is due the fact that the
major contribution in the signal fluctuations come from the
distribution of growing \HII regions in the neutral IGM. The degree of
this non-Gaussianity is also expected to evolve with the changing
state of ionization of the IGM during this period. In this article, we
have explored the possibility of using the bispectrum (the Fourier
equivalent of the three point correlation function) to quantify this
non-Gaussianity in the signal. Unlike the power spectrum, the
bispectrum or any other higher-order Fourier statistic of the signal
helps us to quantify the correlation present in the signal between
different Fourier modes. We have used a variety of triangle
configurations and an ensemble of semi-numerical simulations to study
various aspects of the evolving non-Gaussianity in the 21-cm signal
through the bispectrum. In this section we summarize our findings and
their associated interpretations.

Starting with equilateral $k$-triangles (which are sensitive to the
correlations present in the signal between same/similar $k$ modes), we
observe that the bispectrum is non-zero, negative and increases in
amplitude with a decreasing global neutral fraction (up to $\xb \geq
0.5$) for a wide range of Fourier modes ($0.1 \lesssim k \lesssim
1.0\,{\rm Mpc}^{-1}$). The statistically significant non-zero signal
bispectrum establishes the fact that the EoR 21-cm signal is indeed
highly non-Gaussian. The signal bispectrum is also found to closely
follow the $\xh1$ bispectrum. The matter density and other cross
bispectra between the $\xh1$ and the matter density do not affect the
21-cm bispectrum significantly in this regime. Supported by a toy
model of \HI fluctuations, we find that this behaviour of the 21-cm
bispectrum can be ascribed to the evolution of the size distribution
of the \HII bubbles and the global state of ionization of the IGM. The
behaviour of the signal bispectrum deviates from the predictions of
this toy model during the later stages of the EoR ($\xb \leq 0.5$) and
for triangles with larger $k$ modes ($k \geq 1.0 \,{\rm Mpc}^{-1}$),
though it still largely follows the underlying $\xh1$ bispectrum. This
deviation is possibly due to the fact that by this time, most of the
isolated \HII regions have been connected through ionized tunnels to
form a rather filamentary and large ionized region, which is
intertwined with a similar filamentray and large neutral region. Thus
the toy model bispectrum fails to describe the signal qualitatively at
this stage.

We also demonstrate that the evolution of the redshifted 21-cm signal
from the EoR can be visualized as trajectories in the phase space of
power spectrum and equilateral bispectrum [$P(k) - B(k, k, k)$] with
the evolving global neutral fraction $\xb$. These trajectories have
unique shape and turn around points which can be used to identify and
confirm the detection of the signal. Further as these trajectories are
expected to be dependent on the 21-cm topology (which depends on the
reionization model), thus a joint analysis of the signal involving
both $P(k)$ and $B(k, k, k)$ can be used to rule out certain
reionization models or to put constrain the reionization
parameters. Such joint analysis of the signal could provide a robust
constraint on the signal parameters due to the following fundamental
reasons: a) One would be using two independent statistics of the
signal jointly. b) One of the statistics (the bispectrum) will be able
to take into account the non-Gaussianity present in the signal, which
cannot be captured by the power spectrum.

The bispectra for the isosceles triangles capture the correlation
present in the signal between two different $k$ modes. Like the
equilateral triangle bispectrum, for relatively large length scales
({\it i.e.} $k_1 \leq 0.29\, {\rm Mpc}^{-1}$ and $k_3 \leq 0.6\, {\rm
  Mpc}^{-1} $) it is negative and its amplitude also increases with
decreasing $\xb$ (for $\xb \geq 0.5$). We also observe that, at any
stage of reionization, the amplitude of the isosceles bispectrum is a
maximum for the very squeezed triangles ($\cos{\theta} \approx -1$)
and the amplitude gradually goes down as one approaches the stretched
limit ($\cos{\theta} \approx 1$). In this regime, the signal
bispectrum closely follows the $\xh1$ bispectrum and these behaviours
of the signal bispectrum can be mimicked by our toy model for $\xh1$
fluctuations, thus establishing the fact that in this regime the
non-Gaussianity in the signal is driven by the distribution of
isolated \HII regions in the IGM.

We observe a very interesting feature in the isosceles bispectrum as
we move towards the triangles with intermediate and small length
scales ($k_1 \geq 0.59\, {\rm Mpc}^{-1}$ and $k_3 \geq 1.0\, {\rm
  Mpc}^{-1}$) and specifically at the later stages of reionization
($\xb \leq 0.86$). The signal bispectrum in this regime changes sign
from negative to positive. We also observe that this is the regime
where the amplitude of the $\xh1$ bispectrum becomes smaller than the
matter density bispectrum and the signal bispectrum starts to follow
the matter bispectrum closely. A similar behaviour is consistently
observed when we analyze the bispectrum for asymmetric triangle
configurations ($k_2/k_1 = 2$), involving $k$ modes in the same range
($k_1 \geq 0.58\,{\rm Mpc}^{-1}$, $k_2 \geq 1.16\,{\rm Mpc}^{-1}$ and
$k_3 \geq 1.3\,{\rm Mpc}^{-1} $). We further observe that, for a
single triangle configuration, this sign change is dependent on the
state of ionization of the IGM and appears at consistently smaller
values of $k_3$ for lower $\xb$.

This change in the sign and the associated behaviours of the signal
bispectrum cannot be mimicked by our toy model of $\xh1$ fluctuation,
which consists of randomly distributed non-overlapping ionized spheres
in a uniformly dense IGM. In reality, the \HII regions are neither
spherical nor randomly distributed and also the IGM is not uniformly
dense. The ionized regions are expected to cluster around the
collapsed and bound structures in the IGM and as reionization
progresses towards the percolation regim (i.e. $\xb \lesssim 0.73$)
most of them are also expected to become connected with each other
through ionized tunnels and result in a single filamentary ionized
region. A similar web like topology is observed for the neutral IGM as
well \citep{furlanetto16, bag18}.

Depending on the evolving length (as they do not evolve significantly
in cross-section) and distribution of the ionized tunnels and the
neutral bridges the signal is expected to be correlated between
different Fourier modes. This correlation and associated Fourier modes
are also expected to evolve. Further, the Fourier modes corresponding
to the changing lengths of these neutral bridges will be affected by
the non-Gaussianities of the underlying matter density (as we assume
hydrogen follows the matter distribution). The bispectra for isosceles
and asymmetric $k$ triangle configurations are able to capture this
correlation between small and intermediate length scales. This is why
the signal changes sign and becomes positive in this regime. The
corresponding $k$ modes, where this sign change appears, also
gradually evolve with the evolution in the topology of the ionization
and neutral field. For asymmetric triangles ($k_2/k_1 = 5$ and $10$)
with at least one significantly large $k$ mode ($k_3 \geq 2.4 \,{\rm
  Mpc}^{-1} $), the 21-cm bispectrum stays positive within the entire
range of $\cos{\theta}$, irrespective of the state of ionization of
the IGM and follows closely the shape of the matter bispectrum.

Thus the sign of the EoR 21-cm bispectrum is a very crucial signature
of non-Gaussianity in the signal. A negative bispectrum implies that
non-Gaussianities in the corresponding length scales and at that
specific stage of reionization is being driven by the size
distribution and the topology of the ionized regions. On the other
hand, a positive bispectrum implies that the dominant non-Gaussian
contribution comes from the matter bispectrum and various other cross
bispectra. This change in sign of the signal bispectrum, for a
specific $k$ triangle configuration, as a function of $\cos{\theta}$
can be used as a confirmative test for the detection of the
signal. The $\cos{\theta}$ value where this sign change appears can be
further parametrized to estimate the global neutral fraction of the
IGM. It is very difficult to ascertain whether the signal fluctuations
are mostly driven by the dark matter or the $\xh1$ field by solely
using the power spectrum of the EoR 21-cm signal. However, the sign of
the signal bispectrum can be used a confirmative marker to identify
regimes where the signal fluctuations are driven by the neutral
fraction field.  Finally, as the bispectrum (both shape and amplitude)
and its sign are strongly dependent on the topology of 21-cm signal,
this can be used to distinguish between different EoR source models
that gives rise to different 21-cm topologies.

In the context of the EoR 21-cm bispectrum, one important issue is to
identify the optimal $k$ triangle configurations for which one can
extract maximum amount of information about the signal. To properly
address this issue one needs to estimate bispectra for all possible
triangle configurations (in the $n-\cos{\theta}$ parameter space)
along with an wide range in $k$ amplitudes, which will require a large
amount of computational time. We plan to conduct such an analysis with
a faster bispectrum estimation algorithm in a future follow up
work. Among the different triangle configurations discussed in this
paper, we identify isosceles ($n=1$) and $n=2$ triangles to have
maximum amount of interesting features which can probe the evolution
of the signal with the progressing state of reionization. The
triangles with $n=5$ and $10$ stays mostly unaffected by the evolution
of $\xb$. It might also be unfeasible to probe the bispectrum for
these triangles with SKA1-LOW or HERA, as the $k$ modes probed by
these triangles are very large ($k \geq 2.0 \, {\rm Mpc}^{-1}$) and
may fall outside the sensitivity limits of these telescopes
\citep{koopmans15}.

Note that, in our simulations for the EoR 21-cm signal we have assumed
all halos of mass $\geq 10^9\,{\rm M_{\odot}}$ to contribute in the
ionizing photon production. However, halos in the mass range $10^6
\leq {\rm M_{halo}} \leq 10^9\,{\rm M_{\odot}}$ may also have
significant contribution in the ionization photon budget. Further, the
star formation in these low mass halos may also get supressed due to
thermal feedback once their environment is sufficiently ionized. All
of these together can affect the evolution of the 21-cm topology. Our
simulations also do not take into account the effect of spin
temperature fluctuations due to Lyman-$\alpha$ coupling and X-ray
heating during the early stages of the reionization, which can make
the signal significantly correlated in different Fourier modes and may
have a prominent signature in the signal bispectrum (see
e.g. \citealt{furlanetto06a, pritchard07, fialkov14, fialkov15,
  ghara14, ghara15, watkinson15, ross17} etc). We have also not
included the effect of the ``redshift space distortions'' due to the
matter peculiar velocities (see e.g. \citealt{bharadwaj04, barkana05,
  mao12,majumdar13,jensen13,fialkov15,ghara15,majumdar16}) and the
``light cone effect'' arising due to the finite light travel time (see
e.g. \citealt{barkana06, datta12, datta14, zawada14, laplante14,
  ghara15, majumdar16b, mondal17b}) in our simulated signal. These effects will
introduce anisotropies in the signal along the line-of-sight of a
present day observer and in turn can affect towards the contribution
from the various constituents fields on the EoR 21-cm
bispectrum. Further, we have not discussed the detectability of the
EoR 21-cm bispectrum using various presently operating (e.g. LOFAR)
and upcoming (e.g. HERA and SKA1-LOW) radio interferometers. We plan
to study the signature of these effects on the 21-cm signal bispectrum
along with an analysis to find the optimal $k$ triangle configurations
for which the signal bispectrum will be detectable using different
radio interferometers in follow up future works.

\section*{Acknowledgements}
SM would like to thank Adam Lidz, Eiichiro Komatsu, Chris L. Carilli
and Samir Choudhuri for useful discussions on various aspects of the
bispectrum and Fourier transform. SM and JRP acknowledge financial
support from the European Research Council under ERC grant number
638743-FIRSTDAWN. CAW thanks the Science and Technology Facilities
Council via the SKA-preconstruction-phase-continuation
grant. Simulations used in this work were done at the computational
facilities at the Centre for Theoretical Studies, IIT Kharagpur,
India. A part of this work and related discussions were conducted
during the workshop on ``Cosmic Reionization'' supported by the Munich
Institute for Astro and Particle Physics (MIAPP) of the DFG cluster of
excellence ``Origin and Structure of the Universe''.

\bibliographystyle{mn2e} 
\bibliography{refs}

\begin{thebibliography}{106}
\expandafter\ifx\csname natexlab\endcsname\relax\def\natexlab#1{#1}\fi

\bibitem[{{Ali}, {Bharadwaj} \& {Chengalur}(2008){Ali}, {Bharadwaj}, \&
  {Chengalur}}]{ali08}
{Ali} S.~S., {Bharadwaj} S., {Chengalur} J.~N., 2008, \mnras, 385, 2166

\bibitem[{{Ali} {et~al}\mbox{.}(2015){Ali}, {Parsons}, {Zheng}, {Pober}, {Liu},
  {Aguirre}, {Bradley}, {Bernardi}, {Carilli}, {Cheng}, {DeBoer}, {Dexter},
  {Grobbelaar}, {Horrell}, {Jacobs}, {Klima}, {MacMahon}, {Maree}, {Moore},
  {Razavi}, {Stefan}, {Walbrugh}, \& {Walker}}]{ali15}
{Ali} Z.~S. {et~al.}, 2015, \apj, 809, 61

\bibitem[{{Alvarez} {et~al}\mbox{.}(2006){Alvarez}, {Shapiro}, {Ahn}, \&
  {Iliev}}]{alvarez06}
{Alvarez} M.~A., {Shapiro} P.~R., {Ahn} K., {Iliev} I.~T., 2006, \apjl, 644,
  L101

\bibitem[{{Bag} {et~al}\mbox{.}(2018){Bag}, {Mondal}, {Sarkar}, {Bharadwaj}, \&
  {Sahni}}]{bag18}
{Bag} S., {Mondal} R., {Sarkar} P., {Bharadwaj} S., {Sahni} V., 2018, ArXiv
  e-prints:1801.01116

\bibitem[{{Barkana} \& {Loeb}(2005)}]{barkana05}
{Barkana} R., {Loeb} A., 2005, \apjl, 624, L65

\bibitem[{{Barkana} \& {Loeb}(2006)}]{barkana06}
{Barkana} R., {Loeb} A., 2006, \mnras, 372, L43

\bibitem[{{Barnett} {et~al}\mbox{.}(2017){Barnett}, {Warren}, {Becker},
  {Mortlock}, {Hewett}, {McMahon}, {Simpson}, \& {Venemans}}]{barnett17}
{Barnett} R., {Warren} S.~J., {Becker} G.~D., {Mortlock} D.~J., {Hewett} P.~C.,
  {McMahon} R.~G., {Simpson} C., {Venemans} B.~P., 2017, \aap, 601, A16

\bibitem[{{Becker} {et~al}\mbox{.}(2015){Becker}, {Bolton}, {Madau}, {Pettini},
  {Ryan-Weber}, \& {Venemans}}]{becker15}
{Becker} G.~D., {Bolton} J.~S., {Madau} P., {Pettini} M., {Ryan-Weber} E.~V.,
  {Venemans} B.~P., 2015, \mnras, 447, 3402

\bibitem[{{Becker} {et~al}\mbox{.}(2001){Becker}, {Fan}, {White}, {Strauss},
  {Narayanan}, {Lupton}, {Gunn}, {Annis}, {Bahcall}, {Brinkmann}, {Connolly},
  {Csabai}, {Czarapata}, {Doi}, {Heckman}, {Hennessy}, {Ivezi{\'c}}, {Knapp},
  {Lamb}, {McKay}, {Munn}, {Nash}, {Nichol}, {Pier}, {Richards}, {Schneider},
  {Stoughton}, {Szalay}, {Thakar}, \& {York}}]{becker01}
{Becker} R.~H. {et~al.}, 2001, \aj, 122, 2850

\bibitem[{{Bharadwaj} \& {Ali}(2004)}]{bharadwaj04}
{Bharadwaj} S., {Ali} S.~S., 2004, \mnras, 352, 142

\bibitem[{{Bharadwaj} \& {Ali}(2005)}]{bharadwaj05}
{Bharadwaj} S., {Ali} S.~S., 2005, \mnras, 356, 1519

\bibitem[{{Bharadwaj} \& {Pandey}(2005)}]{bharadwaj05a}
{Bharadwaj} S., {Pandey} S.~K., 2005, \mnras, 358, 968

\bibitem[{{Bharadwaj} \& {Sethi}(2001)}]{bharadwaj01a}
{Bharadwaj} S., {Sethi} S.~K., 2001, Journal of Astrophysics and Astronomy, 22,
  293

\bibitem[{{Bharadwaj} \& {Srikant}(2004)}]{bharadwaj04b}
{Bharadwaj} S., {Srikant} P.~S., 2004, Journal of Astrophysics and Astronomy,
  25, 67

\bibitem[{{Bouwens}(2016)}]{bouwens16}
{Bouwens} R., 2016, in Astrophysics and Space Science Library, Vol. 423,
  Astrophysics and Space Science Library, {Mesinger} A., ed., p. 111

\bibitem[{{Bouwens} {et~al}\mbox{.}(2015){Bouwens}, {Illingworth}, {Oesch},
  {Caruana}, {Holwerda}, {Smit}, \& {Wilkins}}]{bouwens15}
{Bouwens} R.~J., {Illingworth} G.~D., {Oesch} P.~A., {Caruana} J., {Holwerda}
  B., {Smit} R., {Wilkins} S., 2015, \apj, 811, 140

\bibitem[{{Bowman} {et~al}\mbox{.}(2013){Bowman}, {Cairns}, {Kaplan}, {Murphy},
  {Oberoi}, {Staveley-Smith}, {Arcus}, {Barnes}, {Bernardi}, {Briggs}, {Brown},
  {Bunton}, {Burgasser}, {Cappallo}, {Chatterjee}, {Corey}, {Coster},
  {Deshpande}, {deSouza}, {Emrich}, {Erickson}, {Goeke}, {Gaensler},
  {Greenhill}, {Harvey-Smith}, {Hazelton}, {Herne}, {Hewitt},
  {Johnston-Hollitt}, {Kasper}, {Kincaid}, {Koenig}, {Kratzenberg}, {Lonsdale},
  {Lynch}, {Matthews}, {McWhirter}, {Mitchell}, {Morales}, {Morgan}, {Ord},
  {Pathikulangara}, {Prabu}, {Remillard}, {Robishaw}, {Rogers}, {Roshi},
  {Salah}, {Sault}, {Shankar}, {Srivani}, {Stevens}, {Subrahmanyan}, {Tingay},
  {Wayth}, {Waterson}, {Webster}, {Whitney}, {Williams}, {Williams}, \&
  {Wyithe}}]{bowman13}
{Bowman} J.~D. {et~al.}, 2013, \pasa, 30, 31

\bibitem[{{Choudhury}(2009)}]{choudhury09}
{Choudhury} T.~R., 2009, Current Science, 97, 841

\bibitem[{{Choudhury}, {Haehnelt} \& {Regan}(2009){Choudhury}, {Haehnelt}, \&
  {Regan}}]{choudhury09b}
{Choudhury} T.~R., {Haehnelt} M.~G., {Regan} J., 2009, \mnras, 394, 960

\bibitem[{{Choudhury} {et~al}\mbox{.}(2015){Choudhury}, {Puchwein}, {Haehnelt},
  \& {Bolton}}]{choudhury14}
{Choudhury} T.~R., {Puchwein} E., {Haehnelt} M.~G., {Bolton} J.~S., 2015,
  \mnras, 452, 261

\bibitem[{{Datta}, {Choudhury} \& {Bharadwaj}(2007){Datta}, {Choudhury}, \&
  {Bharadwaj}}]{datta07a}
{Datta} K.~K., {Choudhury} T.~R., {Bharadwaj} S., 2007, \mnras, 378, 119

\bibitem[{{Datta} {et~al}\mbox{.}(2014){Datta}, {Jensen}, {Majumdar},
  {Mellema}, {Iliev}, {Mao}, {Shapiro}, \& {Ahn}}]{datta14}
{Datta} K.~K., {Jensen} H., {Majumdar} S., {Mellema} G., {Iliev} I.~T., {Mao}
  Y., {Shapiro} P.~R., {Ahn} K., 2014, \mnras, 442, 1491

\bibitem[{{Datta} {et~al}\mbox{.}(2012){Datta}, {Mellema}, {Mao}, {Iliev},
  {Shapiro}, \& {Ahn}}]{datta12}
{Datta} K.~K., {Mellema} G., {Mao} Y., {Iliev} I.~T., {Shapiro} P.~R., {Ahn}
  K., 2012, \mnras, 424, 1877

\bibitem[{{Davis} {et~al}\mbox{.}(1985){Davis}, {Efstathiou}, {Frenk}, \&
  {White}}]{davis85}
{Davis} M., {Efstathiou} G., {Frenk} C.~S., {White} S.~D.~M., 1985, \apj, 292,
  371

\bibitem[{{Di Matteo} {et~al}\mbox{.}(2002){Di Matteo}, {Perna}, {Abel}, \&
  {Rees}}]{dimatteo02}
{Di Matteo} T., {Perna} R., {Abel} T., {Rees} M.~J., 2002, \apj, 564, 576

\bibitem[{{Dillon} {et~al}\mbox{.}(2014){Dillon}, {Liu}, {Williams}, {Hewitt},
  {Tegmark}, {Morgan}, {Levine}, {Morales}, {Tingay}, {Bernardi}, {Bowman},
  {Briggs}, {Cappallo}, {Emrich}, {Mitchell}, {Oberoi}, {Prabu}, {Wayth}, \&
  {Webster}}]{dillon14}
{Dillon} J.~S. {et~al.}, 2014, \prd, 89, 023002

\bibitem[{{Ewall-Wice} {et~al}\mbox{.}(2014){Ewall-Wice}, {Dillon}, {Mesinger},
  \& {Hewitt}}]{ewallwice14}
{Ewall-Wice} A., {Dillon} J.~S., {Mesinger} A., {Hewitt} J., 2014, \mnras, 441,
  2476

\bibitem[{{Fan}, {Carilli} \& {Keating}(2006){Fan}, {Carilli}, \&
  {Keating}}]{fan06b}
{Fan} X., {Carilli} C.~L., {Keating} B., 2006, \araa, 44, 415

\bibitem[{{Fan} {et~al}\mbox{.}(2003){Fan}, {Strauss}, {Schneider}, {Becker},
  {White}, {Haiman}, {Gregg}, {Pentericci}, {Grebel}, {Narayanan}, {Loh},
  {Richards}, {Gunn}, {Lupton}, {Knapp}, {Ivezi{\'c}}, {Brandt}, {Collinge},
  {Hao}, {Harbeck}, {Prada}, {Schaye}, {Strateva}, {Zakamska}, {Anderson},
  {Brinkmann}, {Bahcall}, {Lamb}, {Okamura}, {Szalay}, \& {York}}]{fan03}
{Fan} X. {et~al.}, 2003, \aj, 125, 1649

\bibitem[{{Fialkov} \& {Barkana}(2014)}]{fialkov14}
{Fialkov} A., {Barkana} R., 2014, \mnras, 445, 213

\bibitem[{{Fialkov}, {Barkana} \& {Cohen}(2015){Fialkov}, {Barkana}, \&
  {Cohen}}]{fialkov15}
{Fialkov} A., {Barkana} R., {Cohen} A., 2015, Physical Review Letters, 114,
  101303

\bibitem[{{Friedrich} {et~al}\mbox{.}(2011){Friedrich}, {Mellema}, {Alvarez},
  {Shapiro}, \& {Iliev}}]{friedrich11}
{Friedrich} M.~M., {Mellema} G., {Alvarez} M.~A., {Shapiro} P.~R., {Iliev}
  I.~T., 2011, \mnras, 413, 1353

\bibitem[{{Furlanetto} \& {Oh}(2016)}]{furlanetto16}
{Furlanetto} S.~R., {Oh} S.~P., 2016, \mnras, 457, 1813

\bibitem[{{Furlanetto}, {Oh} \& {Briggs}(2006){Furlanetto}, {Oh}, \&
  {Briggs}}]{furlanetto06}
{Furlanetto} S.~R., {Oh} S.~P., {Briggs} F.~H., 2006, Physics Reports, 433, 181

\bibitem[{{Furlanetto} \& {Pritchard}(2006)}]{furlanetto06a}
{Furlanetto} S.~R., {Pritchard} J.~R., 2006, \mnras, 372, 1093

\bibitem[{{Furlanetto}, {Zaldarriaga} \& {Hernquist}(2004){Furlanetto},
  {Zaldarriaga}, \& {Hernquist}}]{furlanetto04b}
{Furlanetto} S.~R., {Zaldarriaga} M., {Hernquist} L., 2004, \apj, 613, 1

\bibitem[{{Ghara}, {Choudhury} \& {Datta}(2015){Ghara}, {Choudhury}, \&
  {Datta}}]{ghara14}
{Ghara} R., {Choudhury} T.~R., {Datta} K.~K., 2015, \mnras, 447, 1806

\bibitem[{{Ghara}, {Datta} \& {Choudhury}(2015){Ghara}, {Datta}, \&
  {Choudhury}}]{ghara15}
{Ghara} R., {Datta} K.~K., {Choudhury} T.~R., 2015, \mnras, 453, 3143

\bibitem[{{Ghosh} {et~al}\mbox{.}(2012){Ghosh}, {Prasad}, {Bharadwaj}, {Ali},
  \& {Chengalur}}]{ghosh12}
{Ghosh} A., {Prasad} J., {Bharadwaj} S., {Ali} S.~S., {Chengalur} J.~N., 2012,
  \mnras, 426, 3295

\bibitem[{{Goto} {et~al}\mbox{.}(2011){Goto}, {Utsumi}, {Hattori}, {Miyazaki},
  \& {Yamauchi}}]{goto11}
{Goto} T., {Utsumi} Y., {Hattori} T., {Miyazaki} S., {Yamauchi} C., 2011,
  \mnras, 415, L1

\bibitem[{{Greig} \& {Mesinger}(2015)}]{greig15}
{Greig} B., {Mesinger} A., 2015, \mnras, 449, 4246

\bibitem[{{Greig} \& {Mesinger}(2017)}]{greig17}
{Greig} B., {Mesinger} A., 2017, ArXiv e-prints:1705.03471

\bibitem[{{Greig}, {Mesinger} \& {Koopmans}(2015){Greig}, {Mesinger}, \&
  {Koopmans}}]{greig15b}
{Greig} B., {Mesinger} A., {Koopmans} L.~V.~E., 2015, ArXiv e-prints:1509.03312

\bibitem[{{Harker} {et~al}\mbox{.}(2009){Harker}, {Zaroubi}, {Thomas},
  {Jeli{\'c}}, {Labropoulos}, {Mellema}, {Iliev}, {Bernardi}, {Brentjens}, {de
  Bruyn}, {Ciardi}, {Koopmans}, {Pandey}, {Pawlik}, {Schaye}, \&
  {Yatawatta}}]{harker09}
{Harker} G.~J.~A. {et~al.}, 2009, \mnras, 393, 1449

\bibitem[{{Iliev} {et~al}\mbox{.}(2015){Iliev}, {Santos}, {Mesinger},
  {Majumdar}, \& {Mellema}}]{iliev15}
{Iliev} I., {Santos} M., {Mesinger} A., {Majumdar} S., {Mellema} G., 2015,
  Advancing Astrophysics with the Square Kilometre Array (AASKA14), 7

\bibitem[{{Jeli{\'c}} {et~al}\mbox{.}(2014){Jeli{\'c}}, {de Bruyn}, {Mevius},
  {Abdalla}, {Asad}, {Bernardi}, {Brentjens}, {Bus}, {Chapman}, {Ciardi},
  {Daiboo}, {Fernandez}, {Ghosh}, {Harker}, {Jensen}, {Kazemi}, {Koopmans},
  {Labropoulos}, {Martinez-Rubi}, {Mellema}, {Offringa}, {Pandey}, {Patil},
  {Thomas}, {Vedantham}, {Veligatla}, {Yatawatta}, {Zaroubi}, {Alexov},
  {Anderson}, {Avruch}, {Beck}, {Bell}, {Bentum}, {Best}, {Bonafede},
  {Bregman}, {Breitling}, {Broderick}, {Brouw}, {Br{\"u}ggen}, {Butcher},
  {Conway}, {de Gasperin}, {de Geus}, {Deller}, {Dettmar}, {Duscha},
  {Eisl{\"o}ffel}, {Engels}, {Falcke}, {Fallows}, {Fender}, {Ferrari},
  {Frieswijk}, {Garrett}, {Grie{\ss}meier}, {Gunst}, {Hamaker}, {Hassall},
  {Haverkorn}, {Heald}, {Hessels}, {Hoeft}, {H{\"o}randel}, {Horneffer}, {van
  der Horst}, {Iacobelli}, {Juette}, {Karastergiou}, {Kondratiev}, {Kramer},
  {Kuniyoshi}, {Kuper}, {van Leeuwen}, {Maat}, {Mann}, {McKay-Bukowski},
  {McKean}, {Munk}, {Nelles}, {Norden}, {Paas}, {Pandey-Pommier}, {Pietka},
  {Pizzo}, {Polatidis}, {Reich}, {R{\"o}ttgering}, {Rowlinson}, {Scaife},
  {Schwarz}, {Serylak}, {Smirnov}, {Steinmetz}, {Stewart}, {Tagger}, {Tang},
  {Tasse}, {ter Veen}, {Thoudam}, {Toribio}, {Vermeulen}, {Vocks}, {van
  Weeren}, {Wijers}, {Wijnholds}, {Wucknitz}, \& {Zarka}}]{jelic14}
{Jeli{\'c}} V. {et~al.}, 2014, \aap, 568, A101

\bibitem[{{Jeli{\'c}} {et~al}\mbox{.}(2008){Jeli{\'c}}, {Zaroubi},
  {Labropoulos}, {Thomas}, {Bernardi}, {Brentjens}, {de Bruyn}, {Ciardi},
  {Harker}, {Koopmans}, {Pandey}, {Schaye}, \& {Yatawatta}}]{jelic08}
{Jeli{\'c}} V. {et~al.}, 2008, \mnras, 389, 1319

\bibitem[{{Jensen} {et~al}\mbox{.}(2013{\natexlab{a}}){Jensen}, {Datta},
  {Mellema}, {Chapman}, {Abdalla}, {Iliev}, {Mao}, {Santos}, {Shapiro},
  {Zaroubi}, {Bernardi}, {Brentjens}, {de Bruyn}, {Ciardi}, {Harker},
  {Jeli{\'c}}, {Kazemi}, {Koopmans}, {Labropoulos}, {Martinez}, {Offringa},
  {Pandey}, {Schaye}, {Thomas}, {Veligatla}, {Vedantham}, \&
  {Yatawatta}}]{jensen13}
{Jensen} H. {et~al.}, 2013{\natexlab{a}}, \mnras, 435, 460

\bibitem[{{Jensen} {et~al}\mbox{.}(2013{\natexlab{b}}){Jensen}, {Laursen},
  {Mellema}, {Iliev}, {Sommer-Larsen}, \& {Shapiro}}]{jensen13b}
{Jensen} H., {Laursen} P., {Mellema} G., {Iliev} I.~T., {Sommer-Larsen} J.,
  {Shapiro} P.~R., 2013{\natexlab{b}}, \mnras, 428, 1366

\bibitem[{{Komatsu} {et~al}\mbox{.}(2011){Komatsu}, {Smith}, {Dunkley},
  {Bennett}, {Gold}, {Hinshaw}, {Jarosik}, {Larson}, {Nolta}, {Page},
  {Spergel}, {Halpern}, {Hill}, {Kogut}, {Limon}, {Meyer}, {Odegard}, {Tucker},
  {Weiland}, {Wollack}, \& {Wright}}]{komatsu11}
{Komatsu} E. {et~al.}, 2011, \apjs, 192, 18

\bibitem[{{Koopmans} {et~al}\mbox{.}(2015){Koopmans}, {Pritchard}, {Mellema},
  {Aguirre}, {Ahn}, {Barkana}, {van Bemmel}, {Bernardi}, {Bonaldi}, {Briggs},
  {de Bruyn}, {Chang}, {Chapman}, {Chen}, {Ciardi}, {Dayal}, {Ferrara},
  {Fialkov}, {Fiore}, {Ichiki}, {Illiev}, {Inoue}, {Jelic}, {Jones}, {Lazio},
  {Maio}, {Majumdar}, {Mack}, {Mesinger}, {Morales}, {Parsons}, {Pen},
  {Santos}, {Schneider}, {Semelin}, {de Souza}, {Subrahmanyan}, {Takeuchi},
  {Vedantham}, {Wagg}, {Webster}, {Wyithe}, {Datta}, \& {Trott}}]{koopmans15}
{Koopmans} L. {et~al.}, 2015, Advancing Astrophysics with the Square Kilometre
  Array (AASKA14), 1

\bibitem[{{Kubota} {et~al}\mbox{.}(2016){Kubota}, {Yoshiura}, {Shimabukuro}, \&
  {Takahashi}}]{kubota16}
{Kubota} K., {Yoshiura} S., {Shimabukuro} H., {Takahashi} K., 2016, \pasj, 68,
  61

\bibitem[{{La Plante} {et~al}\mbox{.}(2014){La Plante}, {Battaglia},
  {Natarajan}, {Peterson}, {Trac}, {Cen}, \& {Loeb}}]{laplante14}
{La Plante} P., {Battaglia} N., {Natarajan} A., {Peterson} J.~B., {Trac} H.,
  {Cen} R., {Loeb} A., 2014, \apj, 789, 31

\bibitem[{{Lidz} {et~al}\mbox{.}(2007){Lidz}, {Zahn}, {McQuinn}, {Zaldarriaga},
  {Dutta}, \& {Hernquist}}]{lidz07}
{Lidz} A., {Zahn} O., {McQuinn} M., {Zaldarriaga} M., {Dutta} S., {Hernquist}
  L., 2007, \apj, 659, 865

\bibitem[{{Lidz} {et~al}\mbox{.}(2008){Lidz}, {Zahn}, {McQuinn}, {Zaldarriaga},
  \& {Hernquist}}]{lidz08}
{Lidz} A., {Zahn} O., {McQuinn} M., {Zaldarriaga} M., {Hernquist} L., 2008,
  \apj, 680, 962

\bibitem[{{Majumdar}, {Bharadwaj} \& {Choudhury}(2013){Majumdar}, {Bharadwaj},
  \& {Choudhury}}]{majumdar13}
{Majumdar} S., {Bharadwaj} S., {Choudhury} T.~R., 2013, \mnras, 434, 1978

\bibitem[{{Majumdar} {et~al}\mbox{.}(2016{\natexlab{a}}){Majumdar}, {Datta},
  {Ghara}, {Mondal}, {Choudhury}, {Bharadwaj}, {Ali}, \& {Datta}}]{majumdar16b}
{Majumdar} S., {Datta} K.~K., {Ghara} R., {Mondal} R., {Choudhury} T.~R.,
  {Bharadwaj} S., {Ali} S.~S., {Datta} A., 2016{\natexlab{a}}, Journal of
  Astrophysics and Astronomy, 37, 32

\bibitem[{{Majumdar} {et~al}\mbox{.}(2016{\natexlab{b}}){Majumdar}, {Jensen},
  {Mellema}, {Chapman}, {Abdalla}, {Lee}, {Iliev}, {Dixon}, {Datta}, {Ciardi},
  {Fernandez}, {Jeli{\'c}}, {Koopmans}, \& {Zaroubi}}]{majumdar16}
{Majumdar} S. {et~al.}, 2016{\natexlab{b}}, \mnras, 456, 2080

\bibitem[{{Majumdar} {et~al}\mbox{.}(2014){Majumdar}, {Mellema}, {Datta},
  {Jensen}, {Choudhury}, {Bharadwaj}, \& {Friedrich}}]{majumdar14}
{Majumdar} S., {Mellema} G., {Datta} K.~K., {Jensen} H., {Choudhury} T.~R.,
  {Bharadwaj} S., {Friedrich} M.~M., 2014, \mnras, 443, 2843

\bibitem[{{Mao} {et~al}\mbox{.}(2012){Mao}, {Shapiro}, {Mellema}, {Iliev},
  {Koda}, \& {Ahn}}]{mao12}
{Mao} Y., {Shapiro} P.~R., {Mellema} G., {Iliev} I.~T., {Koda} J., {Ahn} K.,
  2012, \mnras, 422, 926

\bibitem[{{McQuinn} {et~al}\mbox{.}(2007){McQuinn}, {Lidz}, {Zahn}, {Dutta},
  {Hernquist}, \& {Zaldarriaga}}]{mcquinn07}
{McQuinn} M., {Lidz} A., {Zahn} O., {Dutta} S., {Hernquist} L., {Zaldarriaga}
  M., 2007, \mnras, 377, 1043

\bibitem[{{McQuinn} {et~al}\mbox{.}(2006){McQuinn}, {Zahn}, {Zaldarriaga},
  {Hernquist}, \& {Furlanetto}}]{mcquinn06}
{McQuinn} M., {Zahn} O., {Zaldarriaga} M., {Hernquist} L., {Furlanetto} S.~R.,
  2006, \apj, 653, 815

\bibitem[{{Mellema} {et~al}\mbox{.}(2015){Mellema}, {Koopmans}, {Shukla},
  {Datta}, {Mesinger}, \& {Majumdar}}]{mellema15}
{Mellema} G., {Koopmans} L., {Shukla} H., {Datta} K.~K., {Mesinger} A.,
  {Majumdar} S., 2015, Advancing Astrophysics with the Square Kilometre Array
  (AASKA14), 10

\bibitem[{{Mellema} {et~al}\mbox{.}(2013){Mellema}, {Koopmans}, {Abdalla},
  {Bernardi}, {Ciardi}, {Daiboo}, {de Bruyn}, {Datta}, {Falcke}, {Ferrara},
  {Iliev}, {Iocco}, {Jeli{\'c}}, {Jensen}, {Joseph}, {Labroupoulos}, {Meiksin},
  {Mesinger}, {Offringa}, {Pandey}, {Pritchard}, {Santos}, {Schwarz},
  {Semelin}, {Vedantham}, {Yatawatta}, \& {Zaroubi}}]{mellema13}
{Mellema} G. {et~al.}, 2013, Experimental Astronomy, 36, 235

\bibitem[{{Mesinger} \& {Furlanetto}(2007)}]{mesinger07}
{Mesinger} A., {Furlanetto} S., 2007, \apj, 669, 663

\bibitem[{{Mitra}, {Choudhury} \& {Ferrara}(2015){Mitra}, {Choudhury}, \&
  {Ferrara}}]{mitra15}
{Mitra} S., {Choudhury} T.~R., {Ferrara} A., 2015, \mnras, 454, L76

\bibitem[{{Mitra}, {Ferrara} \& {Choudhury}(2013){Mitra}, {Ferrara}, \&
  {Choudhury}}]{mitra13}
{Mitra} S., {Ferrara} A., {Choudhury} T.~R., 2013, \mnras, 428, L1

\bibitem[{Mondal, Bharadwaj \& Datta(2017)Mondal, Bharadwaj, \&
  Datta}]{mondal17b}
Mondal R., Bharadwaj S., Datta K.~K., 2017, \mnras, stx2888

\bibitem[{{Mondal}, {Bharadwaj} \& {Majumdar}(2016){Mondal}, {Bharadwaj}, \&
  {Majumdar}}]{mondal16}
{Mondal} R., {Bharadwaj} S., {Majumdar} S., 2016, \mnras, 456, 1936

\bibitem[{{Mondal}, {Bharadwaj} \& {Majumdar}(2017){Mondal}, {Bharadwaj}, \&
  {Majumdar}}]{mondal17}
{Mondal} R., {Bharadwaj} S., {Majumdar} S., 2017, \mnras, 464, 2992

\bibitem[{{Mondal} {et~al}\mbox{.}(2015){Mondal}, {Bharadwaj}, {Majumdar},
  {Bera}, \& {Acharyya}}]{mondal15}
{Mondal} R., {Bharadwaj} S., {Majumdar} S., {Bera} A., {Acharyya} A., 2015,
  \mnras, 449, L41

\bibitem[{{Morales}(2005)}]{morales05}
{Morales} M.~F., 2005, \apj, 619, 678

\bibitem[{{Ota} {et~al}\mbox{.}(2017){Ota}, {Iye}, {Kashikawa}, {Konno},
  {Nakata}, {Totani}, {Kobayashi}, {Fudamoto}, {Seko}, {Toshikawa}, {Ichikawa},
  {Shibuya}, \& {Onoue}}]{ota17}
{Ota} K. {et~al.}, 2017, \apj, 844, 85

\bibitem[{{Ouchi} {et~al}\mbox{.}(2010){Ouchi}, {Shimasaku}, {Furusawa},
  {Saito}, {Yoshida}, {Akiyama}, {Ono}, {Yamada}, {Ota}, {Kashikawa}, {Iye},
  {Kodama}, {Okamura}, {Simpson}, \& {Yoshida}}]{ouchi10}
{Ouchi} M. {et~al.}, 2010, \apj, 723, 869

\bibitem[{{Paciga} {et~al}\mbox{.}(2013){Paciga}, {Albert}, {Bandura}, {Chang},
  {Gupta}, {Hirata}, {Odegova}, {Pen}, {Peterson}, {Roy}, {Shaw}, {Sigurdson},
  \& {Voytek}}]{paciga13}
{Paciga} G. {et~al.}, 2013, \mnras, 433, 639

\bibitem[{{Parsons} {et~al}\mbox{.}(2014){Parsons}, {Liu}, {Aguirre}, {Ali},
  {Bradley}, {Carilli}, {DeBoer}, {Dexter}, {Gugliucci}, {Jacobs}, {Klima},
  {MacMahon}, {Manley}, {Moore}, {Pober}, {Stefan}, \& {Walbrugh}}]{parsons14}
{Parsons} A.~R. {et~al.}, 2014, \apj, 788, 106

\bibitem[{{Patil} {et~al}\mbox{.}(2017){Patil}, {Yatawatta}, {Koopmans}, {de
  Bruyn}, {Brentjens}, {Zaroubi}, {Asad}, {Hatef}, {Jeli{\'c}}, {Mevius},
  {Offringa}, {Pandey}, {Vedantham}, {Abdalla}, {Brouw}, {Chapman}, {Ciardi},
  {Gehlot}, {Ghosh}, {Harker}, {Iliev}, {Kakiichi}, {Majumdar}, {Mellema},
  {Silva}, {Schaye}, {Vrbanec}, \& {Wijnholds}}]{patil17}
{Patil} A.~H. {et~al.}, 2017, \apj, 838, 65

\bibitem[{{Patil} {et~al}\mbox{.}(2014){Patil}, {Zaroubi}, {Chapman},
  {Jeli{\'c}}, {Harker}, {Abdalla}, {Asad}, {Bernardi}, {Brentjens}, {de
  Bruyn}, {Bus}, {Ciardi}, {Daiboo}, {Fernandez}, {Ghosh}, {Jensen}, {Kazemi},
  {Koopmans}, {Labropoulos}, {Mevius}, {Martinez}, {Mellema}, {Offringa},
  {Pandey}, {Schaye}, {Thomas}, {Vedantham}, {Veligatla}, {Wijnholds}, \&
  {Yatawatta}}]{patil14}
{Patil} A.~H. {et~al.}, 2014, \mnras, 443, 1113

\bibitem[{{Planck Collaboration} {et~al}\mbox{.}(2016){Planck Collaboration},
  {Adam}, {Aghanim}, {Ashdown}, {Aumont}, {Baccigalupi}, {Ballardini},
  {Banday}, {Barreiro}, {Bartolo}, {Basak}, {Battye}, {Benabed}, {Bernard},
  {Bersanelli}, {Bielewicz}, {Bock}, {Bonaldi}, {Bonavera}, {Bond}, {Borrill},
  {Bouchet}, {Boulanger}, {Bucher}, {Burigana}, {Calabrese}, {Cardoso},
  {Carron}, {Chiang}, {Colombo}, {Combet}, {Comis}, {Couchot}, {Coulais},
  {Crill}, {Curto}, {Cuttaia}, {Davis}, {de Bernardis}, {de Rosa}, {de Zotti},
  {Delabrouille}, {Di Valentino}, {Dickinson}, {Diego}, {Dor{\'e}}, {Douspis},
  {Ducout}, {Dupac}, {Elsner}, {En{\ss}lin}, {Eriksen}, {Falgarone}, {Fantaye},
  {Finelli}, {Forastieri}, {Frailis}, {Fraisse}, {Franceschi}, {Frolov},
  {Galeotta}, {Galli}, {Ganga}, {G{\'e}nova-Santos}, {Gerbino}, {Ghosh},
  {Gonz{\'a}lez-Nuevo}, {G{\'o}rski}, {Gruppuso}, {Gudmundsson}, {Hansen},
  {Helou}, {Henrot-Versill{\'e}}, {Herranz}, {Hivon}, {Huang}, {Ili{\'c}},
  {Jaffe}, {Jones}, {Keih{\"a}nen}, {Keskitalo}, {Kisner}, {Knox},
  {Krachmalnicoff}, {Kunz}, {Kurki-Suonio}, {Lagache}, {L{\"a}hteenm{\"a}ki},
  {Lamarre}, {Langer}, {Lasenby}, {Lattanzi}, {Lawrence}, {Le Jeune},
  {Levrier}, {Lewis}, {Liguori}, {Lilje}, {L{\'o}pez-Caniego}, {Ma},
  {Mac{\'{\i}}as-P{\'e}rez}, {Maggio}, {Mangilli}, {Maris}, {Martin},
  {Mart{\'{\i}}nez-Gonz{\'a}lez}, {Matarrese}, {Mauri}, {McEwen}, {Meinhold},
  {Melchiorri}, {Mennella}, {Migliaccio}, {Miville-Desch{\^e}nes}, {Molinari},
  {Moneti}, {Montier}, {Morgante}, {Moss}, {Naselsky}, {Natoli}, {Oxborrow},
  {Pagano}, {Paoletti}, {Partridge}, {Patanchon}, {Patrizii}, {Perdereau},
  {Perotto}, {Pettorino}, {Piacentini}, {Plaszczynski}, {Polastri}, {Polenta},
  {Puget}, {Rachen}, {Racine}, {Reinecke}, {Remazeilles}, {Renzi}, {Rocha},
  {Rossetti}, {Roudier}, {Rubi{\~n}o-Mart{\'{\i}}n}, {Ruiz-Granados},
  {Salvati}, {Sandri}, {Savelainen}, {Scott}, {Sirri}, {Sunyaev}, {Suur-Uski},
  {Tauber}, {Tenti}, {Toffolatti}, {Tomasi}, {Tristram}, {Trombetti},
  {Valiviita}, {Van Tent}, {Vielva}, {Villa}, {Vittorio}, {Wandelt}, {Wehus},
  {White}, {Zacchei}, \& {Zonca}}]{planck16}
{Planck Collaboration} {et~al.}, 2016, \aap, 596, A108

\bibitem[{{Planck Collaboration} {et~al}\mbox{.}(2014){Planck Collaboration},
  {Ade}, {Aghanim}, {Armitage-Caplan}, {Arnaud}, {Ashdown}, {Atrio-Barandela},
  {Aumont}, {Baccigalupi}, {Banday}, \& et~al.}]{planck14}
{Planck Collaboration} {et~al.}, 2014, \aap, 571, A16

\bibitem[{{Pober} {et~al}\mbox{.}(2014){Pober}, {Liu}, {Dillon}, {Aguirre},
  {Bowman}, {Bradley}, {Carilli}, {DeBoer}, {Hewitt}, {Jacobs}, {McQuinn},
  {Morales}, {Parsons}, {Tegmark}, \& {Werthimer}}]{pober14}
{Pober} J.~C. {et~al.}, 2014, \apj, 782, 66

\bibitem[{{Pritchard} \& {Furlanetto}(2007)}]{pritchard07}
{Pritchard} J.~R., {Furlanetto} S.~R., 2007, \mnras, 376, 1680

\bibitem[{{Pritchard} \& {Loeb}(2008)}]{pritchard08}
{Pritchard} J.~R., {Loeb} A., 2008, \prd, 78, 103511

\bibitem[{{Pritchard} \& {Loeb}(2012)}]{pritchard12}
{Pritchard} J.~R., {Loeb} A., 2012, Reports on Progress in Physics, 75, 086901

\bibitem[{{Robertson} {et~al}\mbox{.}(2015){Robertson}, {Ellis}, {Furlanetto},
  \& {Dunlop}}]{robertson15}
{Robertson} B.~E., {Ellis} R.~S., {Furlanetto} S.~R., {Dunlop} J.~S., 2015,
  \apjl, 802, L19

\bibitem[{{Ross} {et~al}\mbox{.}(2017){Ross}, {Dixon}, {Iliev}, \&
  {Mellema}}]{ross17}
{Ross} H.~E., {Dixon} K.~L., {Iliev} I.~T., {Mellema} G., 2017, \mnras, 468,
  3785

\bibitem[{{Saiyad Ali}, {Bharadwaj} \& {Pandey}(2006){Saiyad Ali}, {Bharadwaj},
  \& {Pandey}}]{ali06}
{Saiyad Ali} S., {Bharadwaj} S., {Pandey} S.~K., 2006, \mnras, 366, 213

\bibitem[{{Schmit} \& {Pritchard}(2018)}]{schmit17}
{Schmit} C.~J., {Pritchard} J.~R., 2018, \mnras, 475, 1213

\bibitem[{{Shimabukuro} {et~al}\mbox{.}(2015){Shimabukuro}, {Yoshiura},
  {Takahashi}, {Yokoyama}, \& {Ichiki}}]{shimabukuro15a}
{Shimabukuro} H., {Yoshiura} S., {Takahashi} K., {Yokoyama} S., {Ichiki} K.,
  2015, \mnras, 451, 467

\bibitem[{{Shimabukuro} {et~al}\mbox{.}(2016){Shimabukuro}, {Yoshiura},
  {Takahashi}, {Yokoyama}, \& {Ichiki}}]{shimabukuro16}
{Shimabukuro} H., {Yoshiura} S., {Takahashi} K., {Yokoyama} S., {Ichiki} K.,
  2016, \mnras, 458, 3003

\bibitem[{{Shimabukuro} {et~al}\mbox{.}(2017){Shimabukuro}, {Yoshiura},
  {Takahashi}, {Yokoyama}, \& {Ichiki}}]{shimabukuro16b}
{Shimabukuro} H., {Yoshiura} S., {Takahashi} K., {Yokoyama} S., {Ichiki} K.,
  2017, \mnras, 468, 1542

\bibitem[{{Songaila} \& {Cowie}(2010)}]{songaila10}
{Songaila} A., {Cowie} L.~L., 2010, \apj, 721, 1448

\bibitem[{{Tingay} {et~al}\mbox{.}(2013){Tingay}, {Goeke}, {Bowman}, {Emrich},
  {Ord}, {Mitchell}, {Morales}, {Booler}, {Crosse}, {Wayth}, {Lonsdale},
  {Tremblay}, {Pallot}, {Colegate}, {Wicenec}, {Kudryavtseva}, {Arcus},
  {Barnes}, {Bernardi}, {Briggs}, {Burns}, {Bunton}, {Cappallo}, {Corey},
  {Deshpande}, {Desouza}, {Gaensler}, {Greenhill}, {Hall}, {Hazelton}, {Herne},
  {Hewitt}, {Johnston-Hollitt}, {Kaplan}, {Kasper}, {Kincaid}, {Koenig},
  {Kratzenberg}, {Lynch}, {Mckinley}, {Mcwhirter}, {Morgan}, {Oberoi},
  {Pathikulangara}, {Prabu}, {Remillard}, {Rogers}, {Roshi}, {Salah}, {Sault},
  {Udaya-Shankar}, {Schlagenhaufer}, {Srivani}, {Stevens}, {Subrahmanyan},
  {Waterson}, {Webster}, {Whitney}, {Williams}, {Williams}, \&
  {Wyithe}}]{tingay13}
{Tingay} S.~J. {et~al.}, 2013, \pasa, 30, 7

\bibitem[{{Trenti} {et~al}\mbox{.}(2010){Trenti}, {Stiavelli}, {Bouwens},
  {Oesch}, {Shull}, {Illingworth}, {Bradley}, \& {Carollo}}]{trenti10}
{Trenti} M., {Stiavelli} M., {Bouwens} R.~J., {Oesch} P., {Shull} J.~M.,
  {Illingworth} G.~D., {Bradley} L.~D., {Carollo} C.~M., 2010, \apjl, 714, L202

\bibitem[{{van Haarlem} {et~al}\mbox{.}(2013){van Haarlem}, {Wise}, {Gunst},
  {Heald}, {McKean}, {Hessels}, {de Bruyn}, {Nijboer}, {Swinbank}, {Fallows},
  {Brentjens}, {Nelles}, {Beck}, {Falcke}, {Fender}, {H{\"o}randel},
  {Koopmans}, {Mann}, {Miley}, {R{\"o}ttgering}, {Stappers}, {Wijers},
  {Zaroubi}, {van den Akker}, {Alexov}, {Anderson}, {Anderson}, {van Ardenne},
  {Arts}, {Asgekar}, {Avruch}, {Batejat}, {B{\"a}hren}, {Bell}, {Bell}, {van
  Bemmel}, {Bennema}, {Bentum}, {Bernardi}, {Best}, {B{\^i}rzan}, {Bonafede},
  {Boonstra}, {Braun}, {Bregman}, {Breitling}, {van de Brink}, {Broderick},
  {Broekema}, {Brouw}, {Br{\"u}ggen}, {Butcher}, {van Cappellen}, {Ciardi},
  {Coenen}, {Conway}, {Coolen}, {Corstanje}, {Damstra}, {Davies}, {Deller},
  {Dettmar}, {van Diepen}, {Dijkstra}, {Donker}, {Doorduin}, {Dromer}, {Drost},
  {van Duin}, {Eisl{\"o}ffel}, {van Enst}, {Ferrari}, {Frieswijk}, {Gankema},
  {Garrett}, {de Gasperin}, {Gerbers}, {de Geus}, {Grie{\ss}meier}, {Grit},
  {Gruppen}, {Hamaker}, {Hassall}, {Hoeft}, {Holties}, {Horneffer}, {van der
  Horst}, {van Houwelingen}, {Huijgen}, {Iacobelli}, {Intema}, {Jackson},
  {Jelic}, {de Jong}, {Juette}, {Kant}, {Karastergiou}, {Koers}, {Kollen},
  {Kondratiev}, {Kooistra}, {Koopman}, {Koster}, {Kuniyoshi}, {Kramer},
  {Kuper}, {Lambropoulos}, {Law}, {van Leeuwen}, {Lemaitre}, {Loose}, {Maat},
  {Macario}, {Markoff}, {Masters}, {McFadden}, {McKay-Bukowski}, {Meijering},
  {Meulman}, {Mevius}, {Middelberg}, {Millenaar}, {Miller-Jones}, {Mohan},
  {Mol}, {Morawietz}, {Morganti}, {Mulcahy}, {Mulder}, {Munk}, {Nieuwenhuis},
  {van Nieuwpoort}, {Noordam}, {Norden}, {Noutsos}, {Offringa}, {Olofsson},
  {Omar}, {Orr{\'u}}, {Overeem}, {Paas}, {Pandey-Pommier}, {Pandey}, {Pizzo},
  {Polatidis}, {Rafferty}, {Rawlings}, {Reich}, {de Reijer}, {Reitsma},
  {Renting}, {Riemers}, {Rol}, {Romein}, {Roosjen}, {Ruiter}, {Scaife}, {van
  der Schaaf}, {Scheers}, {Schellart}, {Schoenmakers}, {Schoonderbeek},
  {Serylak}, {Shulevski}, {Sluman}, {Smirnov}, {Sobey}, {Spreeuw}, {Steinmetz},
  {Sterks}, {Stiepel}, {Stuurwold}, {Tagger}, {Tang}, {Tasse}, {Thomas},
  {Thoudam}, {Toribio}, {van der Tol}, {Usov}, {van Veelen}, {van der Veen},
  {ter Veen}, {Verbiest}, {Vermeulen}, {Vermaas}, {Vocks}, {Vogt}, {de Vos},
  {van der Wal}, {van Weeren}, {Weggemans}, {Weltevrede}, {White}, {Wijnholds},
  {Wilhelmsson}, {Wucknitz}, {Yatawatta}, {Zarka}, {Zensus}, \& {van
  Zwieten}}]{haarlem13}
{van Haarlem} M.~P. {et~al.}, 2013, \aap, 556, A2

\bibitem[{{Wang} {et~al}\mbox{.}(2013){Wang}, {Xu}, {An}, {Gu}, {Guo}, {Li},
  {Wang}, {Liu}, {Martineau-Huynh}, \& {Wu}}]{wang2013}
{Wang} J. {et~al.}, 2013, \apj, 763, 90

\bibitem[{{Watkinson} {et~al}\mbox{.}(2017){Watkinson}, {Majumdar},
  {Pritchard}, \& {Mondal}}]{watkinson17}
{Watkinson} C.~A., {Majumdar} S., {Pritchard} J.~R., {Mondal} R., 2017, \mnras,
  472, 2436

\bibitem[{{Watkinson} \& {Pritchard}(2014)}]{watkinson14}
{Watkinson} C.~A., {Pritchard} J.~R., 2014, \mnras, 443, 3090

\bibitem[{{Watkinson} \& {Pritchard}(2015)}]{watkinson15}
{Watkinson} C.~A., {Pritchard} J.~R., 2015, \mnras, 454, 1416

\bibitem[{{White} {et~al}\mbox{.}(2003){White}, {Becker}, {Fan}, \&
  {Strauss}}]{white03}
{White} R.~L., {Becker} R.~H., {Fan} X., {Strauss} M.~A., 2003, \aj, 126, 1

\bibitem[{{Yatawatta} {et~al}\mbox{.}(2013){Yatawatta}, {de Bruyn},
  {Brentjens}, {Labropoulos}, {Pandey}, {Kazemi}, {Zaroubi}, {Koopmans},
  {Offringa}, {Jeli{\'c}}, {Martinez Rubi}, {Veligatla}, {Wijnholds}, {Brouw},
  {Bernardi}, {Ciardi}, {Daiboo}, {Harker}, {Mellema}, {Schaye}, {Thomas},
  {Vedantham}, {Chapman}, {Abdalla}, {Alexov}, {Anderson}, {Avruch}, {Batejat},
  {Bell}, {Bell}, {Bentum}, {Best}, {Bonafede}, {Bregman}, {Breitling}, {van de
  Brink}, {Broderick}, {Br{\"u}ggen}, {Conway}, {de Gasperin}, {de Geus},
  {Duscha}, {Falcke}, {Fallows}, {Ferrari}, {Frieswijk}, {Garrett},
  {Griessmeier}, {Gunst}, {Hassall}, {Hessels}, {Hoeft}, {Iacobelli}, {Juette},
  {Karastergiou}, {Kondratiev}, {Kramer}, {Kuniyoshi}, {Kuper}, {van Leeuwen},
  {Maat}, {Mann}, {McKean}, {Mevius}, {Mol}, {Munk}, {Nijboer}, {Noordam},
  {Norden}, {Orru}, {Paas}, {Pandey-Pommier}, {Pizzo}, {Polatidis}, {Reich},
  {R{\"o}ttgering}, {Sluman}, {Smirnov}, {Stappers}, {Steinmetz}, {Tagger},
  {Tang}, {Tasse}, {ter Veen}, {Vermeulen}, {van Weeren}, {Wise}, {Wucknitz},
  \& {Zarka}}]{yatawatta13}
{Yatawatta} S. {et~al.}, 2013, \aap, 550, A136

\bibitem[{{Yoshiura} {et~al}\mbox{.}(2015){Yoshiura}, {Shimabukuro},
  {Takahashi}, {Momose}, {Nakanishi}, \& {Imai}}]{yoshiura14}
{Yoshiura} S., {Shimabukuro} H., {Takahashi} K., {Momose} R., {Nakanishi} H.,
  {Imai} H., 2015, \mnras, 451, 266

\bibitem[{{Zahn} {et~al}\mbox{.}(2007){Zahn}, {Lidz}, {McQuinn}, {Dutta},
  {Hernquist}, {Zaldarriaga}, \& {Furlanetto}}]{zahn07}
{Zahn} O., {Lidz} A., {McQuinn} M., {Dutta} S., {Hernquist} L., {Zaldarriaga}
  M., {Furlanetto} S.~R., 2007, \apj, 654, 12

\bibitem[{{Zaldarriaga}, {Furlanetto} \& {Hernquist}(2004){Zaldarriaga},
  {Furlanetto}, \& {Hernquist}}]{zaldarriaga04}
{Zaldarriaga} M., {Furlanetto} S.~R., {Hernquist} L., 2004, \apj, 608, 622

\bibitem[{{Zawada} {et~al}\mbox{.}(2014){Zawada}, {Semelin}, {Vonlanthen},
  {Baek}, \& {Revaz}}]{zawada14}
{Zawada} K., {Semelin} B., {Vonlanthen} P., {Baek} S., {Revaz} Y., 2014,
  \mnras, 439, 1615

\bibitem[{{Zheng} {et~al}\mbox{.}(2017){Zheng}, {Wang}, {Rhoads}, {Infante},
  {Malhotra}, {Hu}, {Walker}, {Jiang}, {Jiang}, {Hibon}, {Gonzalez}, {Kong},
  {Zheng}, {Galaz}, \& {Barrientos}}]{zheng17}
{Zheng} Z.-Y. {et~al.}, 2017, \apjl, 842, L22

\end{thebibliography}

\end{document}